\documentclass[11pt,letterpaper,english,3p, preprint]{elsarticle}
\usepackage[T1]{fontenc}
\usepackage[latin9]{inputenc}
\usepackage{verbatim}
\usepackage{mathrsfs}
\usepackage{bm}
\usepackage{amsmath}
\usepackage{amssymb}
\usepackage{stackrel}
\usepackage{graphicx}
\usepackage{esint}

\makeatletter

\pdfpageheight\paperheight
\pdfpagewidth\paperwidth

\numberwithin{equation}{section}

\usepackage{mathrsfs}
\usepackage{amsbsy}

\makeatother

\usepackage{babel}
\begin{document}

\begin{frontmatter}{}

\title{Finite grid instability and spectral fidelity of the electrostatic
Particle-In-Cell algorithm}

\author[lanl]{C.-K. Huang\corref{cor}}

\ead{huangck@lanl.gov}

\author[lanl]{Y. Zeng}

\author[lanl]{Y. Wang}

\author[lanl,ucla]{M.D. Meyers}

\author[lanl]{S. Yi}

\author[lanl]{B. J. Albright}

\cortext[cor]{Corresponding author}

\address[lanl]{Los Alamos National Laboratory, Los Alamos, NM, USA 87545}

\address[ucla]{University of California, Los Angeles, CA, USA 90095}
\begin{abstract}
The origin of the Finite Grid Instability (FGI) is studied by resolving
the dynamics in the 1D electrostatic Particle-In-Cell (PIC) model
in the spectral domain at the single particle level and at the collective
motion level. The spectral fidelity of the PIC model is contrasted
with the underlying physical system or the gridless model. The systematic
spectral phase and amplitude errors from the charge deposition and
field interpolation are quantified for common particle shapes used
in the PIC models. It is shown through such analysis and in simulations
that the lack of spectral fidelity relative to the physical system
due to the existence of aliased spatial modes is the major cause of
the FGI in the PIC model.\end{abstract}
\begin{keyword}
Finite Grid Instability, Particle-In-Cell, spectral fidelity, numerical
instability
\end{keyword}

\end{frontmatter}{}

\newpage{}

\section{Introduction\label{sec:Introduction}}

N-body type problems arise in many disciplines and underpins our understanding
of complex dynamical systems like plasmas and the cosmos. In a typical
N-body problem, the interaction in particle pairs can be of electrostatic,
or electromagnetic, or gravitational in nature and each particle responds
to a force that is the linear superposition of all one-to-one interactions
it receives. Direct calculation of all one-to-one interactions of
$N_{p}$ particles has a computation cost of $O(N_{p}^{2})$, therefore
it is amenable to numerical simulation only when $N_{p}$ is small.
The PIC method \cite{Birdsall-Langdon}, or more generally the particle-mesh
method, is an efficient numerical method that reduces the computation
complexity by introducing a computation grid and taking advantage
of the linearity in the sum of one-to-one interactions. In the PIC
method, the interaction among the particles is mediated by the grid
through the Green\textquoteright s function of the interaction represented
on the grid. The computation complexity is reduced from $O(N_{p}^{2})$
to $O(N_{g})+O(N_{p})$, where $N_{g}$ is the number of grid points.
When the number of particles per cell $N_{p}/N_{g}\gg1$, the gain
in speedup is large ($\sim N_{p}$), therefore the PIC method is a
popular choice in the \textit{ab-initio} numerical simulation of
N-body systems. However, two major problems arise in the PIC method
due to the discrete grid: (1) the use of an Eulerian grid for the
moments of the particle distribution and fields, in conjunction with
individual Lagrangian particles in continuous phase space, implies
an inherent inconsistency; (2) the grid representation of the Green\textquoteright s
function is usually an approximation of the real Green\textquoteright s
function in the continuous space. 

Despite the computation efficiency of the PIC method and its wide-spread
use, especially in plasma physics, common PIC models are vulnerable
to an electrostatic numerical instability known as the Finite Grid
Instability (FGI) \cite{Langdon1970,OKUDA1972} (there is also an
electromagnetic instability known as the numerical Cherenkov instability
\cite{GODFREY1974,Godfrey2013,Xu20132503}). Early practitioners using
the electrostatic PIC model to simulate plasma dynamics observed a
heating effect to the plasma which depends on the numerical parameters,
i.e., the grid size $\Delta x$ and the number of particles per cell
$N_{C}$. This numerical heating has been extensively studied since
the early development of the PIC model and the empirical scaling of
the heating time $\tau_{H}$ of FGI in a thermal plasma, which has
the form $\omega_{p}\tau_{H}\sim(\lambda_{D}/\Delta x)^{2}(N_{D}+N_{C}$),
is summarized in \cite{Birdsall-Langdon}. Here $N_{D}$ is the number
of particles in a Debye length $\lambda_{D}$, $\omega_{p}$ is the
plasma frequency. It is also known that FGI comes from the aliased
modes in the system due to the incompatibility between the Fourier
spectra of the discrete Eulerian and continuous Lagrangian variables.
(Numerical Cherenkov instability may also come from an Eulerian-Lagrangian
mismatch in the convective derivative \cite{GODFREY1974}). The numerical
instabilities have been conventionally analyzed as unphysical resonances
between physical and aliased modes \cite{Birdsall-Langdon,GODFREY1974,Godfrey2013,Xu20132503,Lindman1969,Meyers2015}.
The locations and growth rates of the unstable modes have been solved
for using linear dispersion analysis in limited, yet essential cases,
i.e., for spatially uniform cold or Maxwellian distributions. It is
worth noting that, unlike the two-stream type instabilities, for some
commonly used electrostatic and electromagnetic PIC models, FGI can
arise without the intersection of the physical and aliased modes \cite{Birdsall1980,Meyers2015}.
Analysis for more realistic and nonlinear simulations has not been
carried out. 

Various methods and numerical schemes to mitigate FGI have been proposed,
including introducing grid interlacing \cite{Chen1974} and random
jiggling \cite{Chen1974,Brackbill1994}, employing higher order particle
shapes \cite{Birdsall1980} and temporal/spatial filtering \cite{Lewis1972},
implicit time differencing and enforcing the energy conservation property
of the numerical algorithm between time steps \cite{Lewis1970,Chen2011,Markidis2011,Taitano2013}.
All these techniques have shown great promise with regard to the reduction
of the instability growth rate. However, this is often achieved with
substantial distortion/damping of the meaningful dynamics at the short
wavelength scale or by sacrificing conservation properties (such as
the loss of momentum conservation in an energy conserving algorithm,
which has long been debated in the development of the PIC models \cite{Birdsall-Langdon}).
Recent rigorous work on energy conserving algorithm has led to a large
improvement of the momentum conservation through nonlinear iterations
\cite{Chen2011}. 

The above efforts notwithstanding, the important questions about how
and where FGI arises exactly remain to be answered. Previous works
treat aliased modes as inherent in the system and study their properties
and corresponding mitigation method. However, the origin of the numerical
instabilities is clearly unphysical. Therefore, in principle a simulation
plasma should be contrasted with the underlying continuous system,
which has the same number of particles and particle shape as the simulation
plasma, to determine the origin of the numerical instabilities. We
call latter the physical system in the following, as it obeys a Vlasov
equation for the shaped particles, as long as the same particle shape
is used in defining the charge density from the particle and the electric
field on the particle. 

There are many choices about what to be contrasted between the PIC
and the physical systems. Conservation properties and dispersion relationship
have been used. It should be noted that one direct consequence of
FGI in a PIC model without built-in energy (momentum) conservation
property is, as can be expected, the gross violation of the energy
(momentum) conservation. For this reason, recent efforts have been
devoted to improve the energy and/or momentum conservation of the
numerical scheme in order to control FGI. But it should be emphasized
that conservation laws are desirable when eliminating the FGI, but
they are neither necessary nor sufficient conditions. An isolated
plasma system can exhibit various kinds of physical instabilities
while strictly conserving momentum and energy. Furthermore, as total
energy is a global property of the underlying microscopic processes
and only one constraint on the degrees of freedom (two if total momentum
is also considered) in the phase space, to understand what gives rise
to FGI and its consequences, we need a better resolution into the
dynamics. Linear dispersion, in which the eigen modes with complex
Fourier frequency can be viewed as a way to resolve the (linearized)
dynamics, is a better choice. However, such analysis is limited to
special cases of the particle distribution and small perturbation
amplitude. Insight from such analysis for the improvement to the numerical
scheme is useful but not easy to obtained and applied to more general
situations. Recently, symplectic PIC codes \cite{Xiao2013,Xiao2015}
have also been developed, for which the symplectic structures of the
Hamiltonian system is preserved. The symplectic structure may be a
good choice to contrast the PIC and the physical systems, however,
it is not clear how it is related to the numerical instabilities at
present.

In this paper, we will study the FGI in the 1D electrostatic PIC models
by spectrally resolving the dynamics at the single particle level,
thus allowing us to identify the components in the model that lead
to unphysical instability. The dynamics in PIC result from the superposition
of the pair-wise interactions as in a physical system. The major components
of an electrostatic PIC model --- the charge deposition, the field
interpolation and the particle pusher, all operate on a single particle,
while the field solver can be viewed as operating on the spatial Fourier
modes. Therefore the use of the particle and spectral resolutions
are natural choices for this purpose. Such a representation of the
PIC models is given in section \ref{sec:Spectral-representation-PIC}
and the spectral errors in PIC models are analyzed in section \ref{sec:Spatial-error-PIC}.
As an alternative to the individual particle representation, one can
also choose the modes of the collective particle motion as a representation.
This has the advantage that the plasma dynamics can then be viewed
as the collective wave-particle interactions and such couplings in
a physical system and in a PIC model can also be contrasted. We note
that the deposition, field interpolation and field solver only involve
spatial operations at a fixed time, while the particle update in the
pusher is a temporal operation in continuous space whose stability
and convergence can be verified to rule out its role in FGI. To facilitate
simulation comparison with the physical system, the gridless model
\cite{Vlad2001,Decyk,Evstatiev2013} is used, in which all components
and elementary operations of the physical system are projected onto
the finite Fourier basis. It is demonstrated in section \ref{sec:Comparison-gridless-PIC}
that the lack of spectral fidelity in the deposition, field interpolation
is the major cause of the FGI. Finally we summarize in section \ref{sec:Summary}.

\section{Spectral representation of the PIC model\label{sec:Spectral-representation-PIC}}

\subsection{Charge deposition}

Let's first look at the charge deposition scheme in Fourier space
to understand the effect of aliasing. We will see that the most important
effect in Fourier space is the summation over all Brillouin zones
which is the result of a convolution process between a continuous
spectrum and a periodic spectrum over Brillouin zones. The sampling
needed to go from continuous space to a discrete grid is the cause
of the latter spectrum.

In a grid-based model like PIC, the contribution of a particle at
position $\bm{x_{p}}$ \footnote{Italic bold font is used for vectors.}
and of total charge $Q$ on the density grid $\bm{r_{\rho}}$ is $\rho(\bm{r_{\rho}})=QW(\bm{r_{\rho}},\,\bm{x_{p}})$,
where $\bm{r_{\rho}}$ is a vector on a uniform grid with grid size
$\Delta\bm{x}=(\Delta x,\,\Delta y,\,\Delta z)$. The interpolation
function for the deposition is $W(\bm{r_{\rho}},\,\bm{r})=W(|\bm{r_{\rho}}-\bm{r}|)$.
Note that we have assumed the cell volume $V=\Delta x\Delta y\Delta z=1$
and dropped it for clarity. The difference between a particle shape
function $S(\bm{r})$ and a interpolation function $W(\bm{r_{\rho}},\,\bm{r})$
is discussed in \ref{sub:A1}. In the rest of this section, we will
not distinguish these two and will use $S(\bm{r})$ for clarity, $\rho(\bm{r_{\rho}})=QS_{\rho}(\bm{r_{\rho}}-\bm{x_{p}})$.
We define a transform (note this is not necessarily the proper Discrete
Fourier Transform as $\bm{r_{\rho}}$ may be shifted from the origin
of the coordinate system, as will be shown later),

\begin{equation}
\tilde{\rho}(\bm{k})=\underset{\bm{r_{\rho}}}{\sum}\rho(\bm{r_{\rho}})e^{-i\bm{k}\cdot\bm{r_{\rho}}}=Q\underset{\bm{r_{\rho}}}{\sum}S_{\rho}(\bm{r_{\rho}}-\bm{x_{p}})e^{-i\bm{k}\cdot\bm{r_{\rho}}},\label{eq:transform}
\end{equation}

\noindent which can be viewed as the continuous Fourier transform
of $\rho(\bm{r})\underset{\bm{r_{\rho}}}{\sum}\delta(\bm{r}-\bm{r_{\rho}})$,
where $\rho(\bm{r})=QS_{\rho}(\bm{r}-\bm{x_{p}})$. Then the non-unitary
inverse transform is 

\[
\rho(\bm{r_{\rho}})=\frac{1}{8\pi^{3}}\int_{-\infty}^{\infty}\tilde{\rho}(\bm{k})e^{i\bm{k}\cdot\bm{r_{\rho}}}d\bm{k}.
\]

As $S_{\rho}(\bm{r})$ is a function in continuous space, 
\begin{equation}
S_{\rho}(\bm{r})=\frac{1}{8\pi^{3}}\int_{-\infty}^{\infty}\tilde{S}_{\rho}(\bm{k'})e^{i\bm{k'}\cdot\bm{r}}d\bm{k'}.
\end{equation}

{} 

Then, 

\begin{equation}
\tilde{\rho}(\bm{k})=\frac{Q}{8\pi^{3}}\underset{\bm{r_{\rho}}}{\sum}(\int_{-\infty}^{\infty}\tilde{S}_{\rho}(\bm{k'})e^{i\bm{k'}\cdot(\bm{r_{\rho}}-\bm{x_{p}})}d\bm{k'})e^{-i\bm{k}\cdot\bm{r_{\rho}}}=\frac{Q}{8\pi^{3}}\int_{-\infty}^{\infty}\tilde{S}_{\rho}(\bm{k'})e^{-i\bm{k'}\cdot\bm{x_{p}}}\cdot\underset{\bm{r_{\rho}}}{\sum}e^{i(\bm{k'}-\bm{k})\cdot\bm{r_{\rho}}}\cdot d\bm{k'}.\label{eq:rho-fourier}
\end{equation}

Carrying out the infinite sum (see \ref{sub:A2}), we obtain,

\begin{equation}
\tilde{\rho}(\bm{k})=Q\underset{\mathbf{q}}{\sum}\psi(\mathbf{q}\cdot\bm{k_{g}},\,\Delta\bm{r_{\rho}})\tilde{S}_{\rho}(\bm{k_{q}})e^{-i\bm{k_{q}}\cdot\bm{x_{p}}},\label{eq:deposition-PIC}
\end{equation}

\noindent and 
\[
\tilde{\rho}(\bm{k}+\mathbf{q'}\cdot\bm{k_{g}})=\psi(-\mathbf{q'}\cdot\bm{k_{g}},\,\Delta\bm{r_{\rho}})\tilde{\rho}(\bm{k}),
\]

\noindent where 
\begin{equation}
\psi(\mathbf{q}\cdot\bm{k_{g}},\,\Delta\bm{r_{\rho}})=e^{i(\mathbf{q}\cdot\bm{k_{g}})\cdot\Delta\bm{r_{\rho}}}\label{eq:phase-factor}
\end{equation}
 is a phase factor. $\bm{r_{\rho}}=\bm{r_{f}}+\Delta\bm{r_{\rho}}$
and $\bm{r_{f}}$ is a reference grid that contains the origin of
coordinate system. $\bm{k_{q}}=\bm{k}+\mathbf{q}\cdot\bm{k_{g}}$
is an vector alias of $\bm{k}$ in the $\mathbf{q}$-th Brillouin
zone $[\mathbf{q}\cdot\bm{k_{g}}-\bm{k_{g}}/2,\,\mathbf{q}\cdot\bm{k_{g}}+\bm{k_{g}}/2)$,
$\bm{k_{g}}=\left(2\pi/\Delta x,\,2\pi/\Delta y,\:2\pi/\Delta z\right)$
and 
\[
\mathbf{q}=\left(\begin{array}{ccc}
n\\
 & m\\
 &  & l
\end{array}\right).
\]

\noindent In the case that $\Delta\bm{r_{\rho}}\ne0$, the spectra
in Brillouin zones are not necessarily the same as a result of the
transform defined in Eq. (\ref{eq:transform}).

For a physical system consisting of particles of shape $S_{\rho}(\bm{r})$,
a particle at position $\bm{x_{p}}$ will contribute to the density
as, 

\begin{equation}
\tilde{\rho}(\bm{k})=Q\tilde{S}_{\rho}(\bm{k})e^{-i\bm{k}\cdot\bm{x_{p}}},\label{eq:deposition-gridless}
\end{equation}

\noindent where $\bm{k}\in(-\infty,\,\infty)$. For a gridless model
which does charge deposition in Fourier space using truncated Fourier
basis, Eq. (\ref{eq:deposition-gridless}) is used for a truncated
domain $\bm{k}\in[-\bm{k_{c}},\,\bm{k_{c}}]$, where $\bm{k_{c}}$
is the cut-off wavenumber. To compare with a grid-based model, it
is fair to set $\bm{k_{c}=\bm{k_{g}}/2}$ so both models have the
same spectral resolution.

In Eq. (\ref{eq:deposition-PIC}) for the grid-based model and Eq.
(\ref{eq:deposition-gridless}) for the physical or the gridless model,
one can clearly see that the charge density is from the contribution
$e^{-i\bm{k}\cdot\bm{x_{p}}}$ of a point-charge particle modified
by its shape $\tilde{S}(\bm{k})$. Furthermore, the grid-based model
has the aliasing effect in Fourier space which is absent in the physical
or the gridless model, i.e., the summation over all Brillouin zones.
The Brillouin zones exist due to the discreteness of the grid. The
summation is due to the need to convolve the particle shape's continuous
spectrum with a spectrum that includes Brillouin zones. 

The ratio between $\tilde{\rho}(\bm{k})$ for the grid-based model,
Eq. (\ref{eq:deposition-PIC}), and gridless models, Eq. (\ref{eq:deposition-gridless}),
is, 

\begin{align}
(\tilde{S}_{\rho}(\bm{k}))^{-1}e^{i\bm{k}\cdot\bm{x_{p}}}\underset{\mathbf{q}}{\sum}\psi(\mathbf{q}\cdot\bm{k_{g}},\,\Delta\bm{r_{\rho}})\tilde{S}_{\rho}(\bm{k_{q}})e^{-i\bm{k_{q}}\cdot\bm{x_{p}}} & =(\tilde{S}_{\rho}(\bm{k}))^{-1}\underset{\mathbf{q}}{\sum}\tilde{S}_{\rho}(\bm{k_{q}})e^{-i(\mathbf{q}\cdot\bm{k_{g}})\cdot(\bm{x_{p}-\Delta\bm{r_{\rho}}})}\nonumber \\
 & \equiv G(\bm{x_{p}-\Delta\bm{r_{\rho}}},\,\bm{k};\:\tilde{S}_{\rho}(\bm{k})).\label{eq:phase-func-deposition}
\end{align}

\noindent $G(\bm{x},\,\bm{k};\:\tilde{S}_{\rho}(\bm{k}))$ determines
the spectral error in the deposition for a particular particle shape
$\tilde{S}_{\rho}(\bm{k})$ and will be quantified in section \ref{sec:Spatial-error-PIC}.

Eqs. (\ref{eq:deposition-PIC}) and (\ref{eq:deposition-gridless})
give the density for a single particle. For an infinite number of
particles from a distribution $f(\bm{x_{p}},\bm{v})$, we have,

\begin{equation}
\tilde{\rho}(\bm{k})=\int d\bm{v}d\bm{x_{p}}f(\bm{x_{p}},\bm{v})Q\underset{\mathbf{q}}{\sum}\psi(\mathbf{q}\cdot\bm{k_{g}},\,\Delta\bm{r_{\rho}})\tilde{S}_{\rho}(\bm{k_{q}})e^{-i\bm{k_{q}}\cdot\bm{x_{p}}}=\int d\bm{v}Q\underset{\mathbf{q}}{\sum}\psi(\mathbf{q}\cdot\bm{k_{g}},\,\Delta\bm{r_{\rho}})\tilde{S}_{\rho}(\bm{k_{q}})\tilde{f}(\bm{k_{q}},\bm{v})\label{eq:deposition-PIC-distribution}
\end{equation}

\noindent for the PIC model and, 

\begin{equation}
\tilde{\rho}(\bm{k})=\int d\bm{v}Q\tilde{S}_{\rho}(\bm{k})\tilde{f}(\bm{k},\bm{v})\label{eq:deposition-gridless-distribution}
\end{equation}

\noindent for the physical or gridless model. Apparently Eq. (\ref{eq:deposition-PIC-distribution})
and (\ref{eq:deposition-gridless-distribution}) differ in both amplitude
and phase, however, such differences depend on $\tilde{f}(\bm{k},\bm{v})$
and are difficult to quantify.

\subsection{Field solver}

For a physical system, Poisson's equation and electric field in continuous
space are

\[
\forall\bm{r}:\,\,\,-\nabla^{2}\phi(\bm{r})=\rho(\bm{r}),\:\bm{E}(\bm{r})=-\nabla\phi(\bm{r}).
\]

\noindent In Fourier space, this is completely equivalent to, 
\begin{equation}
\forall\bm{k}:\,\,\,\left|[\bm{k}]\right|^{2}\tilde{\mathbf{\phi}}(\bm{k})=\tilde{\mathbf{\rho}}(\bm{k}),\,\tilde{\bm{E}}(\bm{k})=-i[\bm{k}]\tilde{\mathbf{\phi}}(\bm{k}),
\end{equation}
 where the operator $[\bm{k}]=\bm{k}$.

In a simulation, there are many ways to solve for the field. The common
choices are,
\begin{itemize}
\item Use a spectral solver, 
\end{itemize}
\begin{equation}
\forall\bm{k}\in[-\bm{k_{g}}/2,\,\bm{k_{g}}/2]:\,\,\,\left|[\bm{k}]_{sp}\right|^{2}\tilde{\mathbf{\phi}}(\bm{k})=\tilde{\mathbf{\rho}}(\bm{k}),\,\tilde{\bm{E}}(\bm{k})=-i[\bm{k}]_{sp}\tilde{\mathbf{\phi}}(\bm{k})\label{eq:spectral-field-solve}
\end{equation}

\noindent where $[\bm{k}]_{sp}=\bm{k}$. In this case, $\bm{E}$ and
$\rho$ are collocated on the same grid which is chosen to be the
reference grid, i.e., $\bm{r_{E}}=\bm{r_{\rho}}=\bm{r_{f}}$, $\tilde{\bm{E}}(\bm{k})=\underset{\bm{r_{f}}}{\sum}\bm{E}(\bm{r_{f}})e^{-i\bm{k\cdot\bm{r_{f}}}}$
and $\tilde{\rho}(\bm{k})=\underset{\bm{r_{f}}}{\sum}\rho(\bm{r_{f}})e^{-i\bm{k\cdot\bm{r_{f}}}}$.
We also have $\tilde{\bm{E}}(\bm{k}+\bm{k}_{g})=\tilde{\bm{E}}(\bm{k})$
and $\tilde{\rho}(\bm{k}+\bm{k}_{g})=\tilde{\rho}(\bm{k})$.
\begin{itemize}
\item Use finite difference on Poisson's equation and the potential for
which the spectral representation is,
\begin{equation}
\forall\bm{k}:\,\,\,\left|[\bm{k}]_{fd}\right|^{2}\tilde{\mathbf{\phi}}(\bm{k})=\tilde{\mathbf{\rho}}(\bm{k}),\:\tilde{\bm{E}}(\bm{k})=-i[\bm{k}]_{fd}\tilde{\mathbf{\phi}}(\bm{k}).\label{eq:fd-field-solve}
\end{equation}

\end{itemize}
Here, $[\bm{k}]_{fd}=\left([k_{x}]_{fd},\,[k_{y}]_{fd},\,[k_{z}]_{fd}\right)$,
$[k_{\alpha}]_{fd}$ for $\alpha\in\{x,\,y,\,z\}$ is related to the
Fourier representation of the finite difference operator. $\tilde{\rho}(\bm{k})=\underset{\bm{r_{\rho}}}{\sum}\rho(\bm{r_{\rho}})e^{-i\bm{k\cdot r_{\rho}}}$
and $\tilde{E_{\alpha}}(\bm{k})=\underset{\bm{r}_{E_{\alpha}}}{\sum}E_{\alpha}(\bm{r}_{E_{\alpha}})e^{-i\bm{k\cdot r}_{E_{\alpha}}}$
from the transform defined in Eq. (\ref{eq:transform}), which in
turn gives the usual form of $[k_{\alpha}]_{fd}$, e.g., $[k_{x}]_{fd}=k_{x}\text{sinc}(k_{x}\Delta x/2)$.
As central differencing is typically used, $[k_{x}]_{fd},\,[k_{y}]_{fd},\,[k_{z}]_{fd}$
are purely real functions. Since $E_{\alpha}$ and $\rho$ may be
defined on different grids from the reference grid, so in general
$\tilde{E_{\alpha}}(\bm{k}+\bm{k_{g}})=\psi(-\bm{k_{g}},\,\Delta\bm{r}_{E_{\alpha}})\tilde{E_{\alpha}}(\bm{k})$
, $\tilde{\rho}(\bm{k}+\bm{k_{g}})=\psi(-\bm{k_{g}},\,\Delta\bm{r_{\rho}})\tilde{\rho}(\bm{k})$
and $\Delta\bm{r}_{E_{\alpha}}=\bm{r}_{E_{\alpha}}-\bm{r_{f}}$, $\Delta\bm{r_{\rho}}=\bm{r_{\rho}}-\bm{r_{f}}$.

Another possible choice is to use spectral solver on Poisson's equation
and finite difference on the potential, as done in the code XES1 \cite{Birdsall-Langdon}.
It is clear that none of these choices for the field solver has systematic
phase error, however, using finite difference inevitably introduces
systematic amplitude error.

\subsection{Field interpolation}

In PIC, after the field on the grid $\bm{E}(\bm{r_{E}})$ is calculated,
it is interpolated to the $j$-th particle's position according to 

\begin{equation}
\bm{\mathcal{E}}(\bm{x_{j}})=\underset{\bm{r_{E}}}{\sum}S_{E}(\bm{x_{j}}-\bm{r_{E}})\bm{E}(\bm{r_{E}}),
\end{equation}

\noindent where the particle shape function is $S_{E}(\bm{r})=\int_{-\infty}^{\infty}d\bm{k}\tilde{S}_{E}(\bm{k})e^{i\bm{k}\cdot\bm{r}}/8\pi^{3}$
and may be different than $S_{\rho}(\bm{r})$ in charge deposition.
Then,

\begin{equation}
\bm{\mathcal{E}}(\bm{x_{j}})=\frac{1}{8\pi^{3}}\int_{-\infty}^{\infty}d\bm{k}\tilde{S}_{E}(\bm{k})e^{i\bm{k}\cdot\bm{x_{j}}}\underset{\bm{r_{E}}}{\sum}\bm{E}(\bm{r_{E}})e^{-i\bm{k}\cdot\bm{r_{E}}}=\frac{1}{8\pi^{3}}\int_{-\infty}^{\infty}d\bm{k}\tilde{S}_{E}(\bm{k})e^{i\bm{k}\cdot\bm{x_{j}}}\tilde{\bm{E}}(\bm{k}).
\end{equation}

This means that $\bm{\mathcal{E}}(\bm{k})=\tilde{S}_{E}(\bm{k})\tilde{\bm{E}}(\bm{k})$,
i.e., no aliasing effect for a particular mode $\tilde{\bm{E}}(\bm{k})$.
This is consistent with the idea that only sampling leads to aliasing,
interpolation does not. Note that this view is for single mode analysis,
for the total force on a particle, we need to integrate over all $\bm{k}$,
so the summation over all Brillouin zones reappears. Generally, we
have $\tilde{E_{\alpha}}(\bm{k}+\bm{k_{g}})=\psi(-\bm{k_{g}},\,\Delta\bm{r}_{E_{\alpha}})\tilde{E_{\alpha}}(\bm{k})$
for each component $\alpha=x,\,y,\ z$, so 

\begin{equation}
\mathscr{\mathcal{E}}_{\alpha}(\bm{x_{j}})=\frac{1}{8\pi^{3}}\int_{-\bm{k_{g}/2}}^{\bm{k_{g}/2}}d\bm{k}\tilde{E_{\alpha}}(\bm{k})\underset{\mathbf{q}}{\sum}\psi(-\mathbf{q}\cdot\bm{k_{g}},\,\Delta\bm{r}_{E_{\alpha}})\tilde{S}_{E}(\bm{k_{q}})e^{i\bm{k_{q}}\cdot\bm{x_{j}}}.\label{eq:field-interpolation-PIC}
\end{equation}

In a physical model, the field on the particle can be directly reconstructed
from the Fourier components,

\begin{equation}
\bm{\mathcal{E}}(\bm{x_{j}})=\frac{1}{8\pi^{3}}\int_{-\infty}^{\infty}d\bm{k}\tilde{S}_{E}(\bm{k})e^{i\bm{k}\cdot\bm{x_{j}}}\tilde{\bm{E}}(\bm{k}).
\end{equation}

In a gridless model, only modes from the truncated Fourier basis,
i.e., $\bm{k}\in[-\bm{k_{g}}/2,\,\bm{k_{g}}/2]$, are used, 

\begin{equation}
\bm{\mathcal{E}}(\bm{x_{j}})=\frac{1}{8\pi^{3}}\int_{-\bm{k_{g}/2}}^{\bm{k_{g}/2}}d\bm{k}\tilde{S}_{E}(\bm{k})e^{i\bm{k}\cdot\bm{x_{j}}}\tilde{\bm{E}}(\bm{k}).
\end{equation}

Similar to the charge deposition, the integrands in the above equations
for the PIC, physical (or gridless) model differ by a ratio, 
\begin{align}
(\tilde{S}_{E}(\bm{k}))^{-1}e^{-i\bm{k}\cdot\bm{x_{j}}}\underset{\mathbf{q}}{\sum}\psi(-\mathbf{q}\cdot\bm{k_{g}},\,\Delta\bm{r}_{E_{\alpha}})\tilde{S}_{E}(\bm{k_{q}})e^{i\bm{k_{q}}\cdot\bm{x_{j}}} & =(\tilde{S}_{E}(\bm{k}))^{-1}\underset{\mathbf{q}}{\sum}\tilde{S}_{E}(\bm{k_{q}})e^{i(\mathbf{q}\cdot\bm{k_{g}})\cdot(\bm{x_{j}}-\Delta\bm{r}_{E_{\alpha}})}\nonumber \\
 & =G(\Delta\bm{r}_{E_{\alpha}}-\bm{x_{j}},\,\bm{k};\,\tilde{S}_{E}(\bm{k)}).\label{eq:phase-func-field-interpolation}
\end{align}

\section{Spatial phase and amplitude errors in PIC\label{sec:Spatial-error-PIC}}

Depending on the initial conditions, driving force, equilibrium etc.,
the plasma that we are trying to model may exhibit a host of unstable
modes itself. So to distinguish the physical and numerical instabilities,
one needs to look for the differences between the two systems. As
the spatial components of the PIC loop, i.e., charge deposition, field
solver and field interpolation, are now spectrally resolved, it is
informative to investigate the errors in the spectral domain involved
in these components. In fact, it turns out that all these three components
have errors in the spectral domain. Such errors can be systematic
or random in nature (e.g. from round-off error) and we will only focus
on the former.

It is particularly worthwhile to understand the role that the systematic
spectral errors play in the numerical instabilities. The phase error
plays a crucial role in the development of instabilities, and in many
cases determines whether the system is stable or not. On the other
hand, the amplitude error typically affects the instability growth
rate. In some other cases, e.g., for instabilities with amplitude
threshold or some parametric instabilities, the mode amplitude can
also determine the stability. However, most numerical instabilities
we encounter and need to mitigate in the PIC model are not known to
be of this type.

\subsection{Spectral error for shaped particle}

Next, we will quantify the phase and amplitude errors of the PIC model
compared to the gridless model, which is a computationally feasible
but costly model for approximating the physical system with both spectral
fidelity (up to the cut-off wave-number) and conservative properties. 

For the field solver, it is easy to see from Eqs. (\ref{eq:spectral-field-solve}),
(\ref{eq:fd-field-solve}) that neither the finite difference or spectral
solver introduces systematic phase error. There is systematic amplitude
error in the finite difference solver but none in the spectral solver.

In general, Eq. (\ref{eq:deposition-PIC-distribution}) indicates
that the deposited charge density has systematic amplitude and phase
errors which depend on the particle distribution function. But for
a system with finite number of shaped particles, we can further discern
these errors by resolving the contributions from individual particles
using Eq. (\ref{eq:phase-func-deposition}). As summing each individual
particle's contribution to the charge density is a linear process,
we do not expect such summation to generate additional errors other
than round-off. In addition, field interpolation is naturally written
for one particle in Eq. (\ref{eq:field-interpolation-PIC}). Therefore,
for a single particle, Eqs. (\ref{eq:phase-func-deposition}) and
(\ref{eq:phase-func-field-interpolation}) give the errors that can
be quantified for a specific particle shape. For an orthogonal coordinate
system and corresponding deposition and force interpolation, $G(\bm{x},\,\bm{k};\,\tilde{S}(\bm{k}))=\underset{\alpha=x,y,z}{\prod}g_{\alpha}(\alpha/\Delta\alpha,\,k_{\alpha};\,\tilde{s}_{\alpha}(k_{\alpha}))$
for $\tilde{S}(\bm{k})=\underset{\alpha=x,y,z}{\prod}\tilde{s}_{\alpha}(k_{\alpha})$.
For Cartesian coordinate, we only need to determine the spectral errors
for one particular dimension, e.g., $g(x/\Delta x,\,k;\,\tilde{s}(k))=\tilde{s}(k)^{-1}\underset{q}{\sum}\tilde{s}(k_{q})e^{-iqk_{g}x}$
for $x$ direction.

Since $g(x/\Delta x,\,k;\,\tilde{s}(k))$ is periodic in $x$, 
\[
g(x+1,\,k;\,\tilde{s}(k))=g(x,\,k;\,\tilde{s}(k)),
\]

\noindent and symmetric in $k$ for symmetric shape function $\tilde{s}(-k)=\tilde{s}(k)$,

\[
g(x,\,-k;\,\tilde{s}(-k))=g(x,\,k;\,\tilde{s}(k)),
\]

\noindent one only needs to determine $g(x,\,k;\,\tilde{s}(k))$ for
$x\in[0,1]$ and $k\in[0,k_{g}/2]$. For the most commonly used B-spline
particle shapes $\tilde{s}_{\mu}(k)$ of order $\mu$ and the Gaussian
function $\tilde{s}_{\mathsf{\mathcal{G}}}(k)$, the systematic spectral
errors $g(x,\,k;\,\tilde{s}(k))$ can be analytically quantified.

For B-spline particle shapes, $\tilde{s}_{\mu}(k)=[\text{sinc}(k\Delta x/2)]^{\mu+1}$,
the infinite sum in $g(x,\,k;\,\tilde{s}_{\mu})$ can be evaluated
into a compact form using the Hurwitz-Lerch transcendent $\Phi(z,s,a)=\sum_{n=0}^{\infty}z^{n}(n+a)^{-s}$
to give,

\begin{align}
g(x,\,k,\,\tilde{s}_{\mu}) & =\left(\frac{-k\Delta x}{2\pi}\right){}^{\mu+1}\left[e^{-i2\pi x}\Phi\left((-1)^{\mu+1}e^{-i2\pi x},\,\mu+1,\,1+\frac{k\Delta x}{2\pi}\right)\right.\nonumber \\
 & \left.+\Phi\left((-1)^{\mu+1}e^{i2\pi x},\,\mu+1,\,\frac{-k\Delta x}{2\pi}\right)\right].\label{eq:spect-err-B-spline}
\end{align}

Similarly, for a Gaussian function of width $\sigma\Delta x$, $s_{\mathsf{\mathcal{G}}}(x;\sigma)=e^{-x^{2}/(\sigma\Delta x)^{2}}/(\sqrt{\pi}\sigma\Delta x)$,
$\tilde{s}_{\mathsf{\mathcal{G}}}(k;\sigma)=e^{-(k\sigma\Delta x/2)^{2}}/\sqrt{2\pi}$,
the infinite sum in $g(x,\,k;\,\tilde{s}_{\mathsf{\mathcal{G}}})$
can be evaluated using the elliptic theta function $\vartheta_{3}(u,q)=1+2\sum_{n=1}^{\infty}q^{n^{2}}\cos(2nu)$
to give,

\begin{equation}
g(x,\,k,\,\tilde{s}_{\mathsf{\mathcal{G}}})=\frac{1}{\sqrt{\pi}\sigma}e^{\sigma{}^{2}\left(k\Delta x/2+ix/\sigma^{2}\right)^{2}}\text{\ensuremath{\vartheta_{3}}}\left(k\Delta x/2+ix/\sigma^{2},\,e^{-1/\sigma^{2}}\right).\label{eq:spect-err-Gaussian}
\end{equation}

\begin{figure}
\includegraphics[width=0.45\columnwidth]{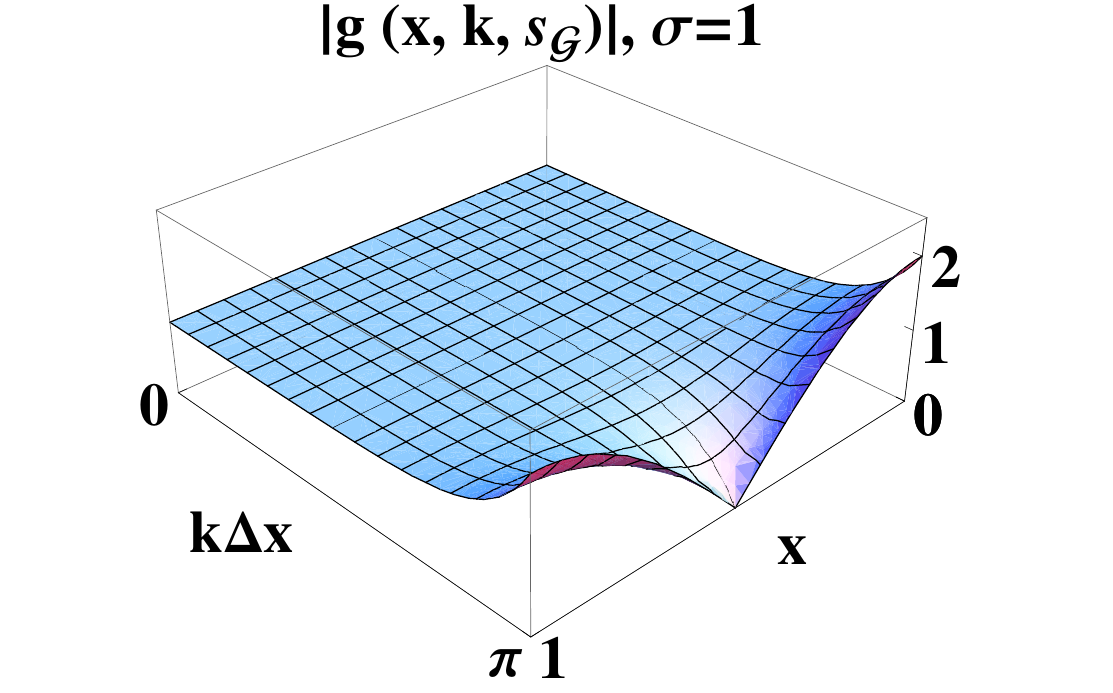}\includegraphics[width=0.45\columnwidth]{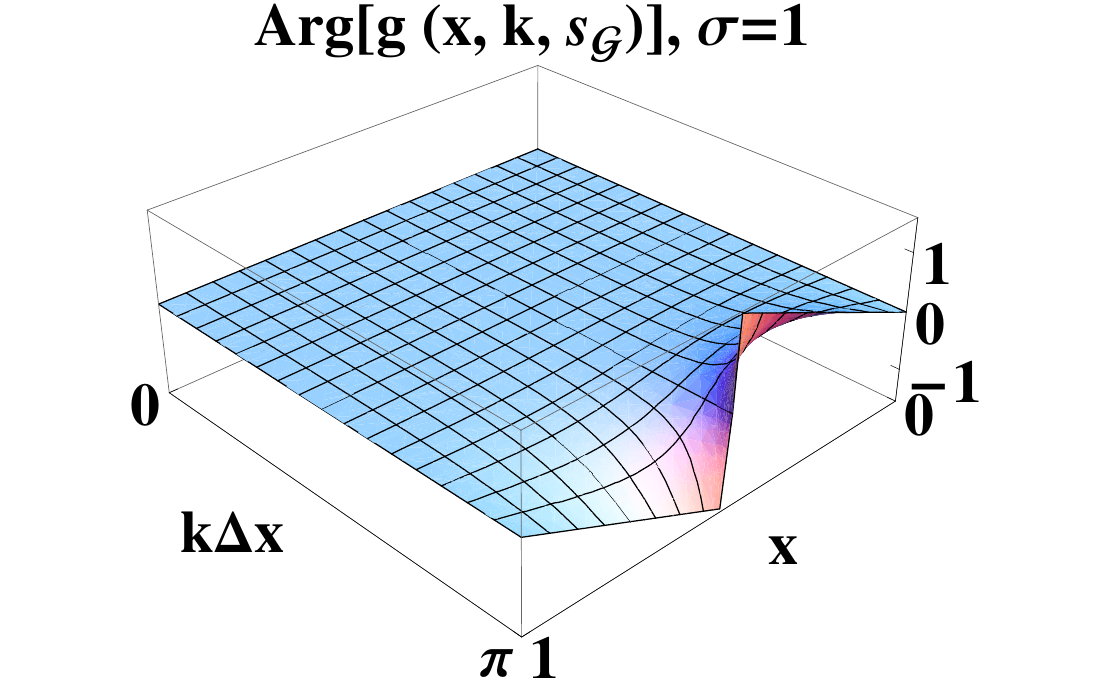}

\includegraphics[width=0.45\columnwidth]{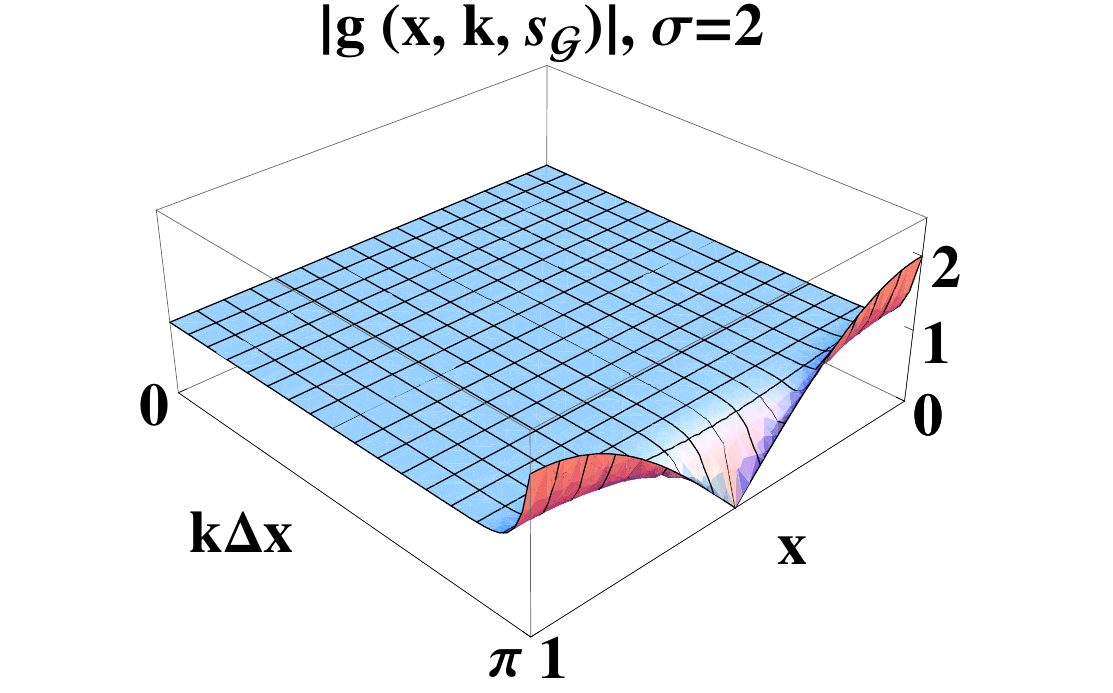}\includegraphics[width=0.45\columnwidth]{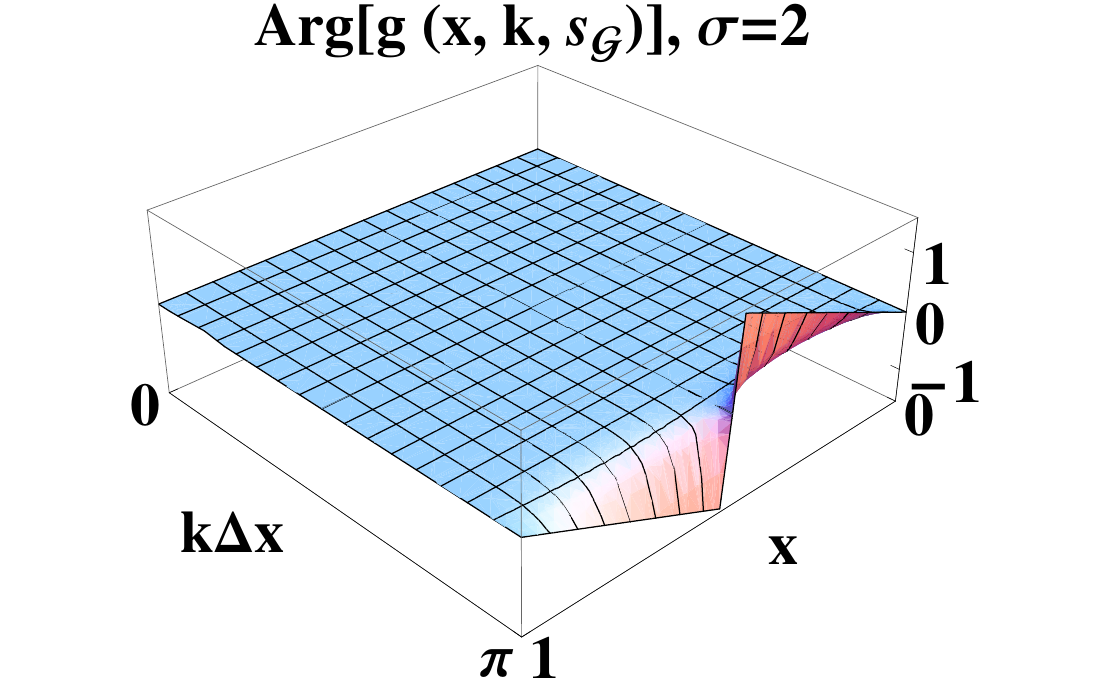}

\includegraphics[width=0.45\columnwidth]{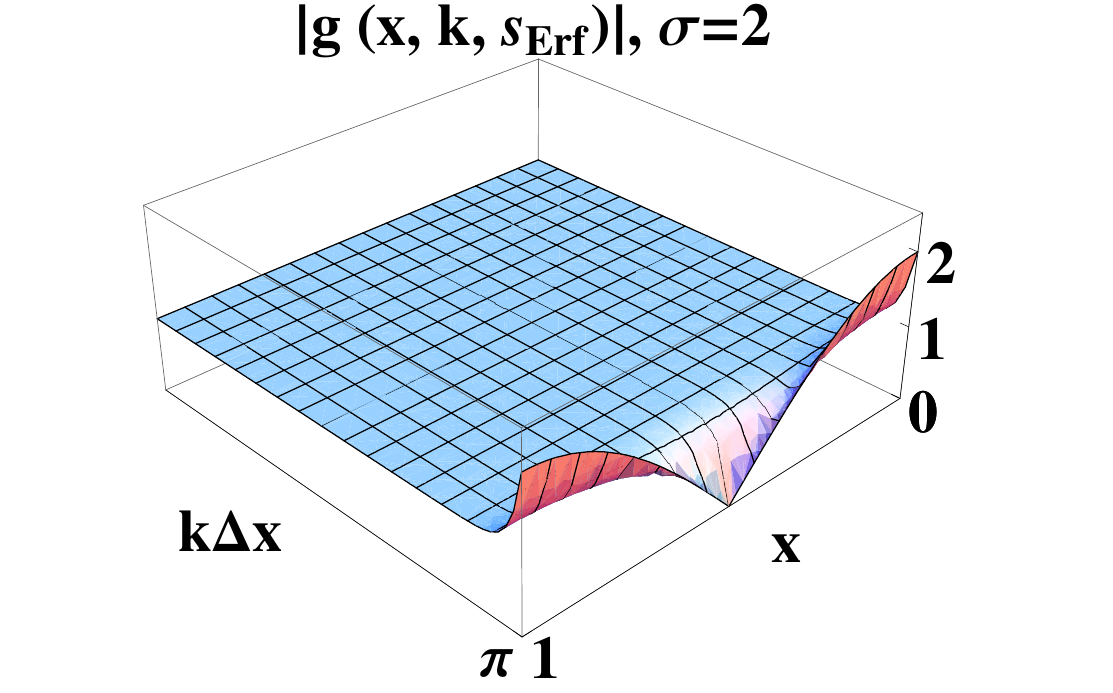}\includegraphics[width=0.45\columnwidth]{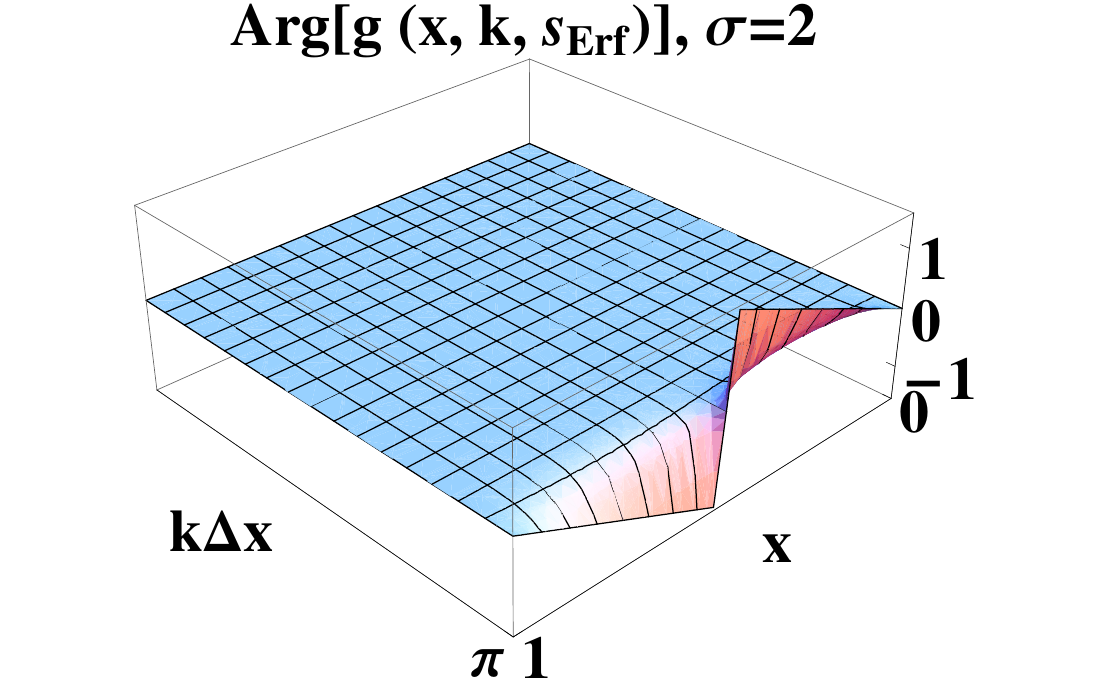}

\caption{The amplitude (left column) and phase (right column) of $g(x,\,k;\,\tilde{s}_{\mathsf{\mathcal{G}}})$
for the Gaussian shape $s_{\mathsf{\mathcal{G}}}(x;\Delta x)$ (top
row), $s_{\mathsf{\mathcal{G}}}(x;2\Delta x)$ (middle row) and of
$g(x,\,k;\,\tilde{s}_{Erf})$ for the $s_{Erf}(x;2\Delta x)$ shape
(bottom row), respectively. \label{fig:Gaussian-spectral-distortion}}
\end{figure}

Although $s_{\mathsf{\mathcal{G}}}(x;\sigma)$ is a valid particle
shape but not a valid interpolation function, one can define a new
function $\tilde{s}(k)=\tilde{s}_{\mathsf{\mathcal{G}}}(k;\sigma)\tilde{s}_{\mu}(k)$
such that the corresponding $s(x)$ is a valid interpolation function.
For example, for $\mu=0$, we have $\tilde{s}_{Erf}(k;\sigma)\equiv\tilde{s}_{\mathsf{\mathcal{G}}}(k;\sigma)\tilde{s}_{0}(k)=e^{-(k\sigma\Delta x/2)^{2}}\text{sinc}(k\Delta x/2)/\sqrt{2\pi}$
and $s_{Erf}(x;\sigma)=\left[\text{Erf}(\frac{1}{2\sigma}-\frac{x}{4\sigma\Delta x})+\text{Erf}(\frac{1}{2\sigma}+\frac{x}{4\sigma\Delta x})\right]/(2\Delta x)$
, where $\text{Erf}(x)$ is the error function. It is not clear whether
$g(x,\,k;\,\tilde{s}_{Erf})$ has a compact analytic form but numerical
evaluation in Fig. \ref{fig:Gaussian-spectral-distortion} shows $g(x,\,k;\,\tilde{s}_{Erf})$
is similar to $g(x,\,k;\,\tilde{s}_{\mathsf{\mathcal{G}}})$. 

\begin{figure}
\includegraphics[width=0.45\columnwidth]{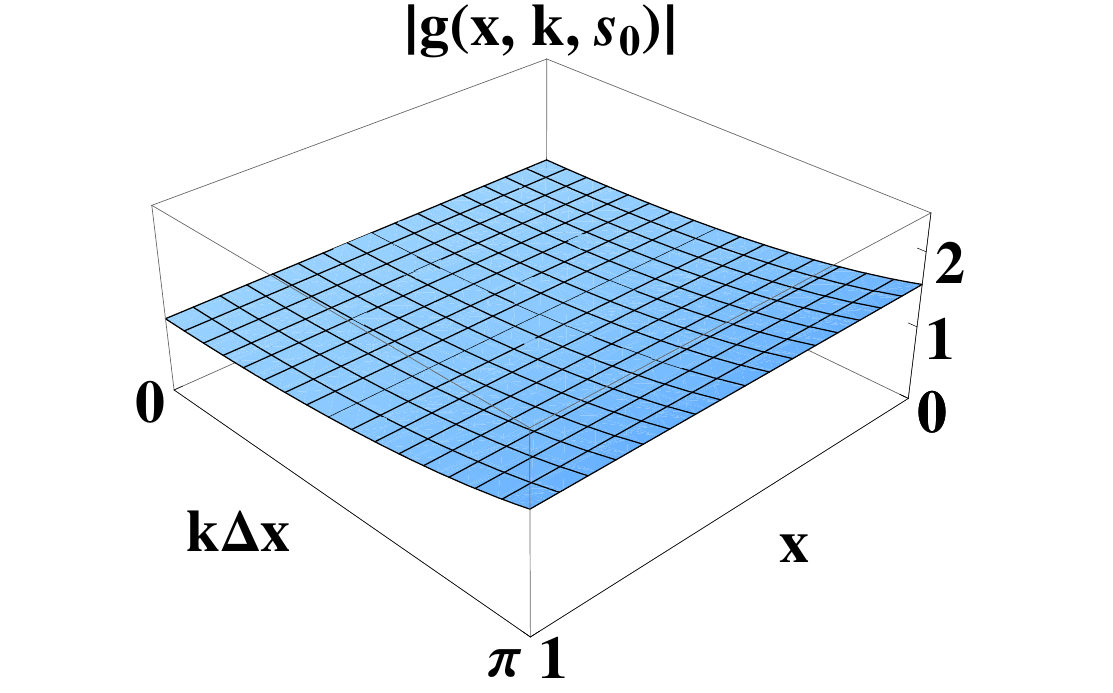}\includegraphics[width=0.45\columnwidth]{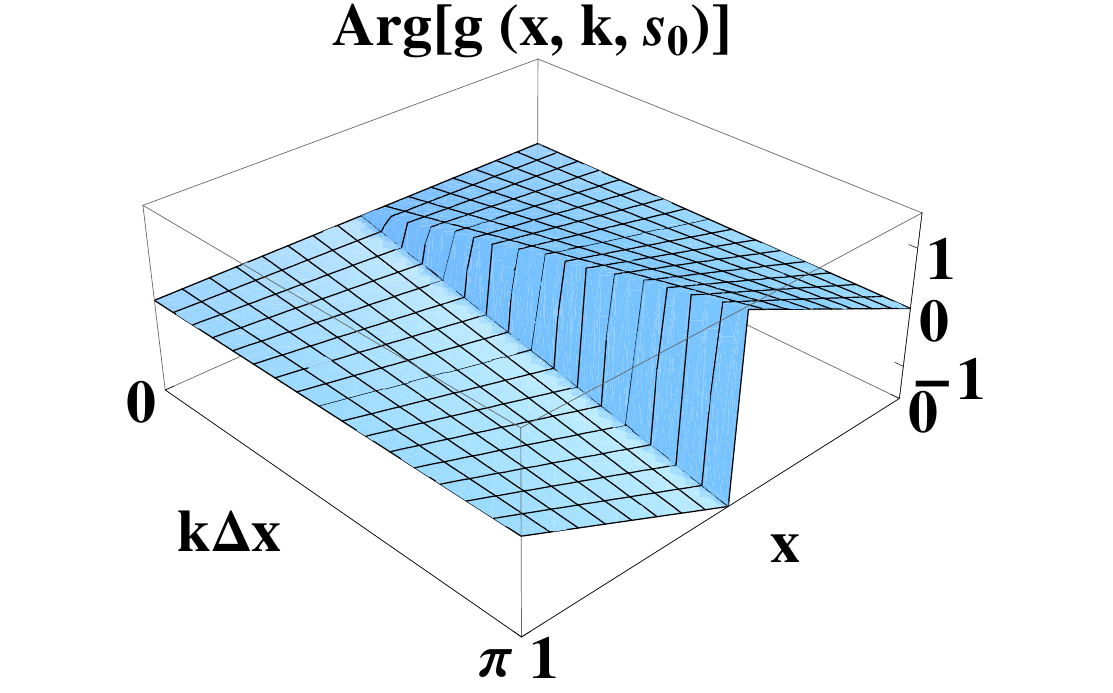}

\includegraphics[width=0.45\columnwidth]{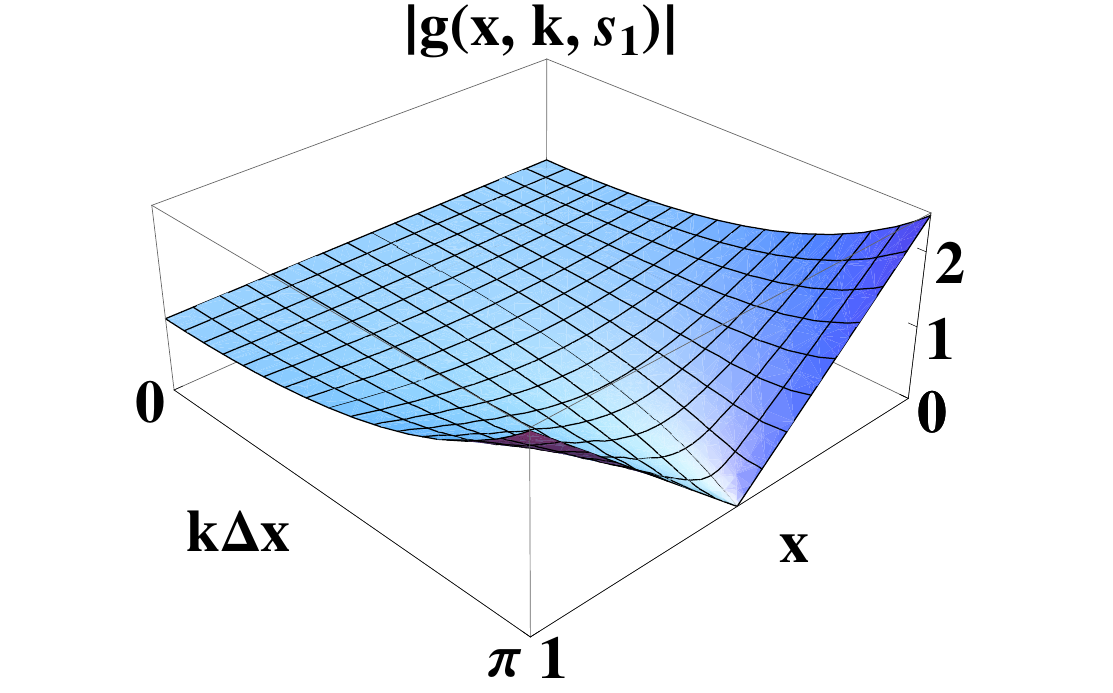}\includegraphics[width=0.45\columnwidth]{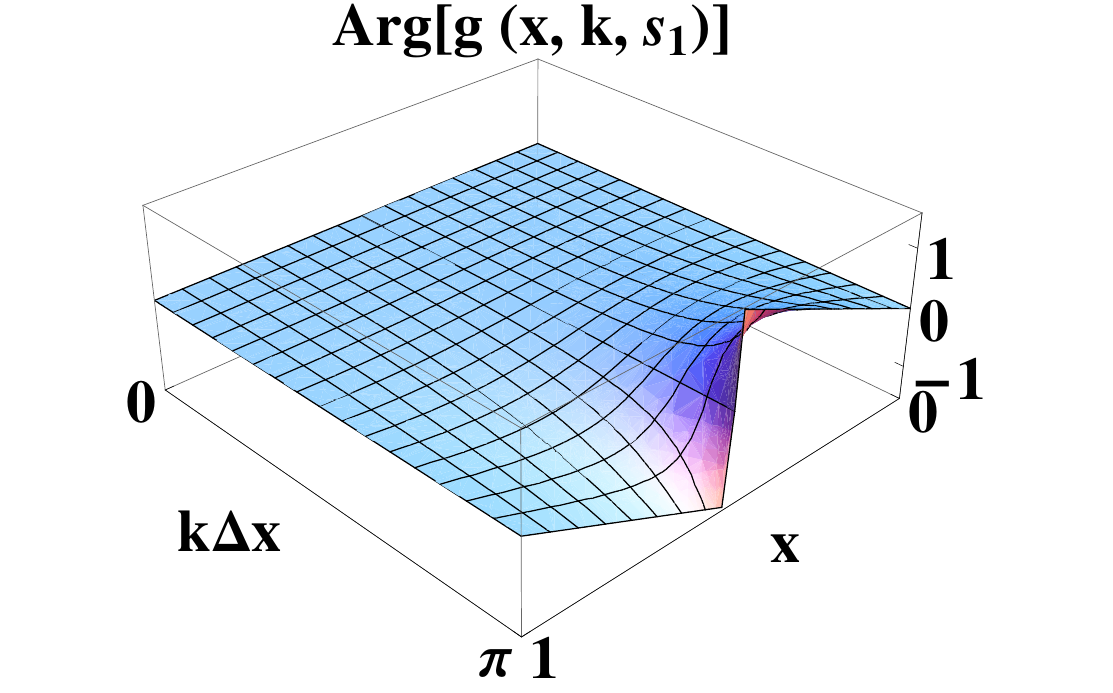}

\includegraphics[width=0.45\columnwidth]{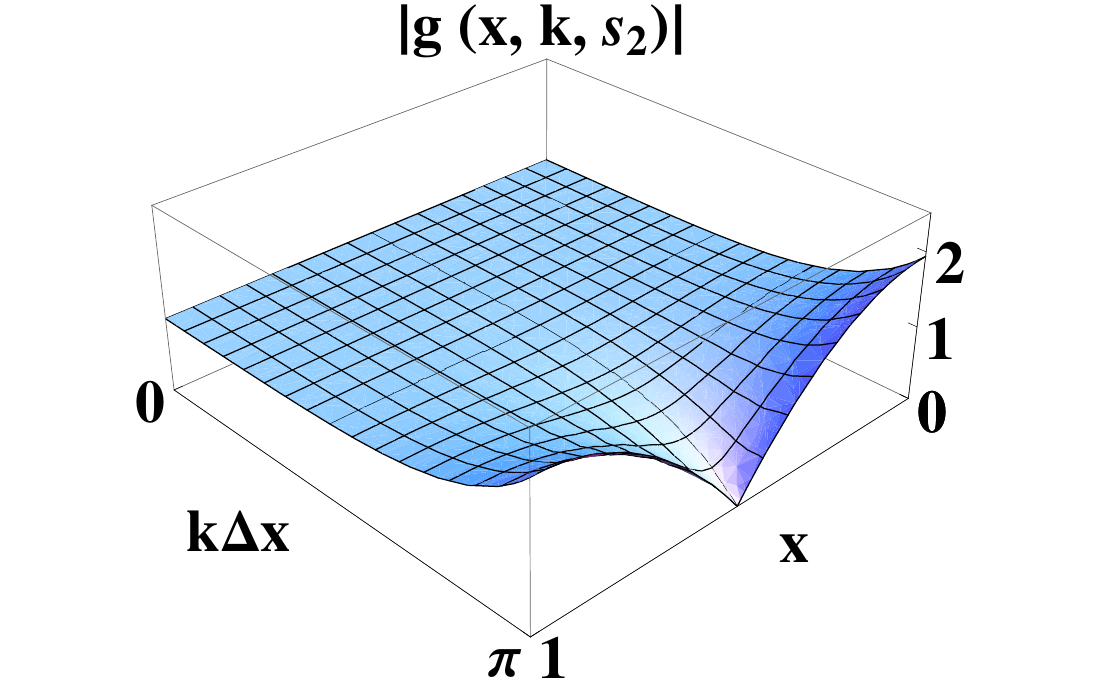}\includegraphics[width=0.45\columnwidth]{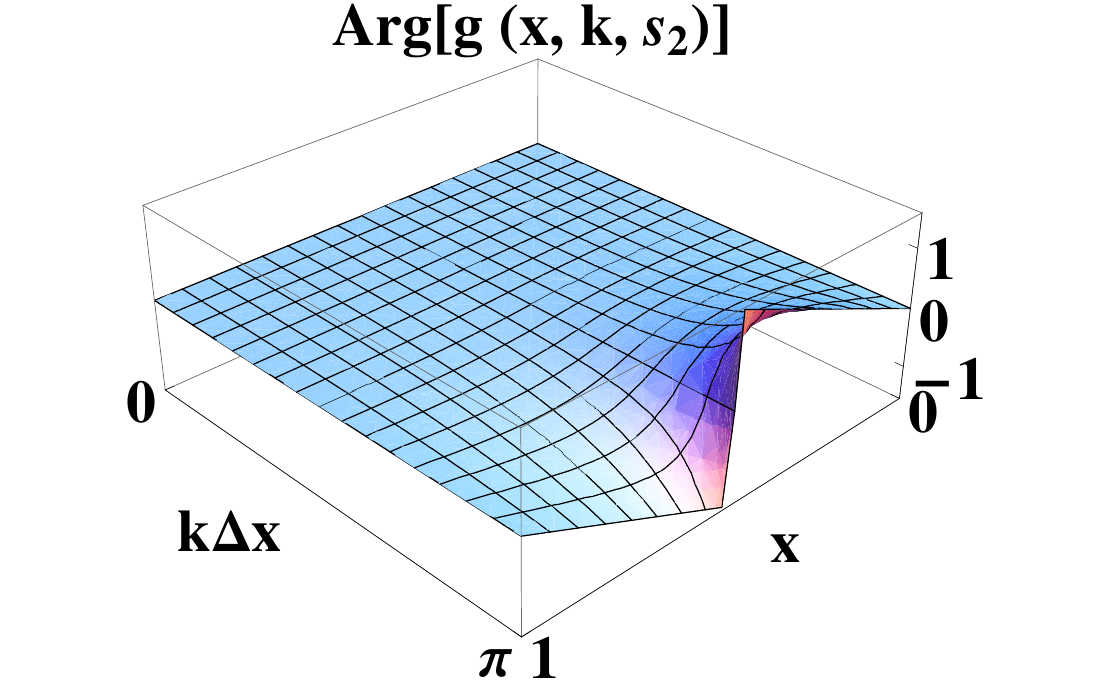}

\includegraphics[width=0.45\columnwidth]{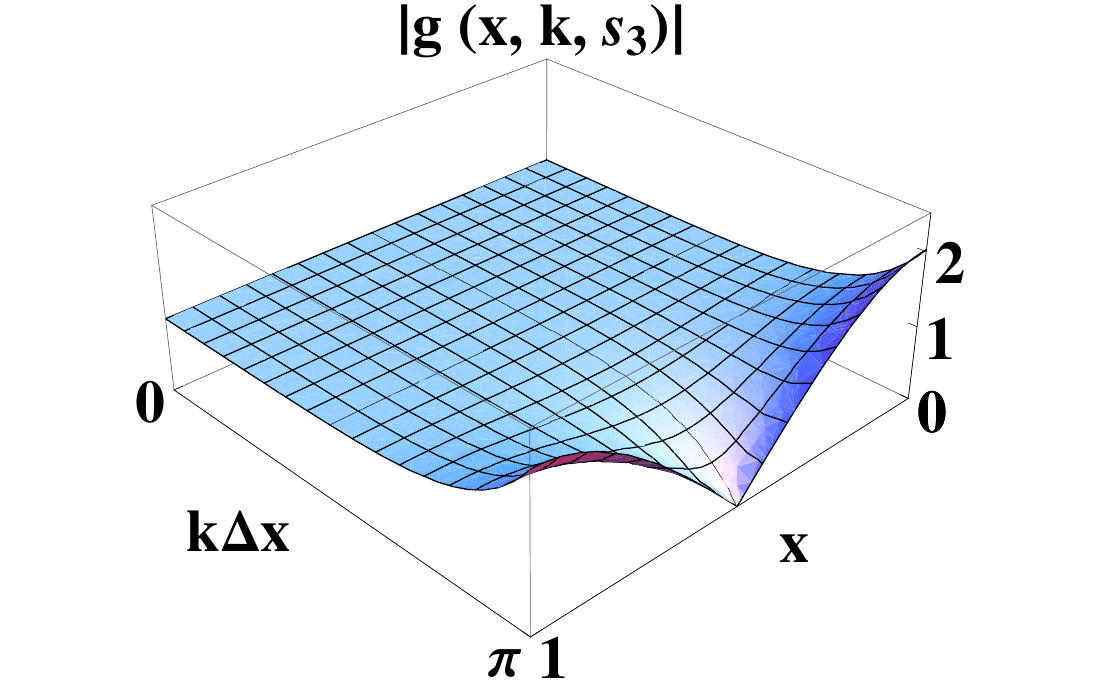}\includegraphics[width=0.45\columnwidth]{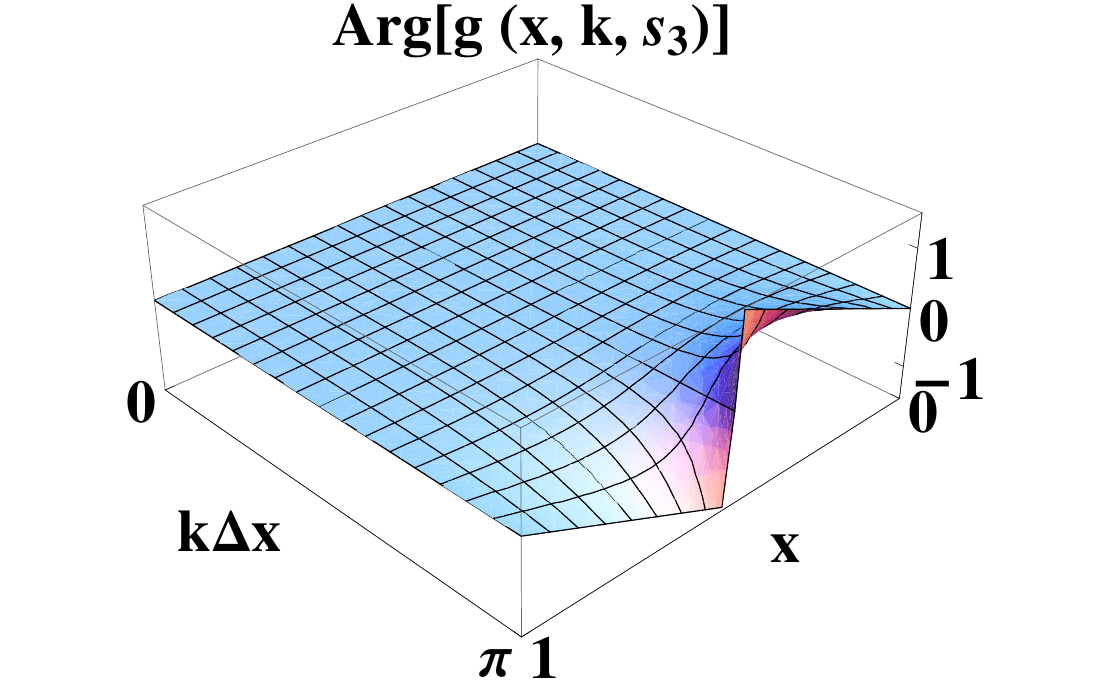}

\caption{The amplitude (left column) and phase (right column) of $g(x,\,k,\,\tilde{s}_{\mu})$
for $\mu=0,\,1,\,2,\,3$, which correspond to nearest grid point,
linear, quadratic, cubic particle shapes, respectively. \label{fig:B-spline-spectral-distortion}}
\end{figure}

The amplitude and phase of $g(x,\,k,\,\tilde{s}_{\mu})$, $g(x,\,k,\,\tilde{s}_{\mathsf{\mathcal{G}}})$
and $g(x,\,k;\,\tilde{s}_{Erf})$ are shown in Figs. \ref{fig:B-spline-spectral-distortion}
and \ref{fig:Gaussian-spectral-distortion}. The general trends in
Figs. \ref{fig:B-spline-spectral-distortion} and \ref{fig:Gaussian-spectral-distortion}
are that (1) larger amplitude and phase errors occur at higher $k$;
and (2) the wider the particle shape, the errors are more concentrated
at high $k$. Both amplitude and phase errors depend on the particle's
position in the cell, except for the amplitude error for $s_{0}$
(nearest grid point). For the phase error, the largest error occurs
when the particle is in the middle of the cell. 

Comparing $g(x,\,k,\,\tilde{s}_{\mu})$ in Fig. \ref{fig:B-spline-spectral-distortion},
another salient observation is that the phase error improves slowly
with the increase of the order of the particle shape. Also the dependence
of the errors on the particle position does not seem to go away quickly
as the order is increased, despite stronger attenuation to the high
$k$ modes. As will be shown at the end of section \ref{sub:linear-FGI},
the spectral errors, especially their dependences on the particle
position, play a key role in determining the growth rate of most unstable
mode in the linear FGI case, while the attenuation to the high $k$
modes due to $\tilde{S}(\bm{k})$ and $\tilde{S}'(\bm{k})$ in the
front factors in Eq. (\ref{eq:phase-func-deposition}) and (\ref{eq:phase-func-field-interpolation})
enlarges the stability domain, but at the expense of accuracy of the
high $k$ modes. The high order B-spline particle shapes offer both
benefits but the latter benefit can also be achieved via a low-pass
filter that is independent of the particle position.

\subsection{Relation between the linear instability and the spectral error\label{sub:linear-FGI}}

The value of $\tilde{s}(k)$ in and outside the fundamental Brillouin
zone $-k_{g}/2<k<k_{g}/2$ play different roles in the linear grid
instability. The latter value is related to the spectral errors $g(x,\,k;\,\tilde{s}(k))$.
In order to distinguish their roles, it is instructive to inspect
the linear dispersion equation Eq. (\ref{eq:cold-dispersion-finite-dt})
in the limit $\Delta t\rightarrow0$ for a 1D cold drifting plasma
with velocity $v_{0}$, 

\begin{equation}
1-\dfrac{1}{k{}^{2}}\underset{q}{\sum}\dfrac{\tilde{s}^{2}(k_{q})k_{q}^{2}}{(\omega-k_{q}v_{0})^{2}}=0,\label{eq:cold-dispersion}
\end{equation}

\noindent where we also assume $\tilde{s}_{\rho}=\tilde{s}_{E}=\tilde{s}$,
$[k]^{2}=k^{2}$ and $\kappa(k_{q})=k_{q}$ for simplicity.

As $\underset{q}{\sum}\tilde{s}(k_{q})e^{-iqk_{g}x}=\tilde{s}(k)g(x/\Delta x,k;\tilde{s}(k))$,
inverse Fourier transform with respect to $x$ gives $\tilde{s}(k_{q})=\mathrm{g}_{q}\tilde{s}(k)$,
where $\mathrm{g}_{\alpha}\equiv\tilde{g}(\alpha,\,k;\:\tilde{s}(k))=\Delta x\int dxe^{-i2\pi\alpha x}g(x,\,k;\:\tilde{s}(k))$
is discrete at $\alpha=q$ due to the periodicity of $g(x,\,k;\:\tilde{s}(k))$
in $x$. For $n$th order B-spline shape, $\mathrm{g}_{q}=(-k/k_{q})^{n+1}$;
for $\tilde{s}_{Erf}(k;\sigma)$ shape, $\mathrm{g}_{q}=-k/k_{q}\cdot e^{\sigma^{2}[(k\Delta x/2)^{2}-(k\Delta x/2+q\pi)^{2}]}$.
Keeping only the 0th and $q$th terms, Eq. (\ref{eq:cold-dispersion})
can be rewritten as, 

\begin{equation}
1-\tilde{s}^{2}(k)\left[\dfrac{1}{\Omega^{2}}+\dfrac{\mathrm{g}_{q}^{2}k_{q}^{2}/k^{2}}{(\Omega-k_{g}v_{0})^{2}}\right]=0,\label{eq:cold-dispersion-2terms}
\end{equation}

\noindent where $\Omega=\omega-kv_{0}$ and $-k_{g}/2<k<k_{g}/2$. 

In Eq. (\ref{eq:cold-dispersion-2terms}), $\tilde{s}(k)$ and $\mathrm{g}_{q}$
affect the instability domain and growth rate differently. For the
model dispersion in Eq. (\ref{eq:cold-dispersion-2terms}), mode stability
requires $\tilde{s}(k)\lesssim k_{g}v_{0}$ (see \ref{sec:A3}). So
for sufficiently small $\tilde{s}(k)$ in the fundamental Brillouin
zone, there is no instability for a particular $k$. However, this
is achieved at the expense of the dispersion accuracy of the physical
modes, since $\tilde{s}(k)$ can be seen as a renormalization of $\Omega$.
While $\mathrm{g}_{q}=\tilde{s}(k_{q})/\tilde{s}(k)$, which is closely
related to the spectral errors $g(x,\,k;\,\tilde{s}(k))$, affects
the alias modes thus the instability only. More specifically, $\mathrm{g}_{q}$
has a large effect on the instability growth rate but little on the
stability domain. $\mathrm{g}_{q}$ is real and can be written as
a sum of the Fourier coefficients of the amplitude and phase components
of $g(x,\,k;\:\tilde{s}(k))$,

\begin{equation}
\mathrm{g}_{q}=\underset{l}{\sum}\mathrm{A}_{l}\mathrm{\Theta}_{q-l}\label{eq:alias-coefficient}
\end{equation}

\noindent where $\mathrm{A}_{l}=\Delta x\int dxe^{-i2\pi lx}|g(x,\,k;\:\tilde{s}(k))|$
and $\mathrm{\Theta}_{q-l}=\Delta x\int dxe^{-i2\pi(q-l)x+i\text{Arg}[g(x,\,k;\:\tilde{s}(k))]}$
. The dominating terms in Eq. (\ref{eq:alias-coefficient}) is $\mathrm{A}_{0}\mathrm{\Theta}_{q}$
and $\mathrm{A}_{q}\mathrm{\Theta}_{0}$ as both $\mathrm{A}_{l}$
and $\mathrm{\Theta}_{l}$ are highly peaked at $l=0$. Therefore
the $q$th alias mode appears mainly because the presence of the amplitude
or phase errors that is position-dependent and modulated at the $q$th
harmonics of the grid. For the common B-spline particle shapes, $\mathrm{A}_{0}\mathrm{\Theta}_{q}$
dominates at high $k$, therefore the amplitude error is more important,
while at lower $k$, $\mathrm{A}_{q}\mathrm{\Theta}_{0}$ dominates,
i.e., the phase error is more important.

The most unstable mode $k^{max}$ in the dispersion Eq. (\ref{eq:cold-dispersion-2terms})
can be obtained from $\tilde{s}(k^{max})\approx k_{g}v_{0}$. Its
normalized growth rate is 
\begin{equation}
\omega_{i}^{max}\approx\dfrac{3^{1/2}}{2\cdot2^{1/3}}\tilde{s}(k^{max})^{1/3}\left(k_{q}^{max}/k^{max}\right)^{2/3}\mathrm{g}_{q}^{2/3}\approx\dfrac{3^{1/2}}{2\cdot2^{1/3}}\left(k_{g}v_{0}\right)^{1/3}\left(1+qk_{g}/k^{max}\right)^{2/3}\mathrm{g}_{q}^{2/3}.\label{eq:max-growth-rate}
\end{equation}
Therefore, smaller $\tilde{s}(k)$ in the fundamental Brillouin zone
(but keeping $\mathrm{g}_{q}$ the same) actually leads to smaller
$k^{max}$ and higher maximum growth rate, while smaller $|\mathrm{g}_{q}|$
(but keeping $\tilde{s}(k)$ in the fundamental Brillouin zone the
same) lowers the maximum growth rate. This indicates that a particle
shape with a high contrast between the fundamental and higher order
Brillouin zones, i.e., small $|\mathrm{g}_{q}|$ that translates to
small position dependent spectral errors, is beneficial when aliasing
is unavoidable.

\subsection{Spectral fidelity of the collective modes}

Now it is clear that the systematic spectral errors in the deposition
and field interpolation occur at the individual particle level, but
the overall errors from an ensemble of plasma particles are less transparent.
To this end, we can analyze how these errors manifest themselves in
the collective motion of a finite number of particles. To simplify
this analysis, an initially cold beam of $N_{p}$ particles moving
at velocity $V_{0}$ is assumed, but the analysis is not restricted
to this velocity distribution. Each particle has charge $Q$, mass
$m$. The system is periodic with length $L$ and $N_{g}$ cells,
$\Delta x=L/N_{g}$ is the cell size. When there is no perturbation,
the $j$-th particle is located at $X_{j}$ and particles are equally
spaced with distance $\Delta X=L/N_{p}$. The particle's position
$x_{j}$ is perturbed at $t=0$ and then evolves self-consistently.
The displacement of the particle is $\delta x(X_{j},t)=x_{j}(t)-X_{j}$
and can be expanded into Fourier modes 
\[
\delta x(X_{j},t)=N_{p}^{-1}\underset{\frac{\pi}{\Delta X}\geqslant|K|}{\sum}\tilde{\delta x}(K,t)e^{iKX_{j}}=2N_{p}^{-1}\underset{\frac{\pi}{\Delta X}\geqslant K\geqslant0}{\sum}A(K,t)\text{cos}\left(KX_{j}+\theta(K,t)\right),
\]
where $\tilde{\delta x}(K,t)=A(K,t)e^{i\theta(K,t)}=\underset{j}{\sum}\delta x(X_{j},t)e^{-iKX_{j}}$,
$A(K,t)$ and $\theta(K,t)$ are the mode amplitude and phase of the
collective motion of these $N_{p}$ particles. The drift motion of
the beam is described by the $K=0$ mode. 

Assuming these particles have shape $\tilde{s}(k)$, their contribution
to the density of mode $k$ in the periodic physical system (or a
gridless model) is

\begin{equation}
\tilde{\rho}(k)=Q\underset{j}{\sum}\tilde{s}(k)e^{-ikx_{j}(t)}=Q\tilde{s}(k)\underset{j}{\sum}e^{-ik\left[X_{j}+2N_{p}^{-1}\underset{\frac{\pi}{\Delta X}\geqslant K\geqslant0}{\sum}A(K,t)\text{cos}\left(KX_{j}+\theta(K,t)\right)\right]},
\end{equation}

\noindent where $|k|<\pi/\Delta x$. the phase factor in this sum
can be written, using the Jacobi-Anger expansion, as 

\begin{align}
 & \,e^{-ik\left[X_{j}+2N_{p}^{-1}\underset{\frac{\pi}{\Delta X}\geqslant K\geqslant0}{\sum}A(K,t)\text{cos}\left(KX_{j}+\theta(K,t)\right)\right]}\nonumber \\
= & \,e^{-ikX_{j}}\cdot\underset{\frac{\pi}{\Delta X}\geqslant K\geqslant0}{\prod}\left\{ \underset{\nu_{K}}{\sum}i^{\nu_{K}}J_{\nu_{K}}\left(-2kA(K,t)/N_{p}\right)\cdot e^{i\nu_{K}\left[KX_{j}+\theta(K,t)\right]}\right\} \nonumber \\
= & \,e^{-ikX_{j}}\cdot\underset{\nu_{0},...,\nu_{K}}{\sum}\left\{ e^{i\sum(\nu_{K}K)\cdot X_{j}+i\sum\nu_{K}\cdot\left[\theta(K,t)+\pi/2\right]}\cdot\underset{\nu_{0},...,\nu_{K}}{\prod}J_{\nu_{K}}\left(-2kA(K,t)/N_{p}\right)\right\} .\label{eq:Jacobi-anger-phase-expansion}
\end{align}

\noindent where $\nu_{K}$ is an integer for each mode $K$ in $\frac{\pi}{\Delta X}\geqslant K\geqslant0$,
$J_{\nu_{K}}(x)$ is the $\nu_{K}$-th Bessel function of the first
kind. $\underset{\nu_{0},...,\nu_{K}}{\sum}$ ($\underset{\nu_{0},...,\nu_{K}}{\prod}$)
is a sum (product) over all combination of $\nu_{K}$s and $\sum\nu_{K}$
is a sum of every $\nu_{K}$ for $\frac{\pi}{\Delta X}\geqslant K\geqslant0$.

Summing Eq. (\ref{eq:Jacobi-anger-phase-expansion}) over $X_{j}$
and using $\stackrel[\nu_{0}=-\infty]{\infty}{\sum}J_{\nu_{0}}(x)=1$
gives, 

\begin{equation}
\tilde{\rho}(k)=N_{p}Q\tilde{s}(k)\underset{\nu_{1},...,\nu_{K}}{\sum}\left\{ \delta\left(k-mk_{P}-\sum\left(\nu_{K}K\right)\right)\cdot e^{i\sum\nu_{K}\cdot\left[\theta(K,t)+\pi/2\right]}\cdot\underset{\nu_{1},...,\nu_{K}}{\prod}J_{\nu_{K}}\left(-2kA(K,t)/N_{p}\right)\right\} ,\label{eq:density-collective-coupling}
\end{equation}

\noindent where $k_{P}=2\pi/\Delta X=N_{p}2\pi/L$ and $m$ is an
integer.

Eq. (\ref{eq:density-collective-coupling}) describes the contribution
from the Fourier modes of the collective motion to the Fourier modes
of the density, where the first, second and third terms in the curly
bracket determining the mode coupling relation, phase and strength,
respectively. The $\pi/2$ in the phase factor comes from the phase
difference between particle's displacement and the density. Note that
Eq. (\ref{eq:density-collective-coupling}) is valid for arbitrary
number of particles and arbitrary spectrum/amplitude of the collective
modes. Due to the finite number of particles in a system, the $mk_{P}$
term in the argument of the delta function introduces a physical coupling
due to the ``particle aliasing'' effect of the collective mode.
Typically we are interested in the case $k\ll k_{P}$, thus this effect
couples a low $k$ mode and a very high $K$ collective mode efficiently
only when $\nu_{K}$ is a small number. Furthermore, the coupling
coefficient $J_{\nu_{K}}\left(-2kA(K,t)/N_{p}\right)$ would be small
when there are sufficient large number of particles, $N_{p}\gg1$,
so practically we may take $m=0$ and ignore this effect in a physical
system for most situations. 

Next we investigate the coupling among collective modes and density
modes only. This can be illustrated by understanding how a single
collective mode couples to the density perturbation first. For a collective
mode spectrum with a single mode $K_{0}$, $A(K_{0})\neq0$, while
$A(K\neq K_{0})=0$. Since $J_{\nu_{K}}(0)\neq0$ only when $\nu_{K}=0$,
so we have $\delta\left(k-\sum\left(\nu_{K}K\right)\right)=\delta\left(k-\nu_{K_{0}}K_{0}\right)$.
This corresponds to the excitation of the fundamental mode $k=K_{0}$
and its harmonics $k=\nu_{K_{0}}K_{0}$ with the mode coupling coefficient
being $s(\nu_{K_{0}}K_{0})e^{i\nu_{K_{0}}\cdot\left[\theta(K_{0},t)+\pi/2\right]}\cdot J_{\nu_{K_{0}}}\left(-2\nu_{K_{0}}K_{0}A(K_{0},t)/N_{p}\right)$.
Similarly, when there are two collective modes $K_{0}$ and $K_{1}$,
their beating will result in density perturbation at mode $k=\nu_{K_{0}}K_{0}+\nu_{K_{1}}K_{1}$
with the coupling coefficient being $s(\nu_{K_{0}}K_{0}+\nu_{K_{1}}K_{1})e^{i[\nu_{K_{0}}\theta(K_{0},t)+\nu_{K_{1}}\theta(K_{1},t)+\pi/2\cdot(\nu_{K_{0}}+\nu_{K_{1}})]}\cdot J_{\nu_{K_{0}}}\left(-2kA(K_{0},t)/N_{p}\right)J_{\nu_{K_{1}}}\left(-2kA(K_{1},t)/N_{p}\right)$.
From this, the coupling for the density perturbation involving more
collective modes can be generalized. 

The density perturbation created by the collective modes in turn drives
more collective modes of the particles through the electric field.
We define the electric field on the particle as a function of $X_{j}$,
\[
\mathcal{E}(X_{j},t)=E(x_{j}(t))=N_{g}^{-1}\underset{k}{\sum}\tilde{s}(k)\tilde{E}(k)e^{ikx_{j}(t)}=N_{g}^{-1}\underset{k}{\sum}\frac{-i}{[k]}\tilde{s}(k)\tilde{\rho}(k)e^{ikx_{j}(t)},
\]
whose Fourier transform with regard to $X_{j}$ is, 

\[
\mathcal{E}(K,t)=N_{g}^{-1}\underset{j}{\sum}\underset{k}{\sum}\frac{-i}{[k]}\tilde{s}(k)\tilde{\rho}(k)e^{i[kx_{j}(t)-KX_{j}]}.
\]
Here, $[k]$ is the effective operator from the field solver. Similar
to the phase factor for the density in Eq. (\ref{eq:Jacobi-anger-phase-expansion}),
the phase factor in the above equation can be expanded and the sum
over $X_{j}$ can be carried out to give, 

\begin{align}
\mathcal{E}(K,t) & =\frac{N_{p}}{N_{g}}\underset{k}{\sum}\frac{-i}{[k]}\tilde{s}(k)\tilde{\rho}(k)\underset{\nu_{1},...,\nu_{K'}}{\sum}\Biggl\{\delta\left(K-k-mk_{P}-\sum\left(\nu_{K'}K'\right)\right)e^{i\sum\nu_{K'}\cdot\left[\theta(K',t)+\pi/2\right]}\nonumber \\
 & \cdot\left.\underset{\nu_{1},...,\nu_{K'}}{\prod}J_{\nu_{K'}}\left(2kA(K',t)/N_{p}\right)\right\} .\label{eq:field-collective-coupling}
\end{align}

The relation between $\mathcal{E}(K,t)$ and $\tilde{\rho}(k)$ (or
$E(k)$) depends on the argument of the delta function and the sum
over $k$, therefore it is more complicated than that between $\tilde{\rho}(k)$
and $\tilde{\delta x}(K,t)$ in Eq. (\ref{eq:density-collective-coupling}).
Again we may ignore the particle aliasing effect in the collective
mode when $N\gg1$. Then for the simplest case of a collective motion
with a single mode $K_{0}$, a non-zero value of $\underset{\nu_{1},...,\nu_{K'}}{\prod}J_{\nu_{K'}}\left(2kA(K',t)/N_{p}\right)$
requires that $\nu_{K'\ne K_{0}}=0$, therefore, $\sum\nu_{K'}=\nu_{K_{0}}$,
$\sum\left(\nu_{K'}K'\right)=\nu_{K_{0}}K_{0}$. Coupling to $\mathcal{E}(K,t)$
thus comes from $\tilde{\rho}(k=K-\nu_{K_{0}}K_{0})$, which is not
necessarily the harmonics of $K_{0}$. When the collective motion
involves two modes $K_{0}$ and $K_{1}$, coupling to $\mathcal{E}(K,t)$
can come from $\tilde{\rho}(k=K-\nu_{K_{0}}K_{0}-\nu_{K_{1}}K_{1})$
for any $\nu_{K_{0}}$ and $\nu_{K_{1}}$. 

Finally, the evolution of the collective modes is determined by the
equation of motion 
\begin{equation}
d(\delta x(K,t))/dt^{2}=\frac{Q}{m}\mathcal{E}(K,t),\label{eq:eq-motion}
\end{equation}
which is an ordinary differential equation (ODE). Clearly, regardless
of whether it is written in a continuous or discrete time variable,
this step does not introduce spectral errors in $K$-space.

Now, Eqs. (\ref{eq:density-collective-coupling}), (\ref{eq:field-collective-coupling})
and (\ref{eq:eq-motion}) are the coupled equations describing the
evolution of the collective modes in the physical system of a cold
beam. When the discrete time version of Eq. (\ref{eq:eq-motion})
is used, these equations represent a discrete-time analog of the physical
system. If Eq. (\ref{eq:eq-motion}) is discretized using a scheme
consistent with the underlying ODE, i.e., the solution of the discretized
equation approaches that of the ODE when the discretization step $\Delta t\rightarrow0$,
then a numerical instability, if occurring in this discrete-time system,
will exhibit dependence on $\Delta t$. Therefore, a convergence test
can discern a numerical instability due to discretization in time.
Early study of FGI already established that the numerical heating
rate is independent of the time step, therefore time discretization
cannot be the reason of FGI. From the single particle analysis of
the systematic spectral errors presented earlier in this section,
we can see that another possibility of the numerical instability in
PIC is the spectral errors in the deposition and field interpolation
due to the aliasing effect. In fact, for the PIC model, Eqs. (\ref{eq:density-collective-coupling})
and (\ref{eq:field-collective-coupling}) take on the following forms,

\begin{align}
\tilde{\rho}(k) & =N_{p}Q\underset{q}{\sum}\tilde{s}(k_{q})\underset{\nu_{1},...,\nu_{K}}{\sum}\Biggl\{\delta\left(k_{q}-mk_{P}-\sum\left(\nu_{K}K\right)\right)e^{i\sum\nu_{K}\cdot\left[\theta(K,t)+\pi/2\right]}\nonumber \\
 & \cdot\left.\underset{\nu_{1},...,\nu_{K}}{\prod}J_{\nu_{K}}\left(-2k_{q}A(K,t)/N_{p}\right)\right\} ,\label{eq:density-collective-coupling-PIC}
\end{align}

\begin{eqnarray}
\mathcal{E}(K,t) & = & \frac{N_{p}}{N_{g}}\underset{q}{\sum}\underset{|k|\le k_{g}/2}{\sum}\frac{-i}{[k_{q}]}\tilde{\rho}(k_{q})\cdot\tilde{s}(k_{q})\cdot\underset{\nu_{1},...,\nu_{K'}}{\sum}\Biggl\{\delta\left(K-k_{q}-mk_{P}-\sum\left(\nu_{K'}K'\right)\right)\nonumber \\
 & \cdot & \left.e^{i\sum\nu_{K'}\cdot\left[\theta(K',t)+\pi/2\right]}\underset{\nu_{1},...,\nu_{K'}}{\prod}J_{\nu_{K'}}\left(2k_{q}A(K',t)/N_{p}\right)\right\} ,\label{eq:field-collective-coupling-PIC}
\end{eqnarray}

\noindent where $k$ has been replaced by its alias $k_{q}$ and a
sum of all aliases is taken.

Compared with Eqs. (\ref{eq:density-collective-coupling}) and (\ref{eq:field-collective-coupling}),
Eqs. (\ref{eq:density-collective-coupling-PIC}) and (\ref{eq:field-collective-coupling-PIC})
lack the spectral fidelity to the physical system, i.e., they contain
not only phase and amplitude errors, but also unphysical mode couplings.
As will be shown in the next section, such lack of spectral fidelity
is the major cause of the finite grid instability.

\section{Comparison of the gridless model and the PIC models\label{sec:Comparison-gridless-PIC}}

In this section, we compare the 1D non-relativistic electrostatic
simulation results of a numerical plasma that is susceptible to the
finite grid instability, for various PIC models and from the gridless
model. The simplest simulation is that of a single-mode electrostatic
oscillation in a cold plasma with a drift. While this mode will generate
high harmonics, such system should be stable for a long time. During
this period, the plasma should remain cold, i.e., the local momentum
spread is zero. The three spatial components of the numerical scheme
--- the charge deposition, the field solver and the field interpolation,
are implemented in the PIC/finite-difference mode and/or the gridless
mode.  The mode history from the simulations are compared, the focus
here is not the accuracy of the simulation but to understand the roles
of the amplitude and phase errors in the finite grid instability.

The simulation box has dimensionless size $L=2\pi c/\omega_{p}$ and
$N=33$ cells. The initial plasma density $n_{0}$ is constant across
the box. $N_{C}=300$ particles per cell is used for the electrons
and the ions constitute a uniform neutralizing background. The electrons
are equally spaced and a sinusoidal displacement $\delta x(x_{p0})=AL\cos(2\pi Mx_{p0}/L)/(2\pi M)$
is applied to initialize a density perturbation, where $M$ is the
mode number and $x_{p0}$ is the unperturbed particle position. The
density perturbation is given by $n_{1}(x)/n_{0}=-\partial\delta x/\partial x=A\sin(2\pi Mx/L)$
for point particles and the actual amplitude will be smaller by $s(k=2\pi M/L)$
when a shaped particle is used in the simulation. The electron initial
velocity is the sum of the drift velocity $V_{0}=0.01c$ and perturbed
velocity $\delta V=AL\omega_{p}\text{sin}(2\pi Mx/L)/(2\pi M)$, where
$c$ is the speed of light. 

\begin{figure}
\centering\includegraphics[height=0.15\paperheight]{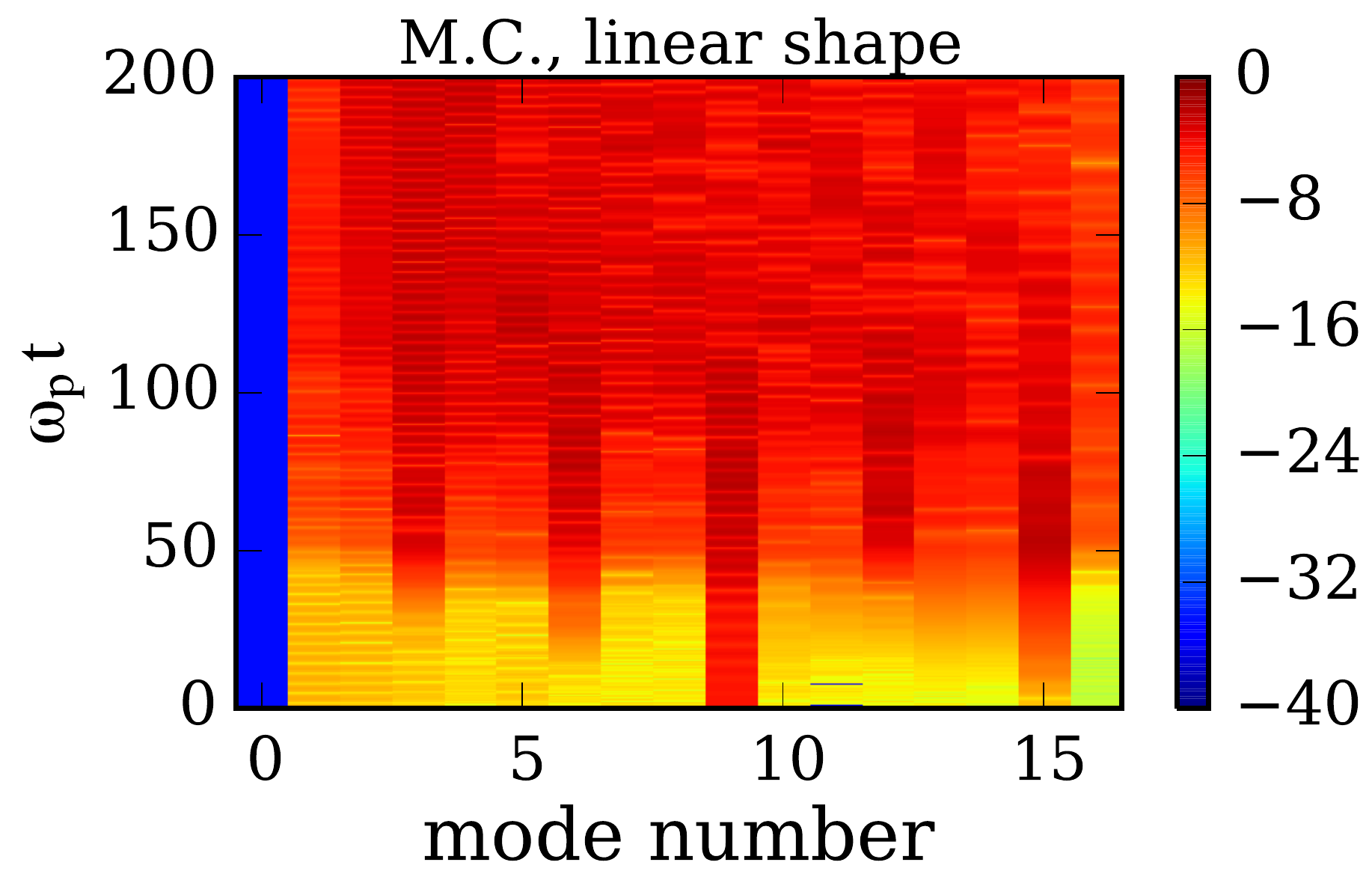}\includegraphics[height=0.145\paperheight]{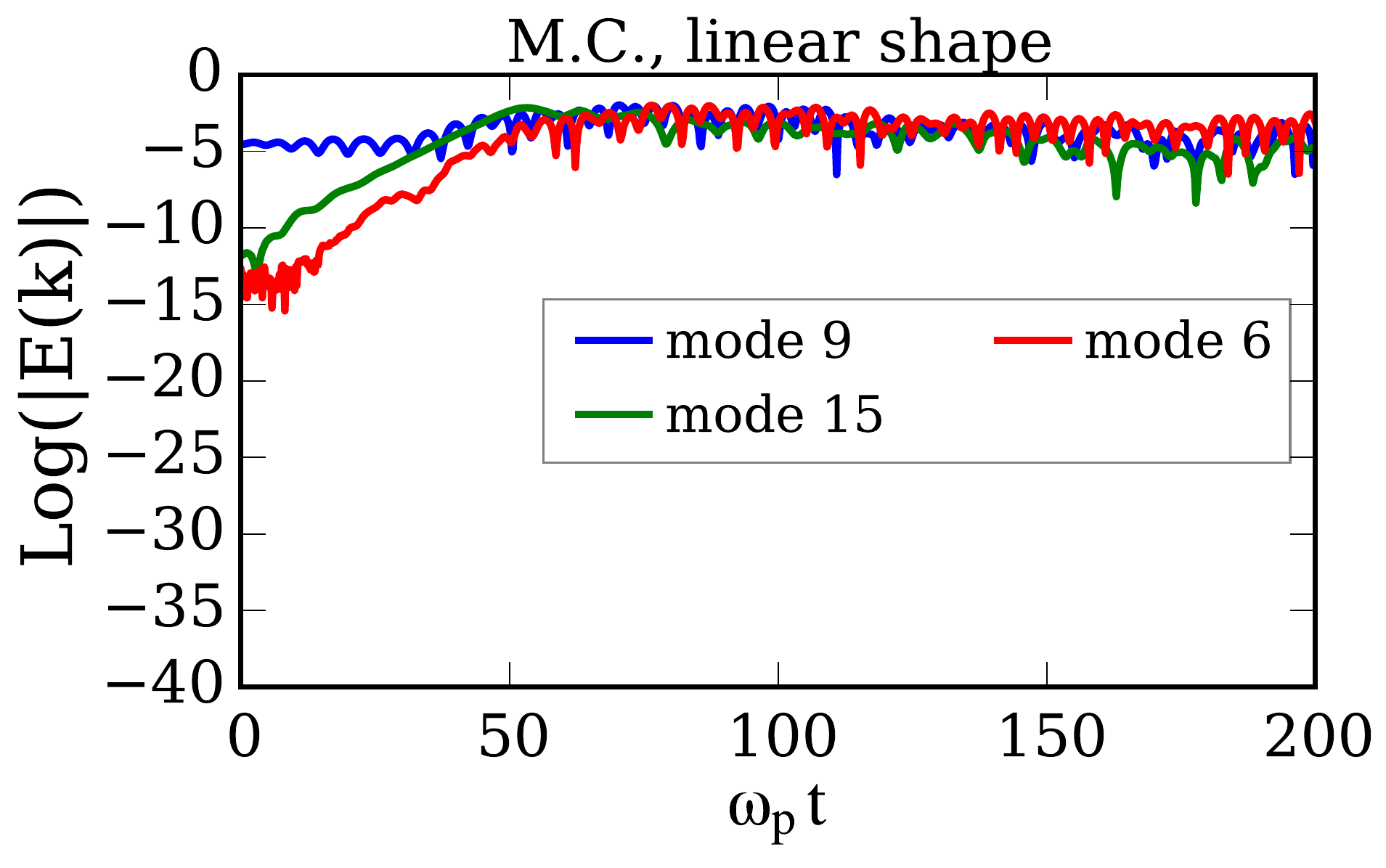}

\includegraphics[height=0.15\paperheight]{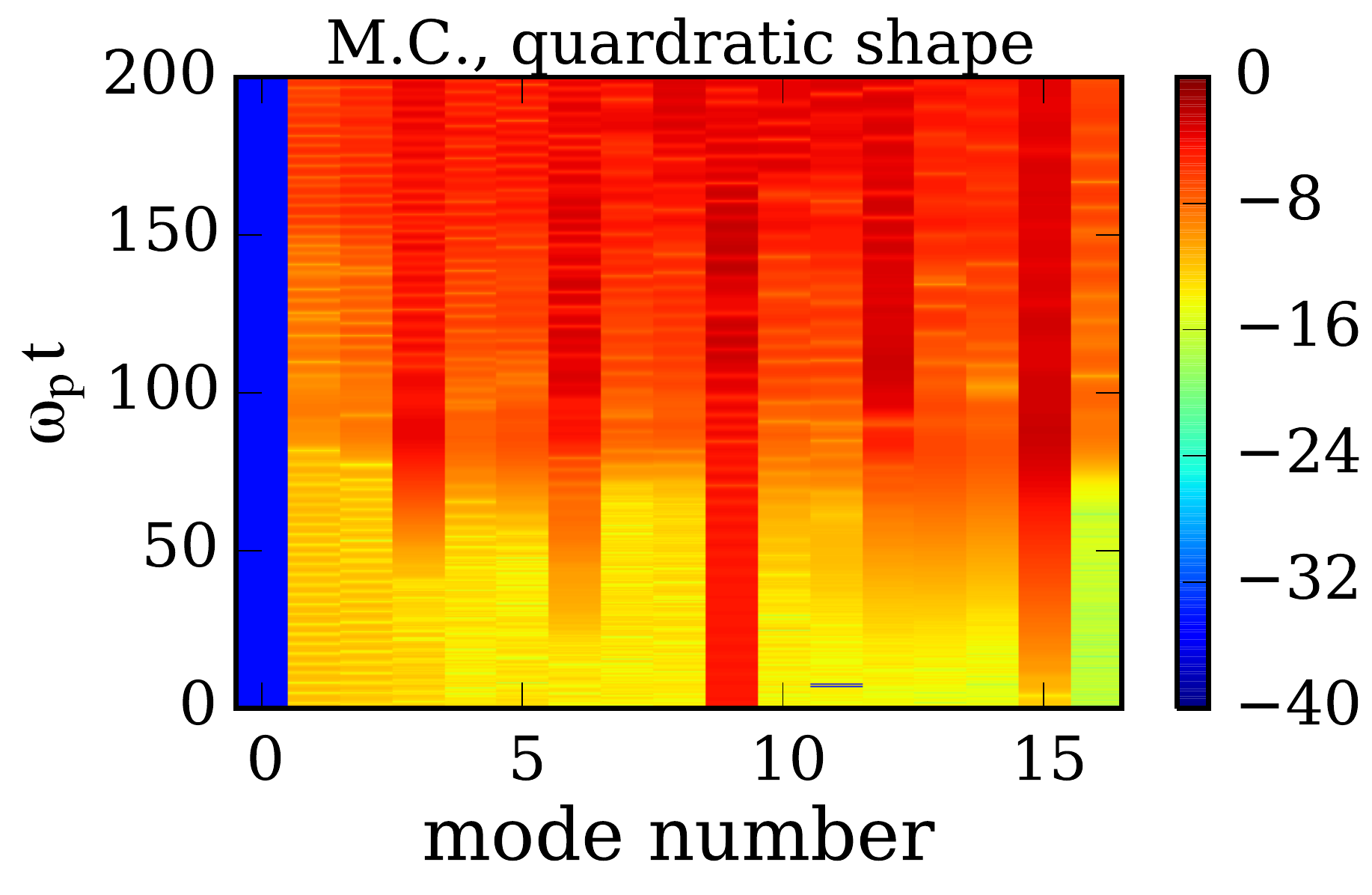}\includegraphics[height=0.145\paperheight]{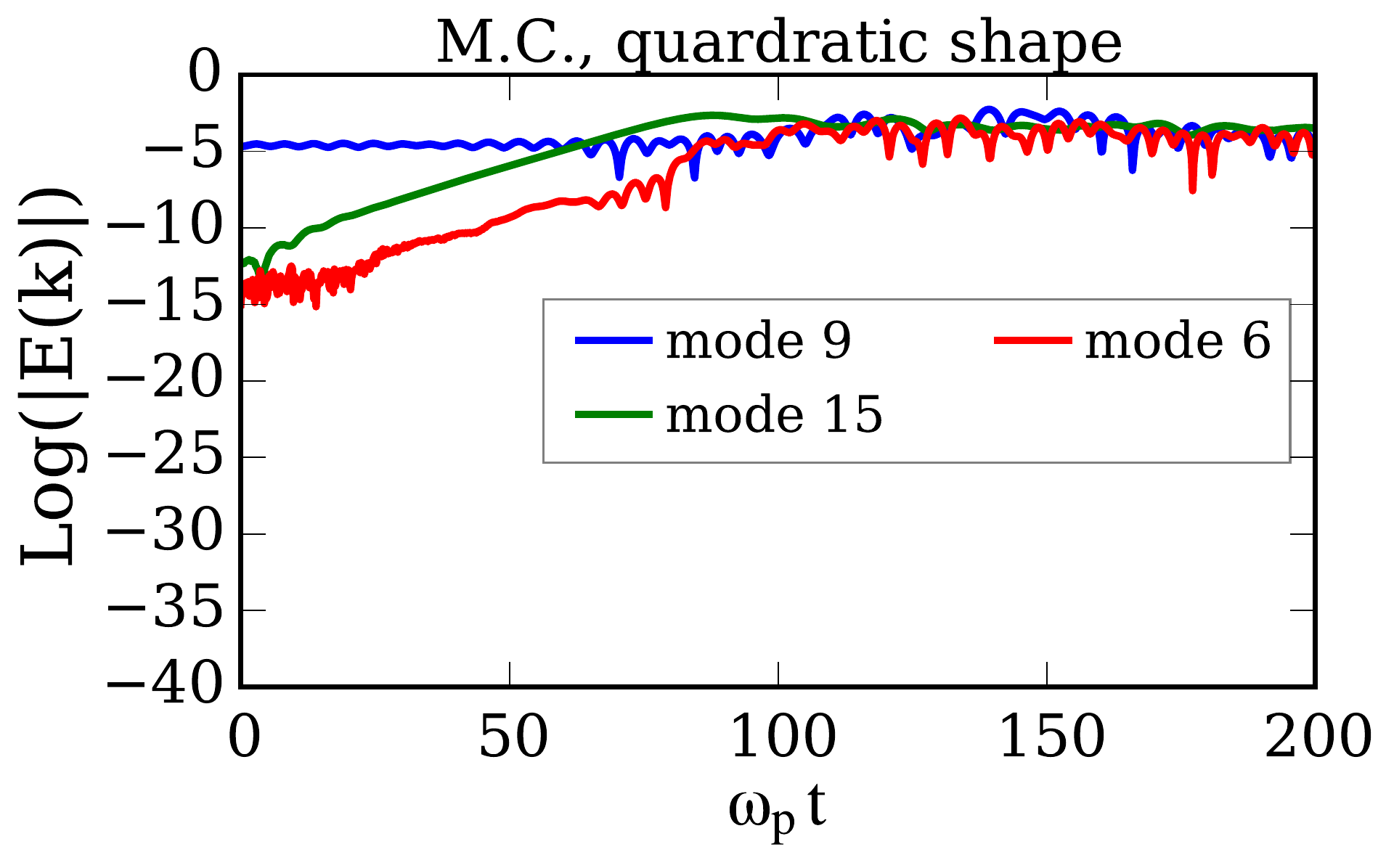}

\includegraphics[height=0.15\paperheight]{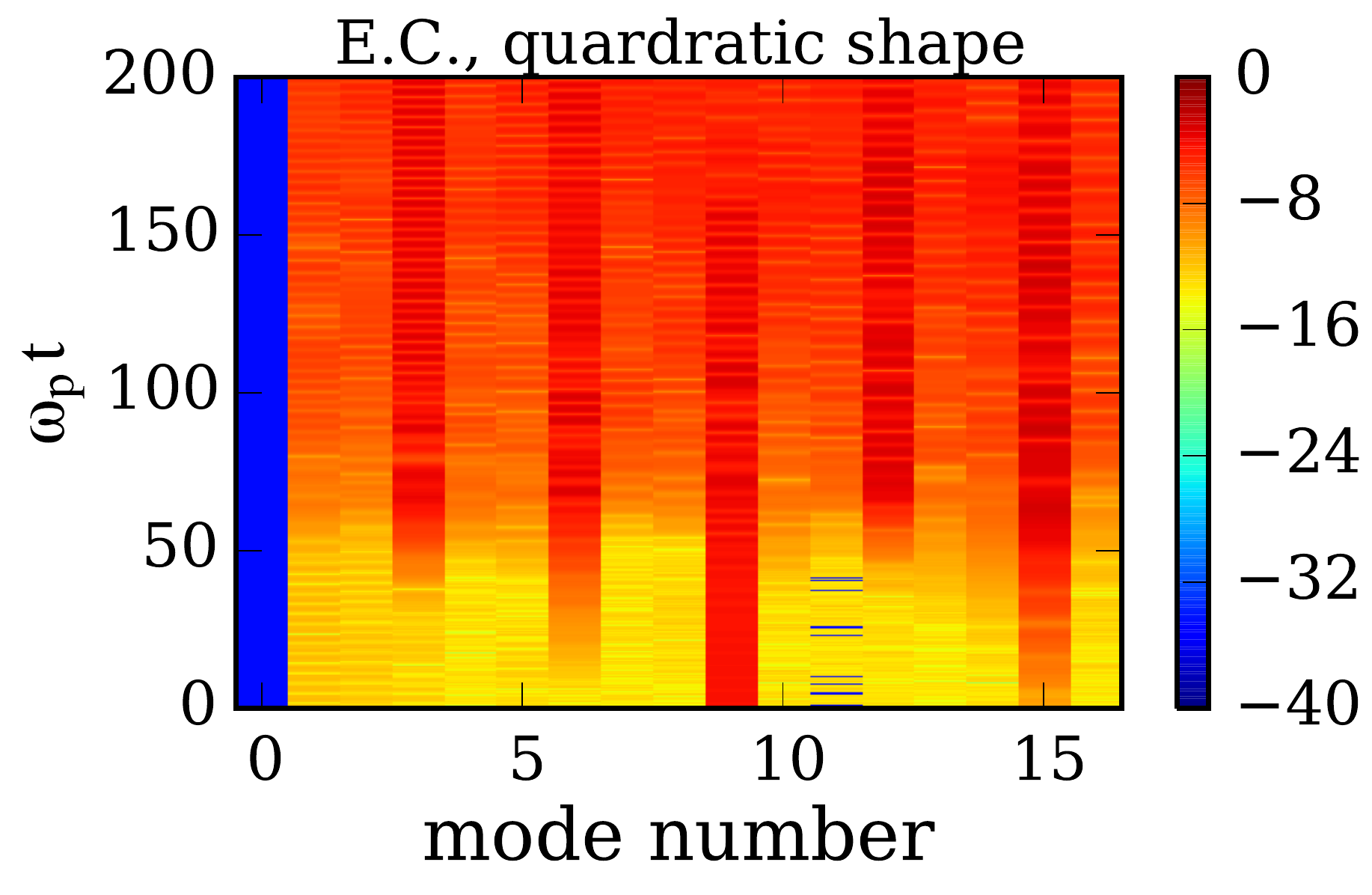}\includegraphics[height=0.145\paperheight]{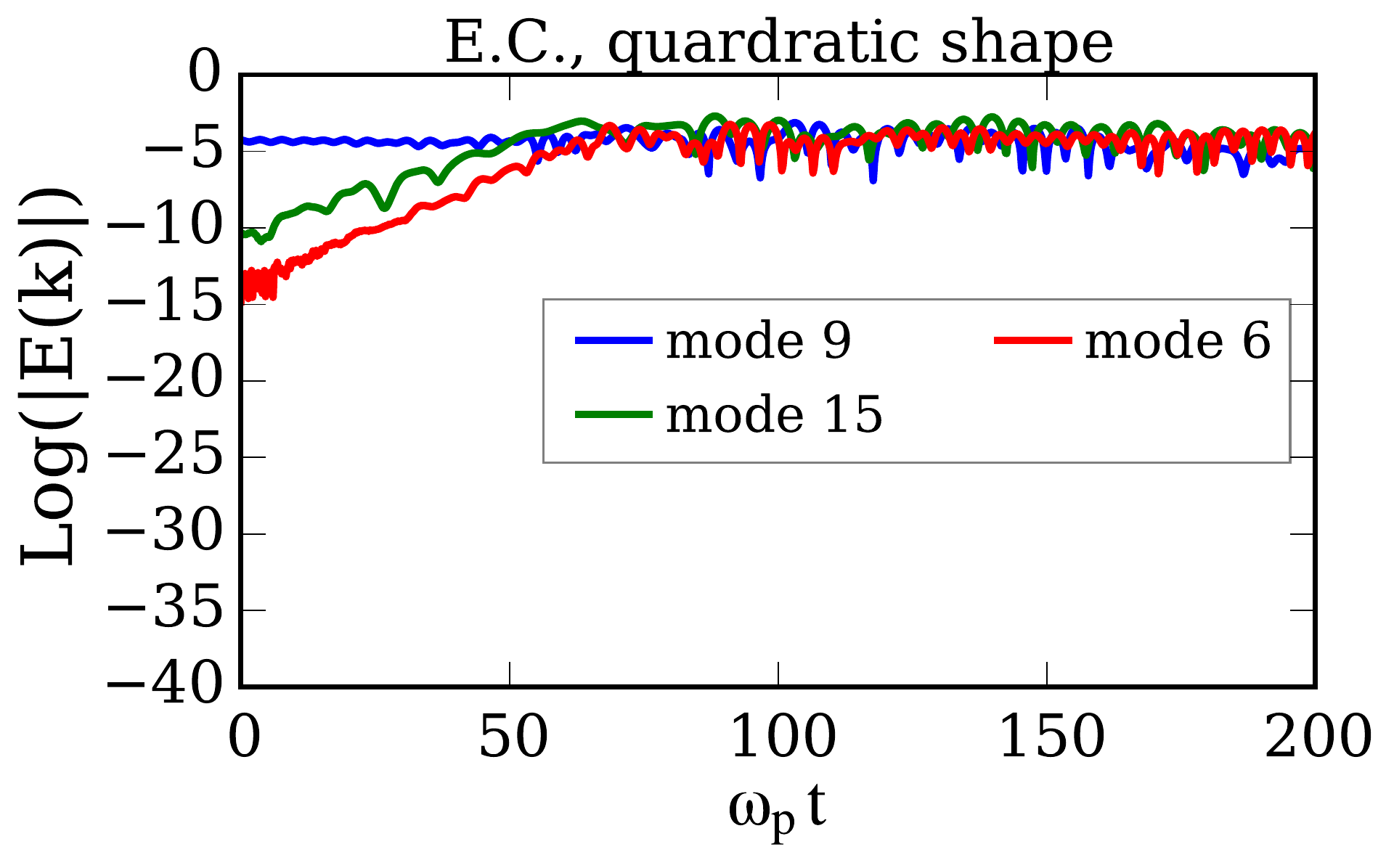}

\includegraphics[height=0.15\paperheight]{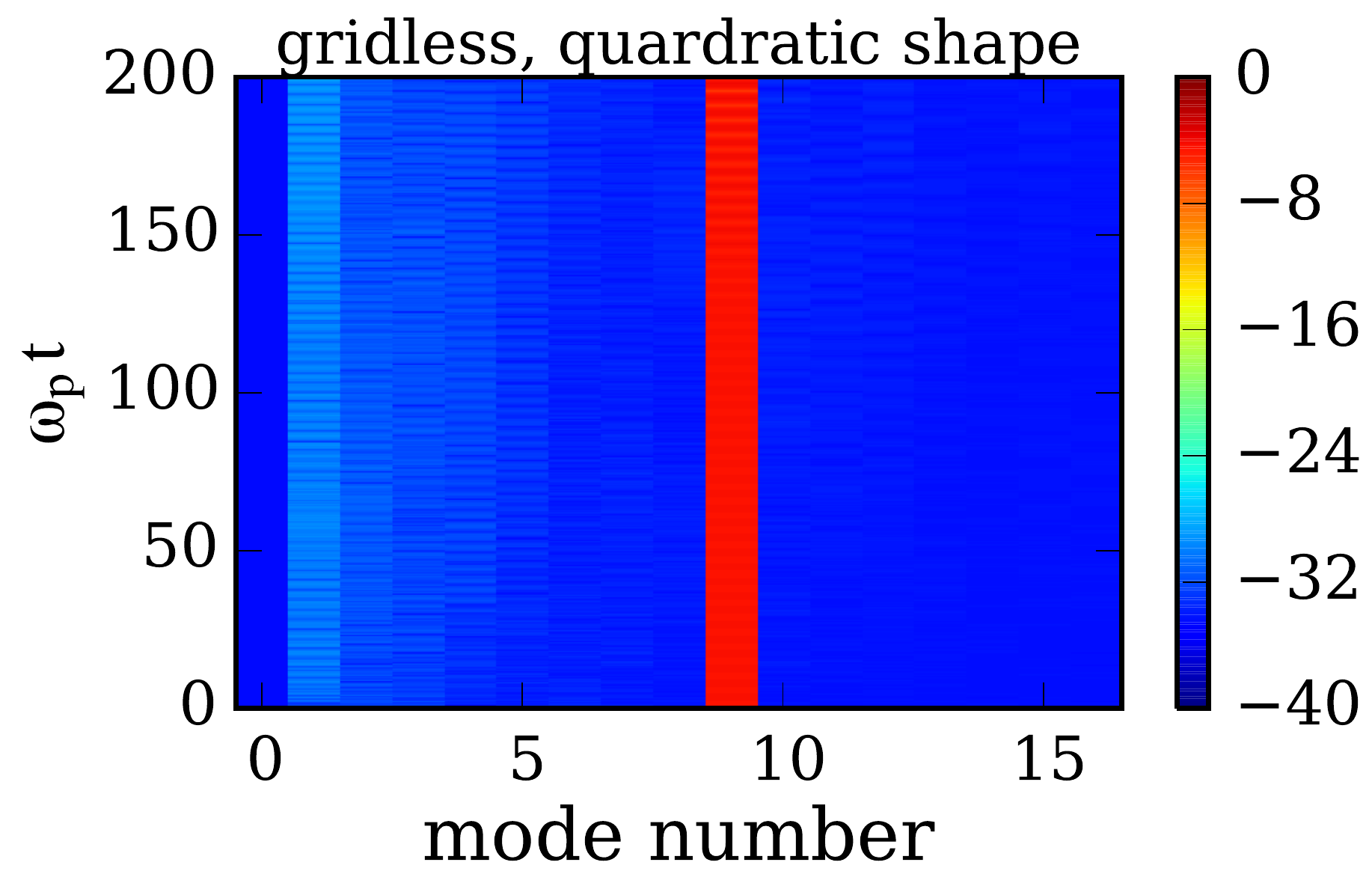}\includegraphics[height=0.145\paperheight]{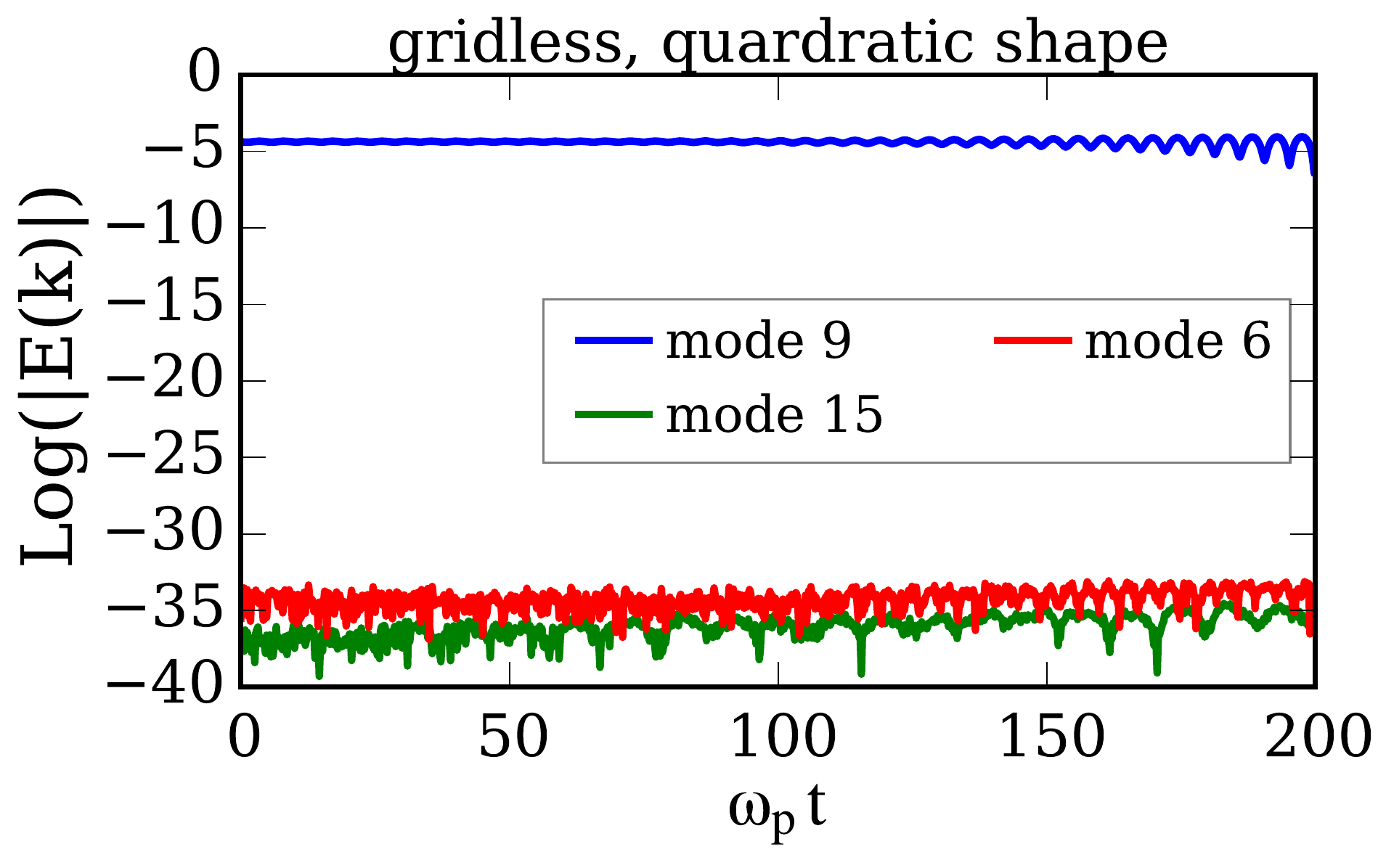}

\caption{Simulation mode spectrum from various models. Left column: full mode
spectrum $\text{ln}(|E(k)|)$ as a function of time. Right column:
evolution of the mode amplitude for the perturbation --- mode $k=9$,
alias of its second harmonic --- mode $k=15(=33-9\times2)$ and the
beat between them --- mode $k=6(=15-9)$. The four rows from top to
bottom correspond to (1) the momentum conserving (M.C.) PIC model
with linear particle shape; (2) the momentum conserving PIC model
with quadratic particle shape; (3) the energy conserving (E.C.) PIC
model with quadratic particle shape; and (4) the gridless model with
quadratic particle shape. \label{fig:mode-spectrum}}
\end{figure}

\begin{figure}
\begin{minipage}[b][0.18\paperwidth][c]{0.32\columnwidth}%
\includegraphics[width=1\columnwidth,height=0.115\paperheight]{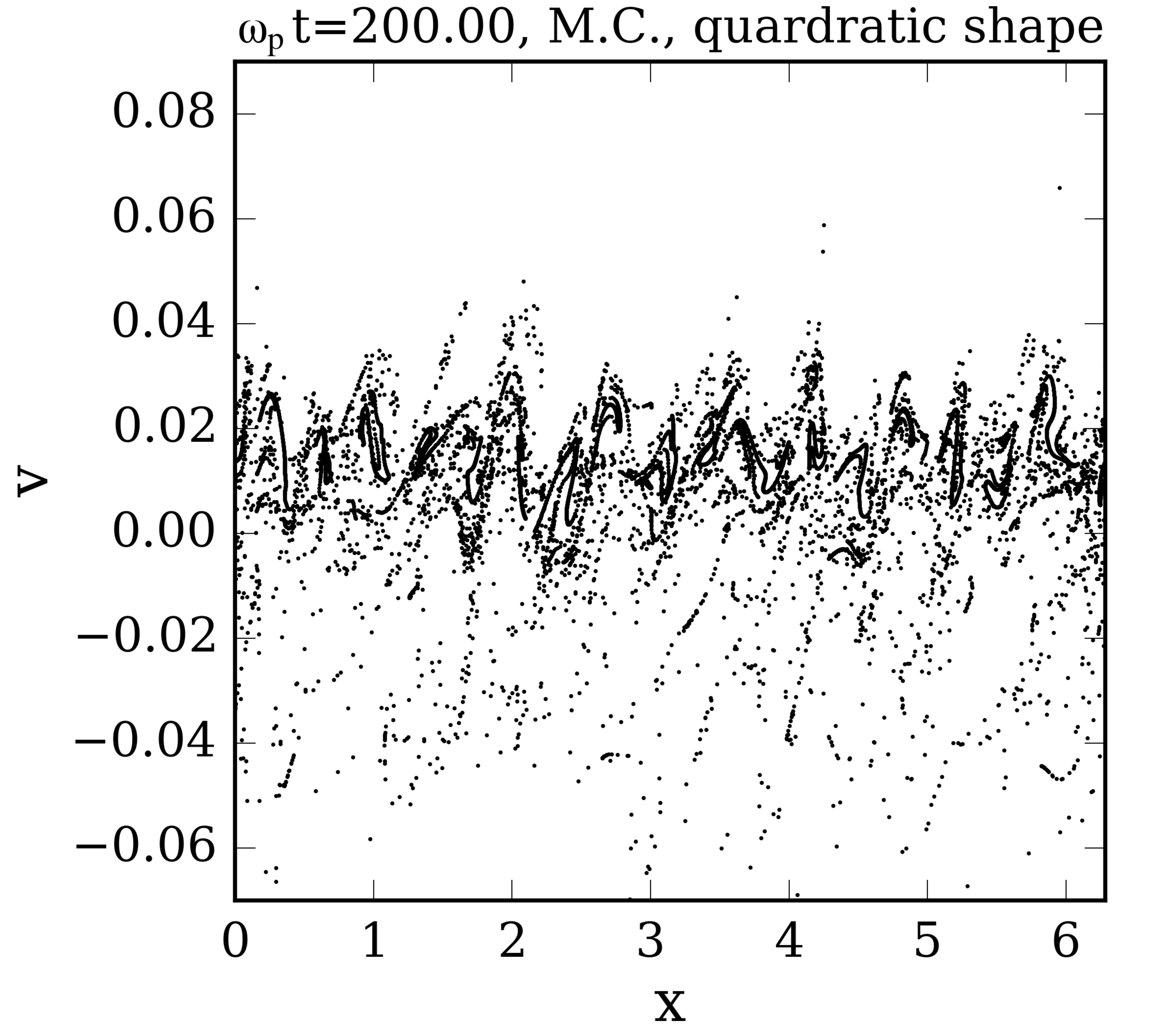}%
\end{minipage}%
\begin{minipage}[b][0.18\paperwidth][c]{0.67\columnwidth}%
\includegraphics[width=0.5\columnwidth,height=0.13\paperheight]{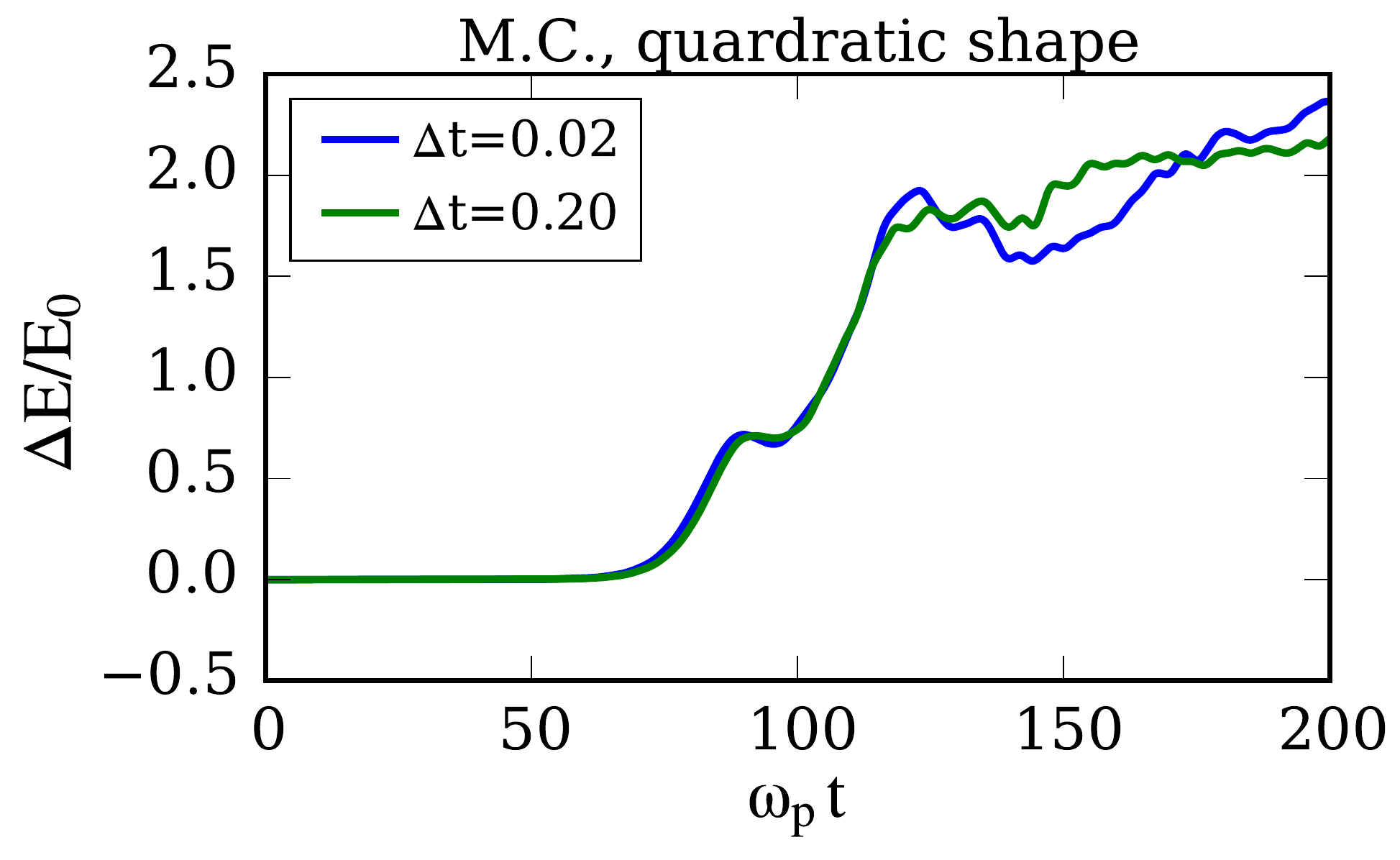}\includegraphics[width=0.5\columnwidth,height=0.13\paperheight]{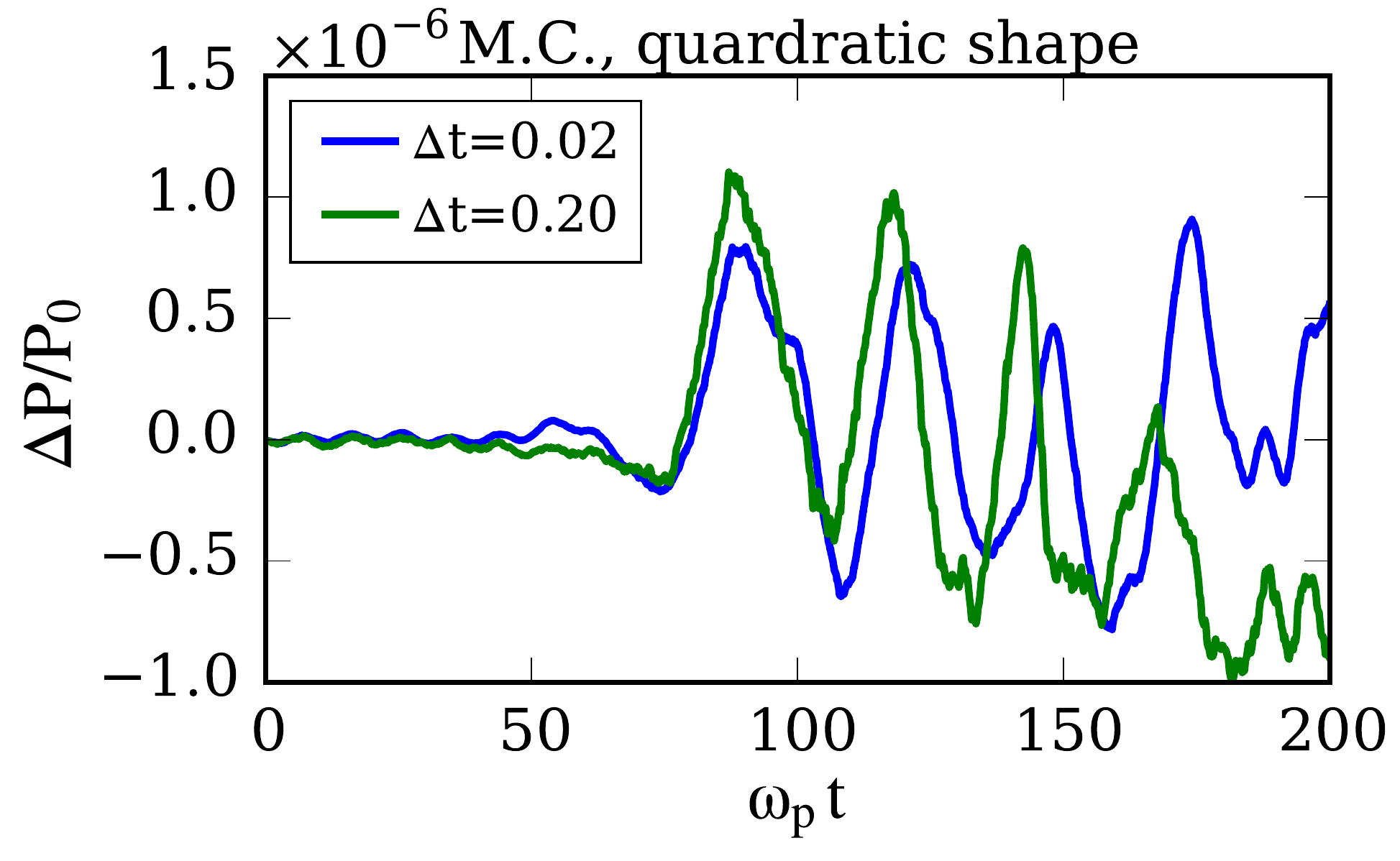}%
\end{minipage}

\begin{minipage}[b][0.18\paperwidth][c]{0.32\columnwidth}%
\includegraphics[width=1\columnwidth,height=0.115\paperheight]{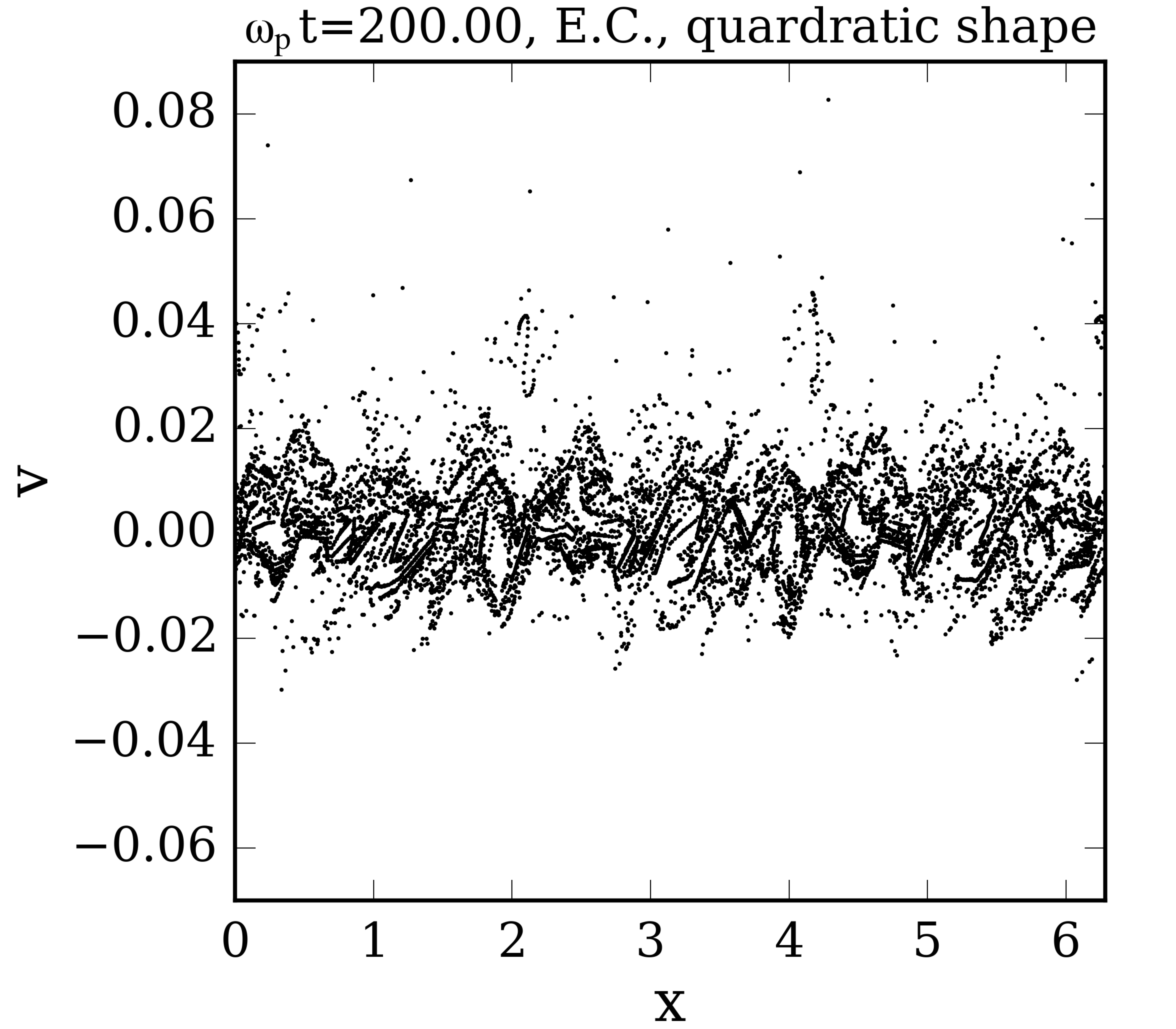}%
\end{minipage}%
\begin{minipage}[b][0.18\paperwidth][c]{0.67\columnwidth}%
\includegraphics[width=0.5\columnwidth,height=0.13\paperheight]{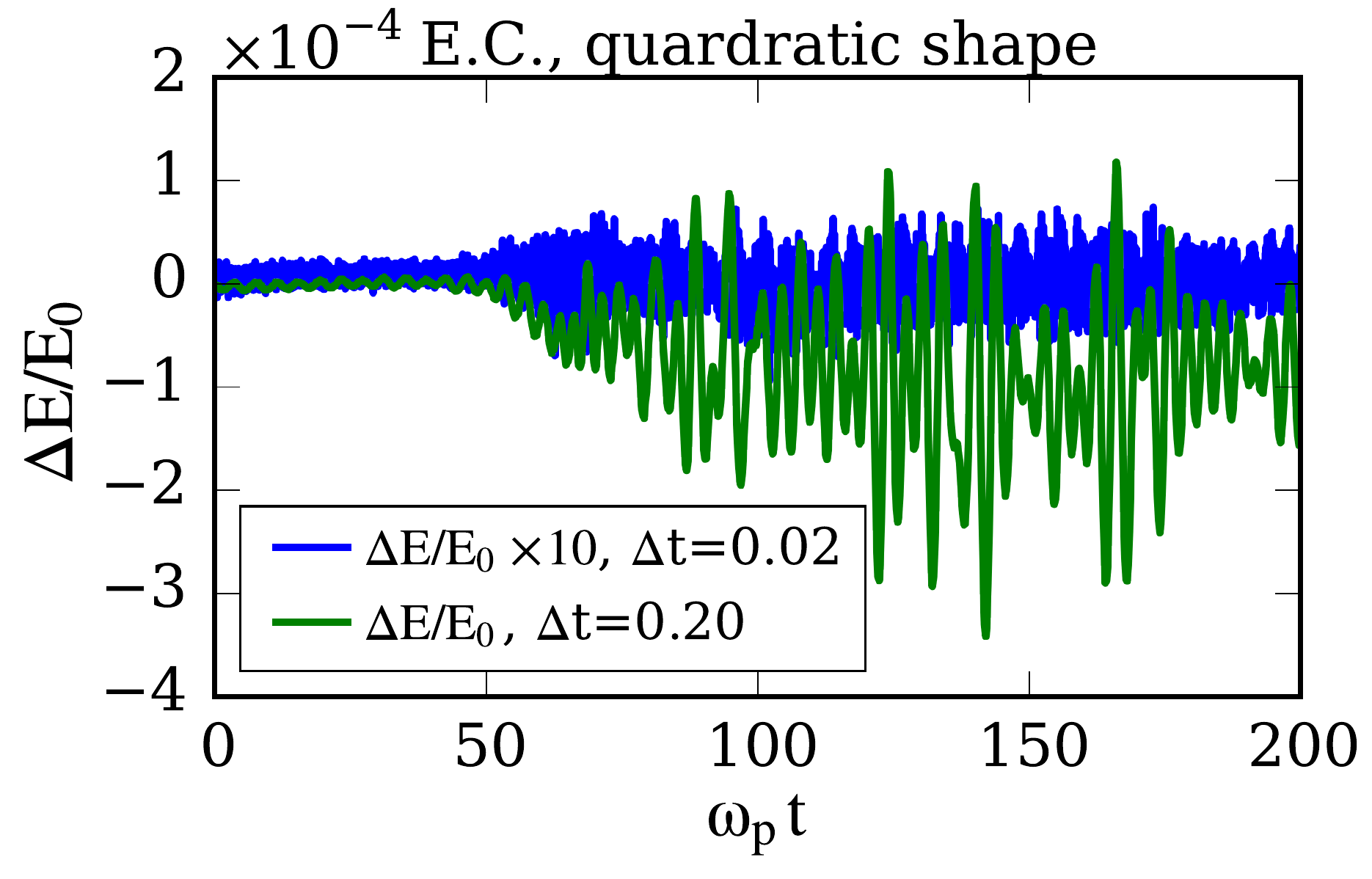}\includegraphics[width=0.5\columnwidth,height=0.13\paperheight]{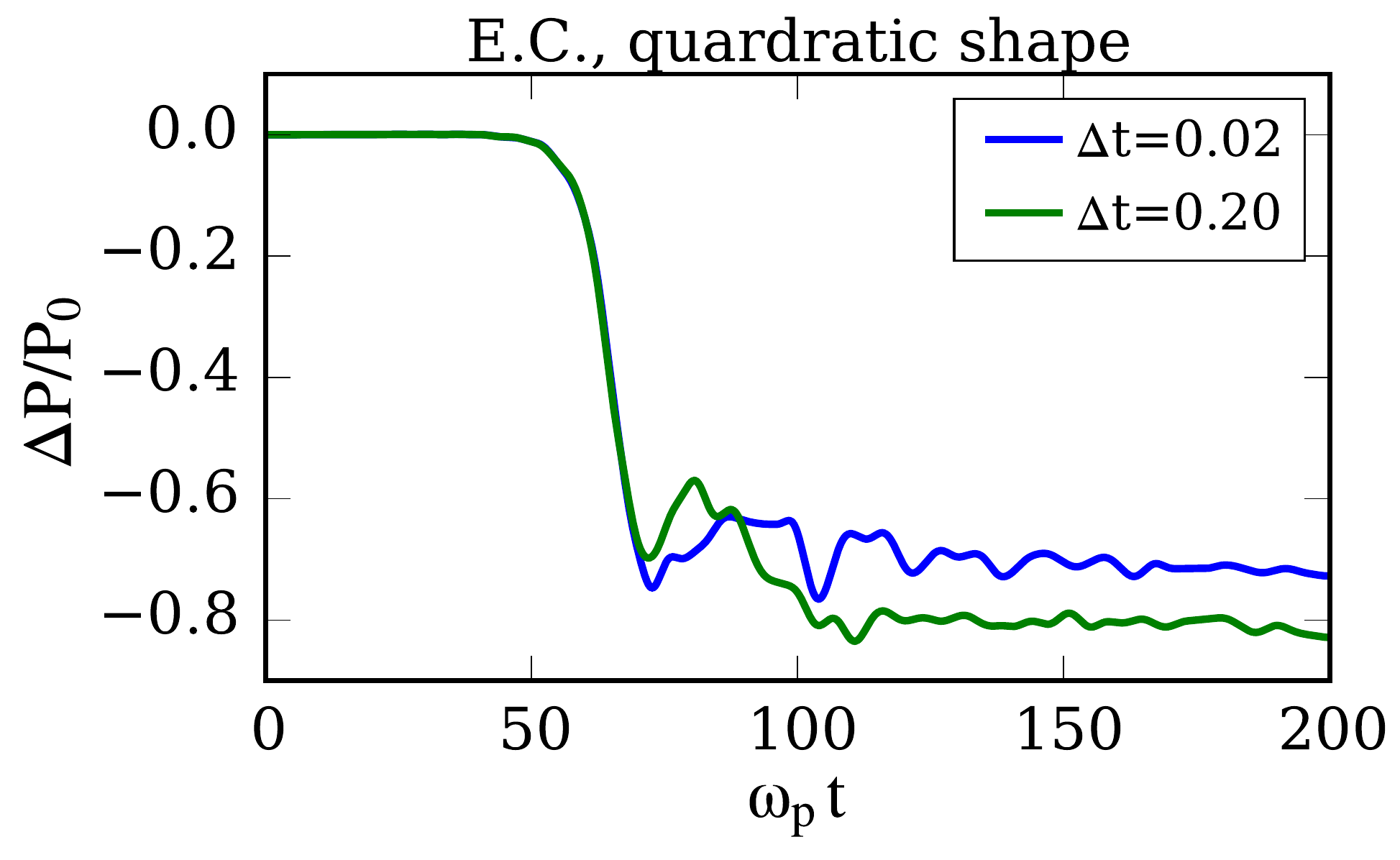}%
\end{minipage}

\begin{minipage}[b][0.18\paperwidth][c]{0.32\columnwidth}%
\includegraphics[width=1\columnwidth,height=0.115\paperheight]{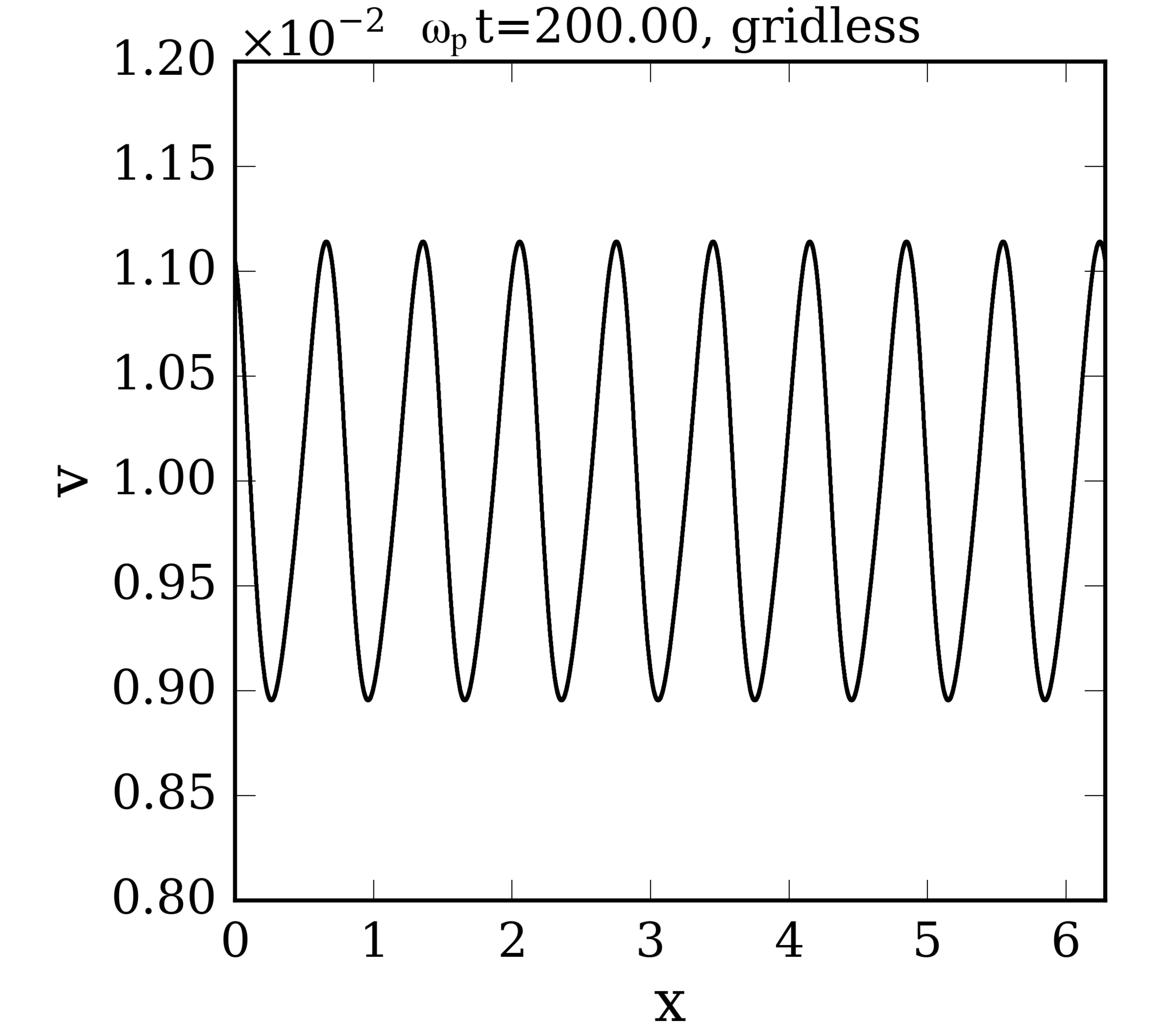}%
\end{minipage}%
\begin{minipage}[b][0.18\paperwidth][c]{0.67\columnwidth}%
\includegraphics[width=0.5\columnwidth,height=0.13\paperheight]{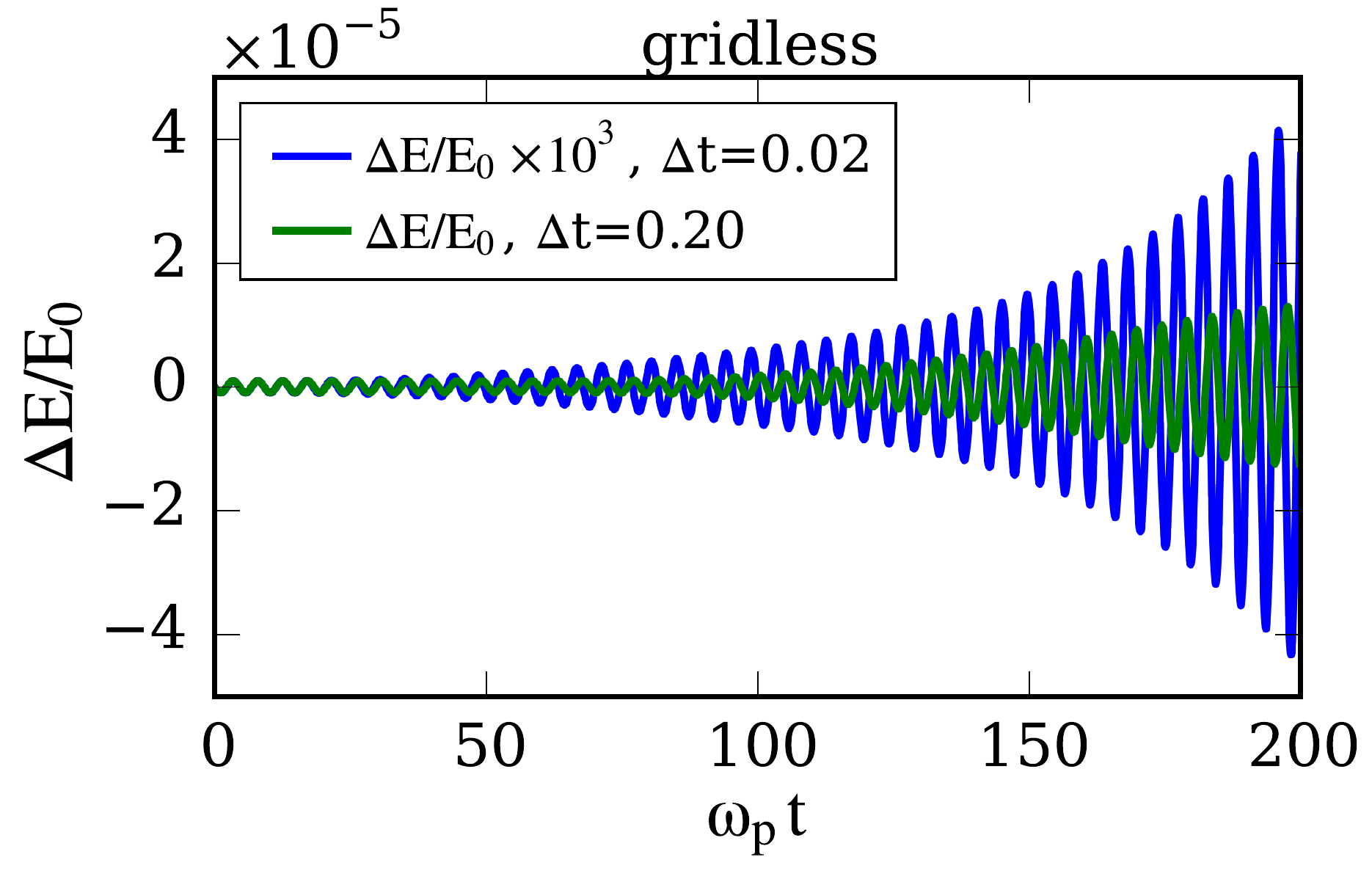}\includegraphics[width=0.5\columnwidth,height=0.13\paperheight]{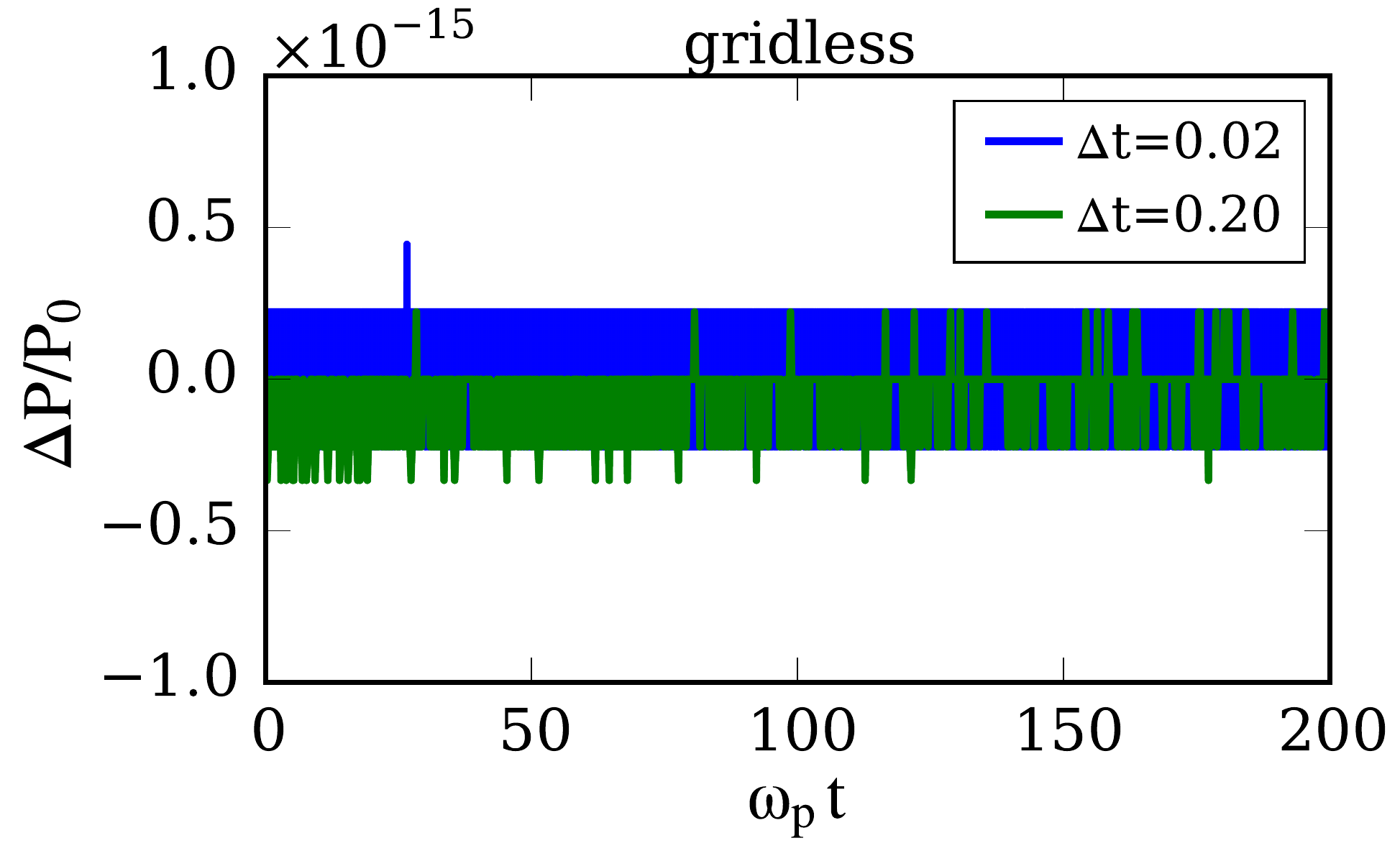}%
\end{minipage}

\caption{Electron phase space at the end of the simulation (left column), energy
(middle column) and momentum (right column) conservation from various
models. The energy and momentum conservation are shown for $\omega_{p}\Delta t=0.2$
(green curves) and $\omega_{p}\Delta t=0.02$ (blue curves). The three
rows from top to bottom correspond to (1) the momentum conserving
(M.C.) PIC model with quadratic particle shape; (2) the energy conserving
(E.C.) PIC model with quadratic particle shape; and (3) the gridless
model with quadratic particle shape. \label{fig:phase-space-engergy-momentum}}
\end{figure}

Results are compared in Fig. \ref{fig:mode-spectrum} for four simulations
with $\omega_{p}\Delta t=0.2$: (1) the momentum conserving PIC model
\cite{Birdsall-Langdon} with linear particle shape; (2) the momentum
conserving PIC model with quadratic particle shape; (3) the energy
conserving PIC model \cite{Lewis1970} with quadratic particle shape;
and (4) the electrostatic gridless model \cite{Decyk,Evstatiev2013}
with quadratic particle shape. For all PIC models compared, the initial
small-amplitude single-mode perturbation ($A\sim10^{-2}$ for mode
$K=9$) excites all the modes in the system at a level of $\sim10^{-6}$
immediately after the first step. Then, it is observed that mode $k=15$,
which is an alias mode from the second harmonics $K=18$ of the initial
perturbation in the adjacent Brillouin zone, grows exponentially in
all PIC models. The growth rate is higher for the lower order particle
shape used in the momentum conserving PIC models. Although the initial
amplitude of mode $k=15$ is higher in the energy conserving PIC model
with quadratic particle shape due to the lower order, i.e., linear,
particle shape used in field interpolation, the growth rate is about
the same as in the momentum conserving PIC with quadratic shape. Following
the growth of mode $k=15$, mode $k=6$ also grows at a similar rate,
which appears to be a result of the beating between modes $k=9$ and
$k=15$. Other exponential growing modes before $\omega_{p}t<50$,
e.g., modes $k=3$ and $k=12$ may also result from beating. These
modes grow to an amplitude comparable to the initial perturbation
during $\omega_{p}t\sim50-100$, after which a wide spectrum of modes
are excited. As a result of the interaction with a wide spectrum of
waves, the phase space plots in Fig. \ref{fig:phase-space-engergy-momentum}
show heated electron distributions in all PIC models, while the momentum
(energy) is relatively constant in the momentum (energy) conserving
model by design. Comparing with the gridless model for which the perturbation
remains single-mode and the electrons stay cold during the simulation,
it can be seen that the development of the wide-band spectrum and
the heating of the electrons are clearly numerical effects. Such numerical
artifacts result from the lack of spectral fidelity in the PIC model.

It should be noted that the gridless model improves not only the spectral
accuracy of the dynamics, but also the energy and momentum conservation
of the simulation model. In Fig. \ref{fig:phase-space-engergy-momentum},
the gross loss of energy (momentum) conservation in the momentum (energy)
conserving PIC models are correlated with the saturation of the unstable
mode growth. Furthermore, despite the $2^{nd}$ order accurate leap-frog
scheme used for the time advance in all models, only the gridless
model exhibits energy conservation (and possibly for momentum conservation)
that scales like $O(\Delta t^{3})$ in this numerical instability
test. For the gridless model, the energy conservation is improved
by about a factor of 1000 when the time step is decreased by a factor
of 10 to $\omega_{p}\Delta t=0.02$, while the momentum conservation
is already at the machine precision for $\omega_{p}\Delta t=0.2$
thus does not improve further. For the energy conserving model, the
heating in phase space, mode history and growth rate are essentially
unchanged when the time step is decreased to $\omega_{p}\Delta t=0.02$,
while the momentum conservation is only slightly improved and the
energy conservation is improved by about a factor of 10. The heating
and unstable mode growth behave similarly in the momentum conserving
model, but neither the energy or momentum conservation improves as
time step is decreased to $\omega_{p}\Delta t=0.02$. The superior
conservative properties of the gridless model may be due to the fact
that it can be derived from the action principle using a truncated
Fourier series \cite{Evstatiev2013} and the resulting finite dimension
dynamical system possesses spectral fidelity to the physical system.
The advantage of the variational models \cite{Evstatiev2013,Xiao2013,Xiao2015}
and the spectral fidelity in conservation properties and numerical
stability deserve further study but will not be pursued here.

\begin{figure}
\begin{minipage}[b][0.18\paperwidth][c]{0.67\columnwidth}%
\includegraphics[width=0.5\columnwidth,height=0.125\paperheight]{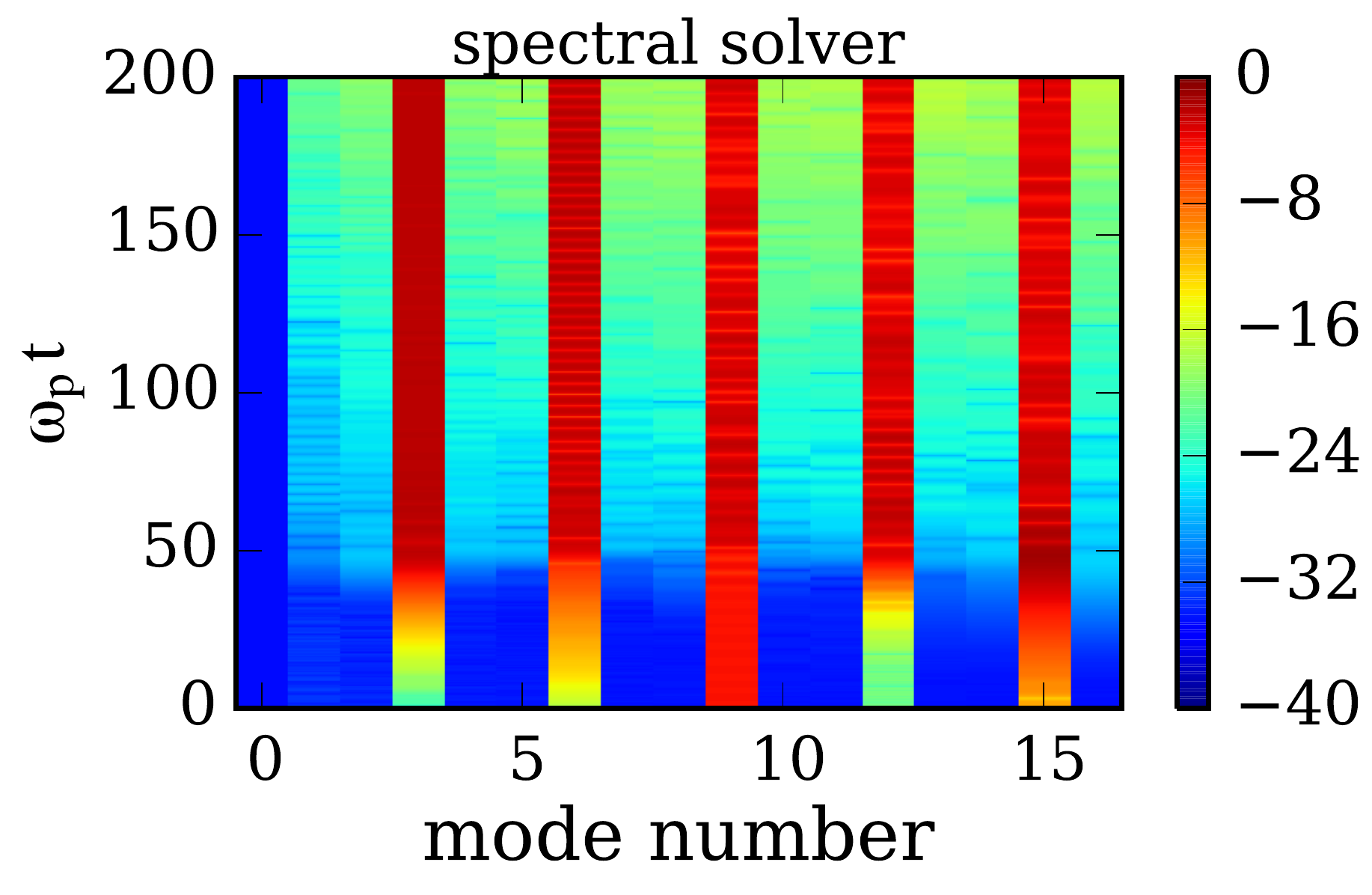}\includegraphics[width=0.5\columnwidth,height=0.125\paperheight]{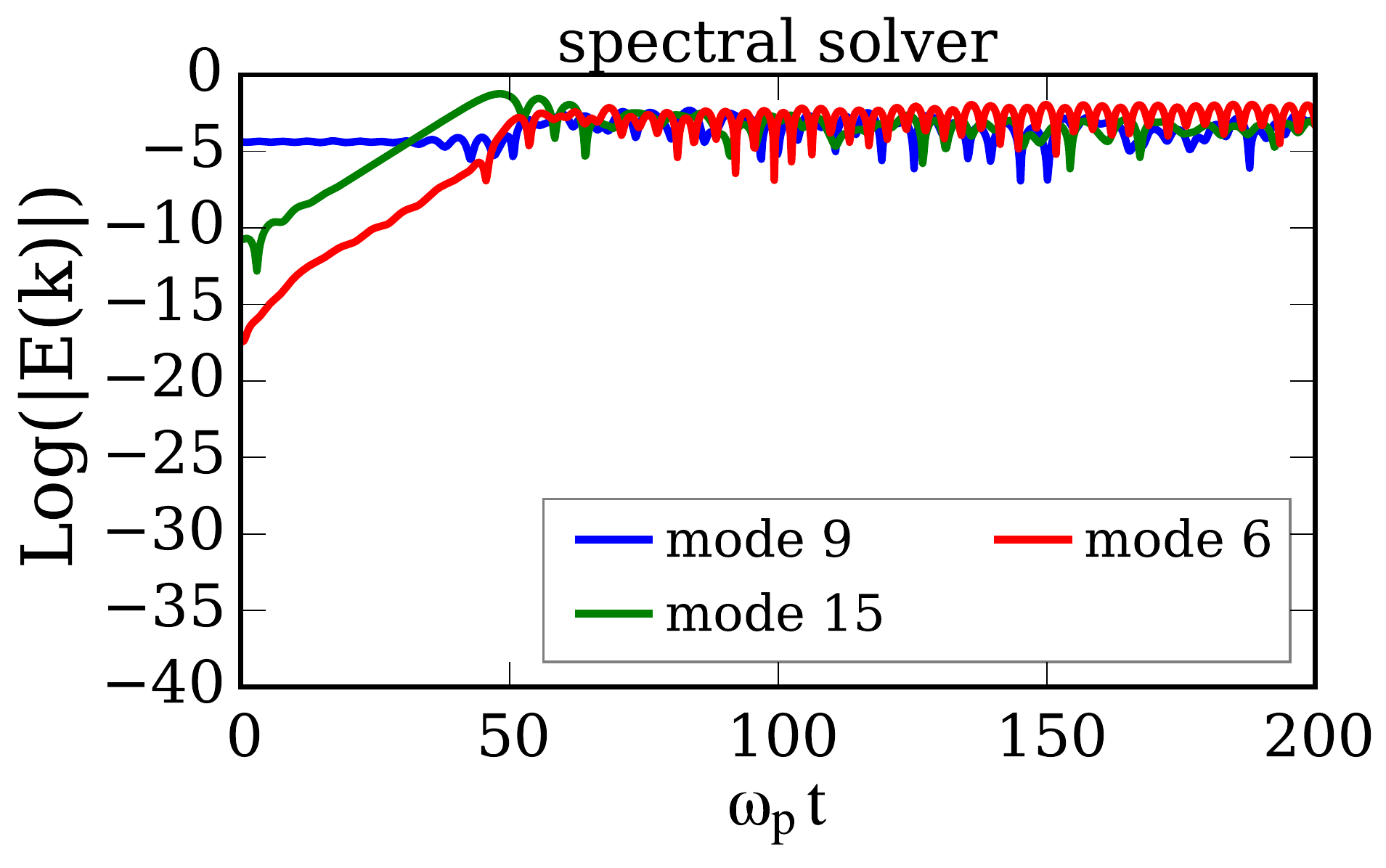}%
\end{minipage}%
\begin{minipage}[b][0.18\paperwidth][c]{0.32\columnwidth}%
\includegraphics[width=1\columnwidth,height=0.12\paperheight]{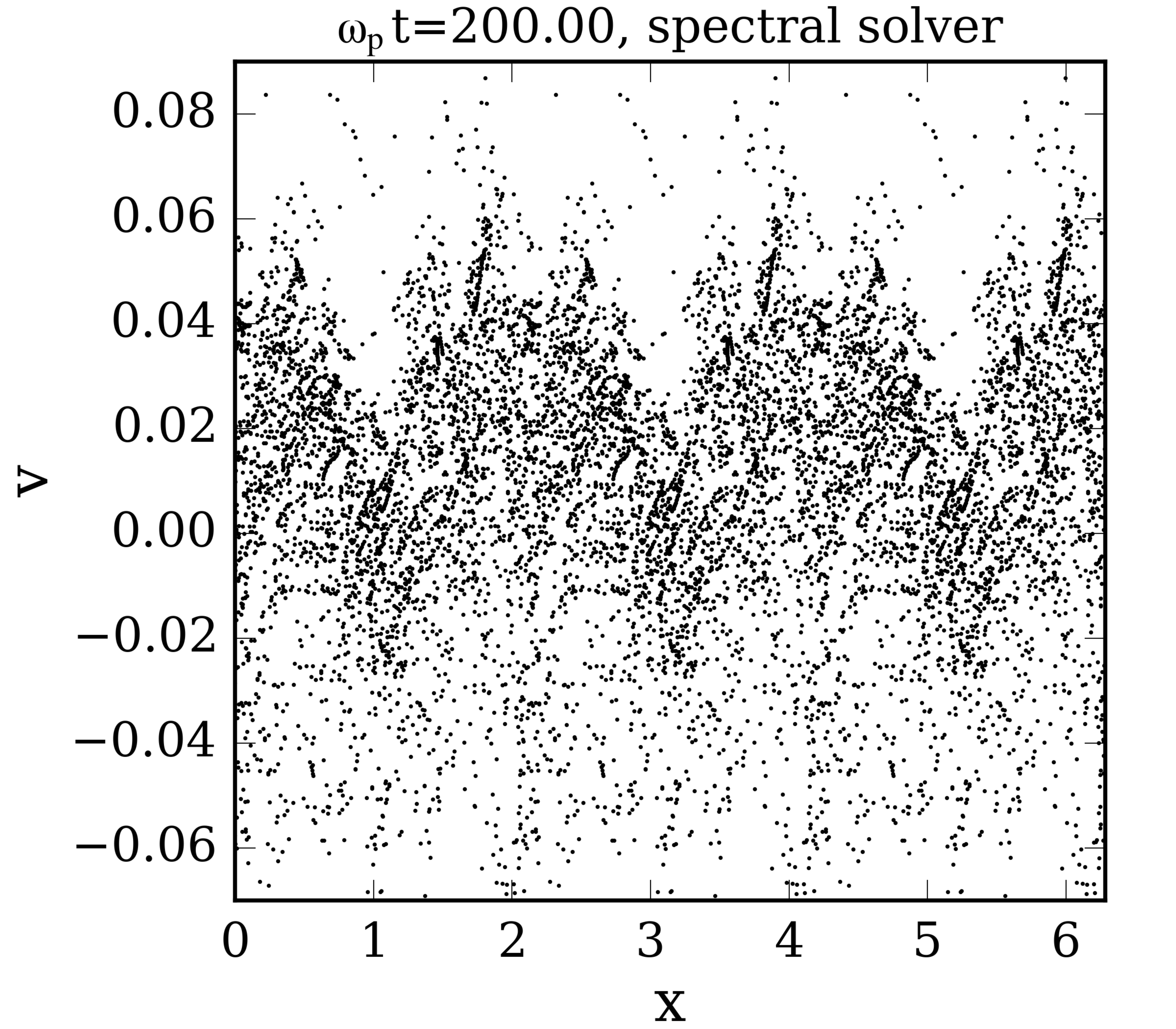}%
\end{minipage}

\begin{minipage}[b][0.18\paperwidth][c]{0.67\columnwidth}%
\includegraphics[width=0.5\columnwidth,height=0.125\paperheight]{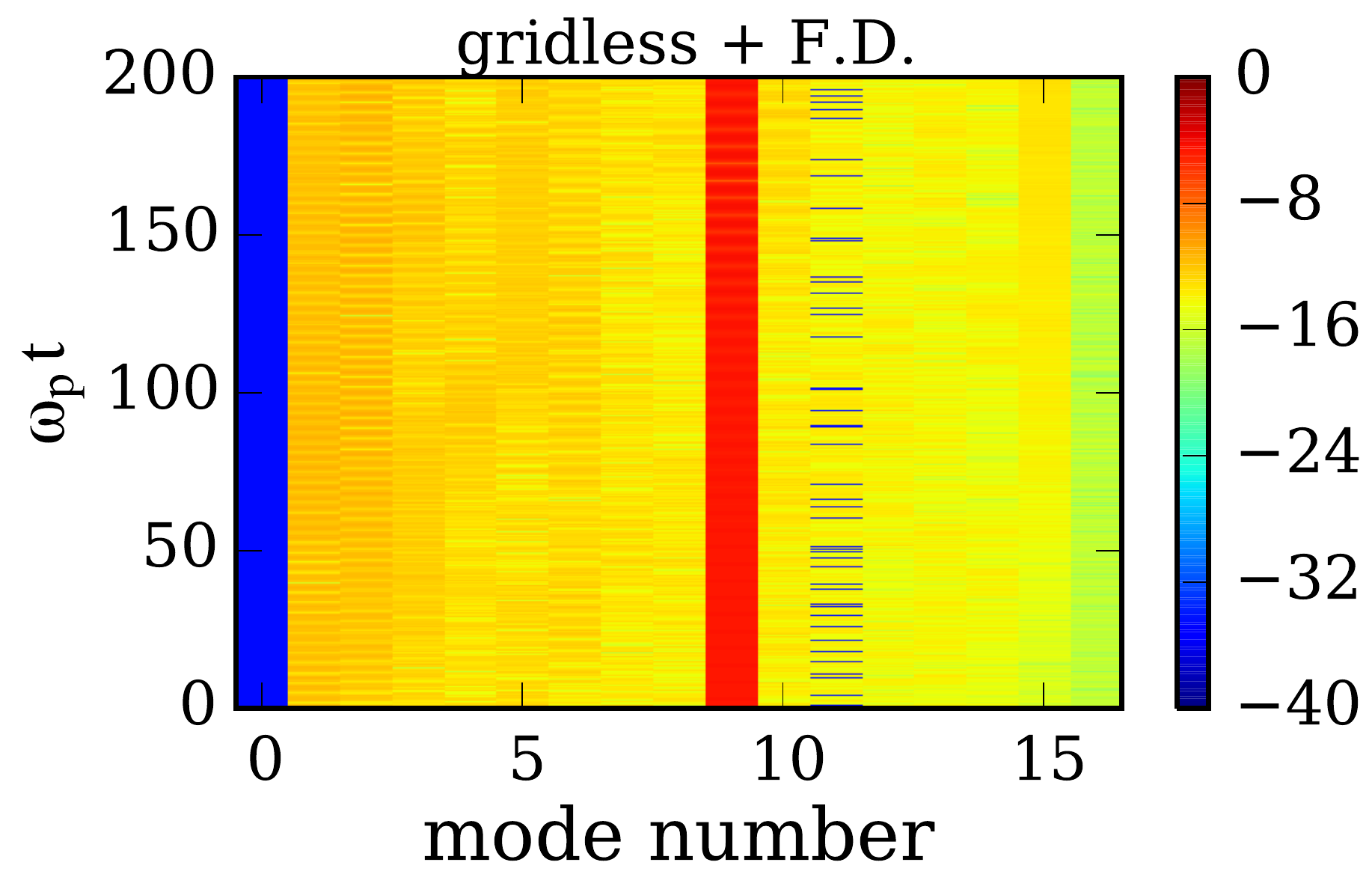}\includegraphics[width=0.5\columnwidth,height=0.125\paperheight]{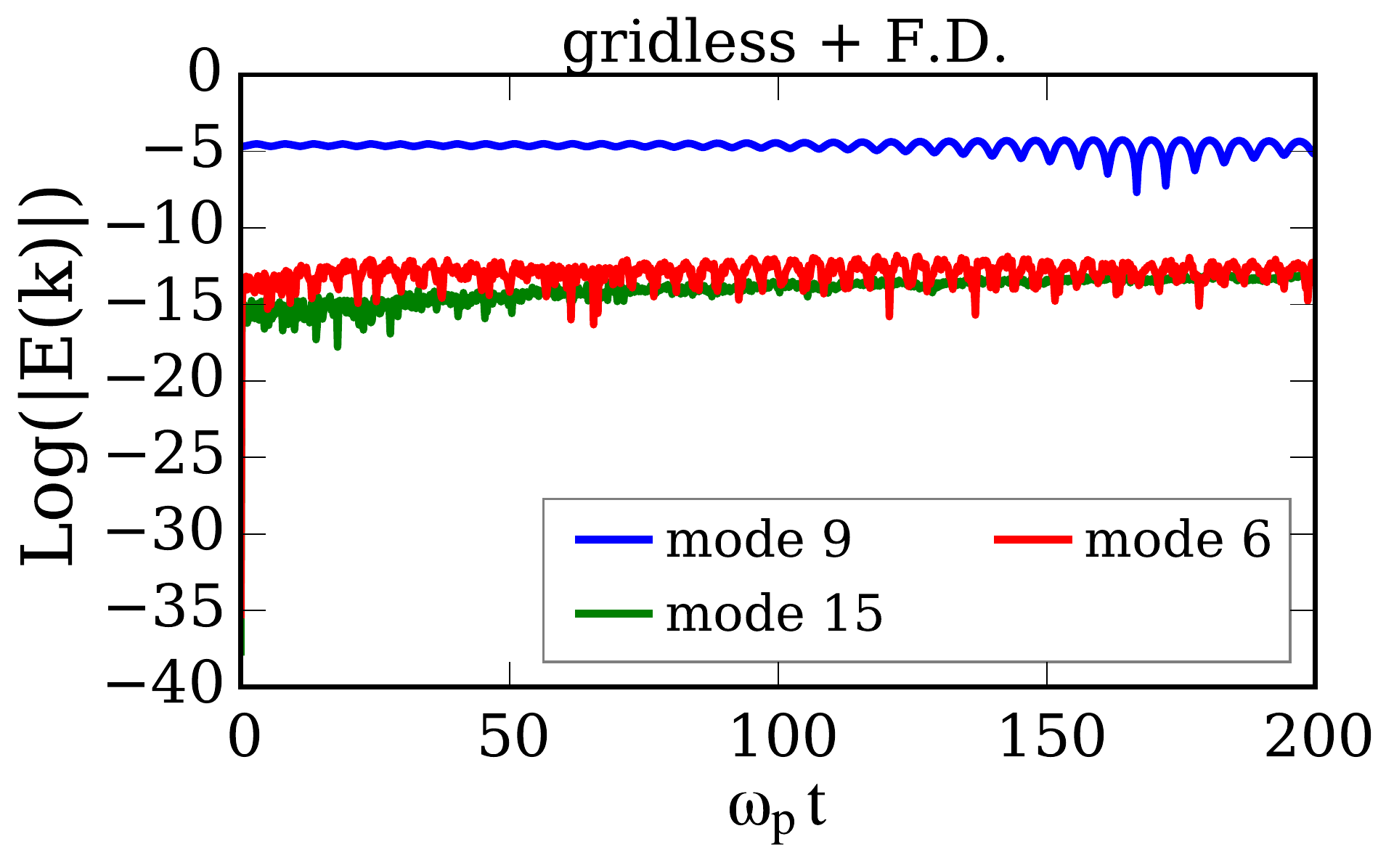}%
\end{minipage}%
\begin{minipage}[b][0.18\paperwidth][c]{0.32\columnwidth}%
\includegraphics[width=1\columnwidth,height=0.12\paperheight]{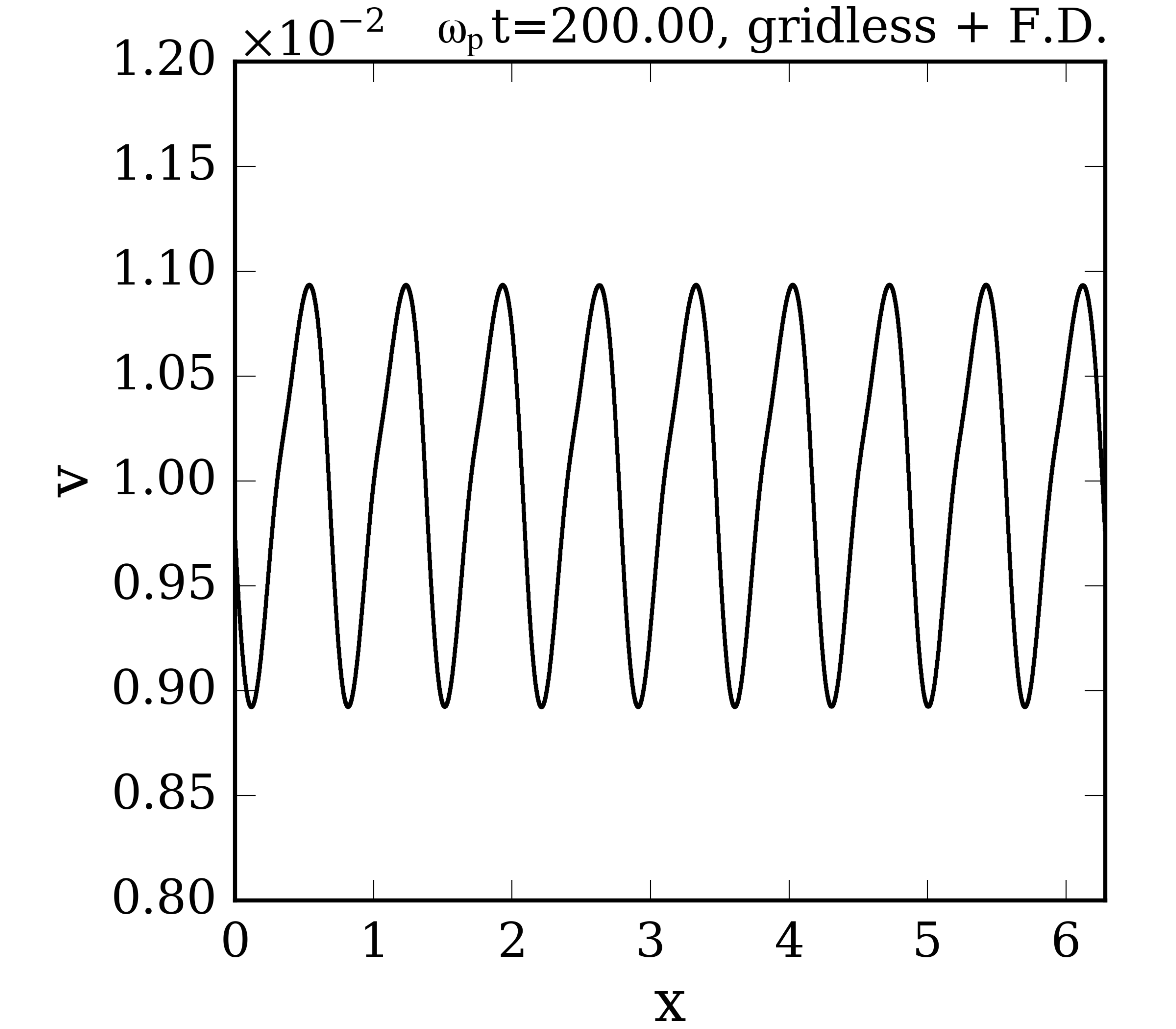}%
\end{minipage}

\caption{The mode spectrum (left column), amplitudes of mode $9,\,15,\,6$
(middle column), and electron phase space from a PIC model (right
column) with (1) a spectral solver (top row); (2) gridless deposition
and field interpolation (bottom row). \label{fig:PIC-mods}}

\end{figure}

To understand the relative importance of the spectral fidelity in
the deposition/interpolation and the field solver, in particular the
role of amplitude and phase errors, simulations are carried out using
the momentum conserving PIC model with the deposition/interpolation
and the field solver implemented in gridless mode respectively. The
results are shown in Fig. \ref{fig:PIC-mods}. Operating the field
solver in the gridless mode effectively turns the finite difference
PIC model into a spectral PIC model, which improves upon the systematic
amplitude error and random spectral errors in a finite difference
solver. The benefit is seen in Fig. \ref{fig:PIC-mods} where the
dominate modes are those separated modes (mode $3,\,6,\,9,\,12,\,15$)
up to $\omega_{p}t=200$, but the onset of whole spectrum excitation,
similar to those shown in Fig. \ref{fig:mode-spectrum}, is delayed
to $\omega_{p}t\sim400$ in a longer simulation (not shown). The growth
rate of mode $15$ is about the same as in the case with a finite
difference solver and the electron phase space shows similar heating.
Thus a spectral solver that improves amplitude and random errors does
not qualitatively change the characteristics of the simulation. Upon
adopting the gridless deposition and field interpolation (but keeping
the finite difference field solver), which improves both the systematic
phase and amplitude errors, the results are essentially similar to
those of the gridless model, with a higher floor of the background
modes. This confirms that the major loss of the spectral fidelity
in PIC occurs in the deposition and the field interpolation where
their phase errors play important roles in FGI. 

The dynamics shown in Fig. \ref{fig:mode-spectrum} involves nonlinear
coupling of a large range of modes as the system evolves, but at the
early time the unstable dynamics grows predominately from the interaction
between density mode $k=\pm15$ and the collective mode $K=18$ which
is excited by the initial perturbation at $K=9$. To quantitatively
determine the behavior of such interaction, we can set $\delta x(K,t)=Ae^{i\omega_{r}t}=A_{0}e^{i(\omega t+\theta_{0})}$,
where $\omega=\omega_{r}-i\omega_{i}$, $\delta x(K,0)=A_{0}e^{i\theta_{0}}$
and $A=A_{0}e^{\omega_{i}t}$. Inserting this into the coupled Eqs.
(\ref{eq:eq-motion}), (\ref{eq:density-collective-coupling-PIC})
and (\ref{eq:field-collective-coupling-PIC}), and ignoring the particle
aliasing effect then $N\gg1$, one obtains,

\begin{align}
-A_{0}\omega^{2}e^{i(\omega t+\theta_{0})} & =\frac{N_{p}^{2}Q^{2}}{N_{g}m}\underset{q'}{\sum}\underset{q}{\sum}\frac{-i}{[k_{q'}]}\tilde{s}(k_{q})\cdot\tilde{s}(k_{q'})\delta\left(k_{q}-\nu_{K}K\right)\delta\left(K-k_{q'}-\nu_{K'}K\right)\nonumber \\
 & \cdot e^{i(\nu_{K}+\nu_{K'})\cdot(\omega_{r}t+\theta_{0}+\pi/2)}J_{\nu_{K}}\left(-2k_{q}A_{0}e^{\omega_{i}t}/N_{p}\right)J_{\nu_{K'}}\left(2k_{q'}A_{0}e^{\omega_{i}t}/N_{p}\right),\label{eq:nonlinear-dispersion}
\end{align}

This can be simplified by equating the time varying phase on both
sides of Eq. (\ref{eq:nonlinear-dispersion}) and setting the arguments
of the delta functions to zero, i.e., using the following relationships
for the mode coupling,

\begin{align}
\nu_{K}+\nu_{K'} & =1,\nonumber \\
q & =q',\nonumber \\
\nu_{K} & =k_{q}/K,\nonumber \\
\nu_{K},q & \in\mathbf{Z},\label{eq:coupling-relation}
\end{align}

\noindent to give, 

\begin{equation}
\omega^{2}e^{\omega_{i}t}\Delta n_{K}^{0}=-2K\underset{\nu_{K}=k_{q}/K}{\sum}\frac{\tilde{s}^{2}(\nu_{K}K)}{[\nu_{K}K]}J_{\nu_{K}}\left(-\nu_{K}e^{\omega_{i}t}\Delta n_{K}^{0}\right)J_{(1-\nu_{K})}\left(\nu_{K}e^{\omega_{i}t}\Delta n_{K}^{0}\right).\label{eq:nonlinear-dispersion-simplified}
\end{equation}

\noindent where the normalization $n\rightarrow n/n_{p}$, $\omega\rightarrow\omega/\omega_{p}$,
$t\rightarrow\omega_{p}t$, $Q\rightarrow N_{g}/N_{p}$, $Q/m\rightarrow1$,
is used and $\Delta n_{K}^{0}=2KA_{0}/N_{p}$ is the normalized initial
density perturbation due to the collective mode $K$. The summation
in Eq. (\ref{eq:nonlinear-dispersion-simplified}) is over all integers
$\nu_{K}$ and $q$ that satisfy $\nu_{K}=k_{q}/K=(k+qk_{g})/K.$
For the above example, $k=\pm15$, $K=18$ and $k_{g}=33$, we can
rewrite the infinite summation to be,

\[
\underset{\nu_{K}=k_{q}/K}{\sum}=\underset{l\in\mathbf{Z}}{\sum},
\]
with $q=6l\mp1$, $\nu_{K}=11l\mp1$. For $|x|\lesssim0.7$, $|J_{\nu_{K}}(-\nu_{K}x)J_{1-\nu_{K}}(\nu_{K}x)|$
drops quickly as $|\nu_{K}|$ increases, so in practice only the $l=0$
term needs to be included in the above sum until the coupling is nonlinear. 

Thus, Eq. (\ref{eq:nonlinear-dispersion-simplified}) becomes,

\begin{equation}
\omega^{2}e^{\omega_{i}t}\Delta n_{K}^{0}\approx-2K\left[\dfrac{\tilde{s}^{2}(-K)}{[-K]}J_{-1}\left(e^{\omega_{i}t}\Delta n_{K}^{0}\right)J_{2}\left(-e^{\omega_{i}t}\Delta n_{K}^{0}\right)+\dfrac{\tilde{s}^{2}(K)}{[K]}J_{1}\left(-e^{\omega_{i}t}\Delta n_{K}^{0}\right)J_{0}\left(e^{\omega_{i}t}\Delta n_{K}^{0}\right)\right].\label{eq:nonlinear-dispersion-sim}
\end{equation}

The first (second) term on the right hand side results from the coupling
through mode $k=15$ ($k=-15$). The coupling from the second term
is much stronger than that from the first term through mode when the
perturbation amplitude is small, i.e., when $e^{\omega_{i}t}\Delta n_{K}^{0}\ll1$.
In fact the second term leads to absolute instability ($\omega_{r}=0$),
while the first term leads to oscillatory solution ($\omega_{i}=0$).
Therefore, an initially small perturbation grows exponentially until
saturation and then oscillates at that level. The linear growth rate
can be obtained by dropping the first term and Taylor expanding the
Bessel functions in the second term, which gives,

\begin{equation}
\omega_{i}\approx\tilde{s}(K)\sqrt{\left|K/[K]\right|},\label{eq:growth-rate}
\end{equation}
which does not depend on the drift velocity $V_{0}$. When $e^{\omega_{i}t}\Delta n_{K}^{0}\approx1.84$,
the RHS of (\ref{eq:nonlinear-dispersion-sim}) is zero, therefore
the normalized instability saturation time is 
\begin{equation}
t_{sat}\approx\text{ln}(1.84/\Delta n_{K}^{0})/\omega_{i}.\label{eq:satuation-time}
\end{equation}
Note that the linear FGI in section \ref{sub:linear-FGI} is different
than the case discussed here. For the former case, unstable modes
grow independently from particle noise. Each unstable mode involves
the fundamental of the collective motion that can be resolved by the
grid (harmonics have been dropped in the linear analysis), and the
corresponding density perturbation which also contains alias modes
that cause the instability. For the latter case, the collective motion
cannot be resolved by the grid. Therefore the density mode contains
alias modes only and the instability can be more severe than the former
case, e.g., for the simulation with large amplitude perturbation.
Also, unlike the former FGI case which is convective and has stability
domains, the latter instability is absolute and has no stability domain
as long as $K>k_{g}$ and $\tilde{s}(K)>0$.

\begin{figure}
\centering\includegraphics[width=0.6\columnwidth]{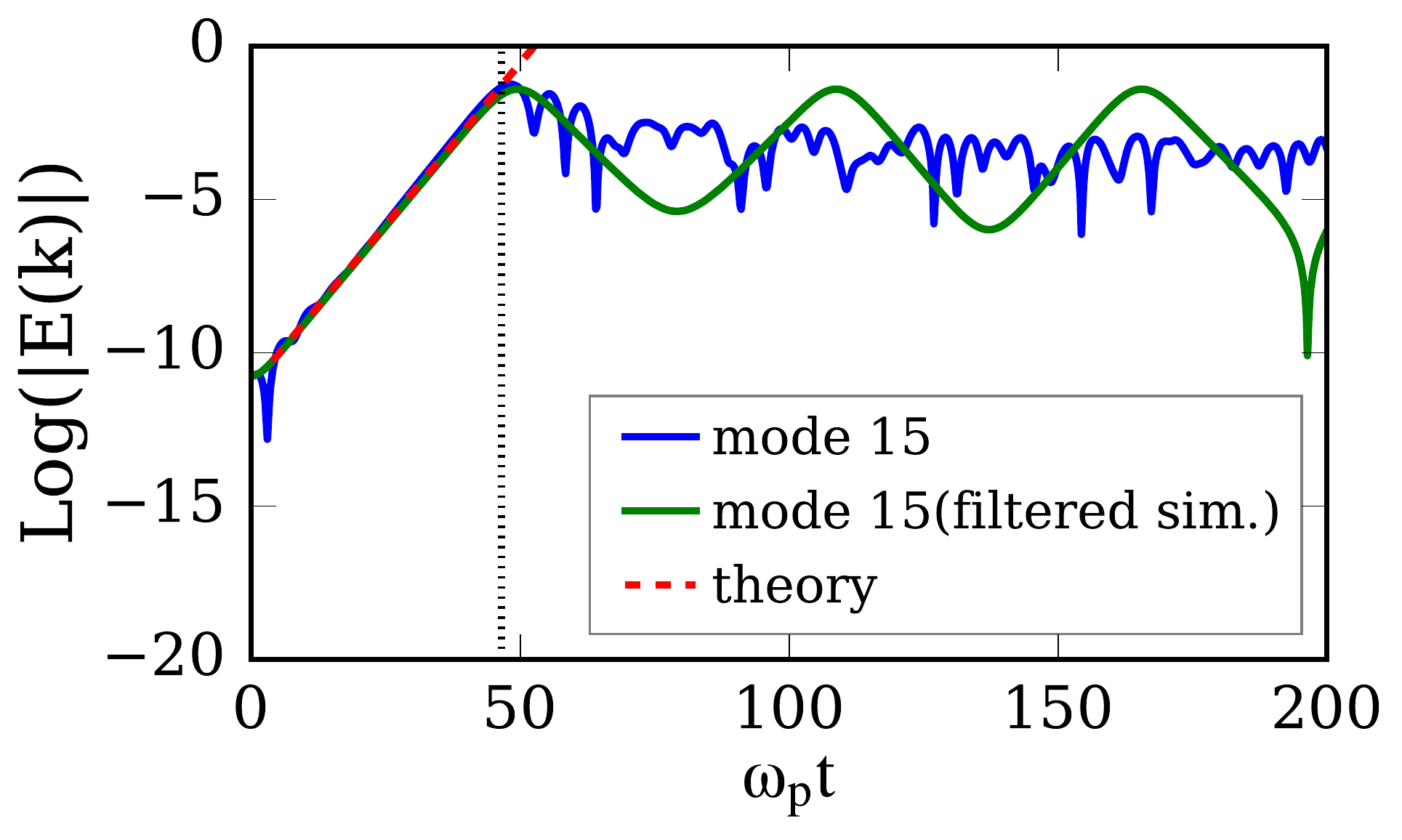}

\caption{Comparison of the FGI growth (red, dashed, $\omega_{i}\approx0.21$)
and the saturation time (black, dotted, $t_{sat}\approx46.5$) predicted
by Eq (\ref{eq:growth-rate}) and Eq. (\ref{eq:satuation-time}),
respectively, with $ln|E(k=15)|$ in a PIC simulation using a 2nd
order particle shape and a spectral field solver (green curve). Eq.
(\ref{eq:cold-dispersion-2terms}) gives $\omega_{i}\approx0.04$.
The simulation is initialized with a perturbation in the collective
mode $K=18$ such that $\Delta n_{K}^{0}=10^{-4}$. Only the collective
mode $K=18$ and the electric field modes $k=\pm15$ are kept in the
simulation. Other numerical parameters are the same as in the simulation
in Fig. \ref{fig:PIC-mods} that employs a spectral solver, no mode
filtering and an initial perturbation in mode $K=9$ such that $\Delta n_{K}^{0}=10^{-2}$
(blue curve).\label{fig:sim-comparision}}
\end{figure}

The linear growth rate from Eq. (\ref{eq:growth-rate}), $\omega_{i}\approx0.21$,
and the instability saturation time from Eq. (\ref{eq:satuation-time}),
$t_{sat}\approx46.5$, are compared to the simulation shown in Fig.
\ref{fig:sim-comparision}, in which only collective mode $K=18$
and electric field modes $k=\pm15$ are kept and an initial perturbation
in the mode $K=18$ is applied. Solving Eq. (\ref{eq:cold-dispersion-2terms})
for modes $k=\pm15$ gives $\omega_{i}\approx0.04$ which is much
smaller than the actual grow rate. This simulation result is also
in close agreement with the example shown in Fig. \ref{fig:PIC-mods},
for the dynamics of mode $k=15$ in the linear growth stage up to
the saturation of the instability, indicating the early unstable dynamics
is due to the coupling between $K=18$ and $k=\pm15$ modes. However,
the detail dynamics after saturation in the latter simulation involves
more mode coupling, thus requiring a more complete treatment similar
to the above analysis. 

\section{Summary\label{sec:Summary}}

In this paper, the origin of the FGI is studied by employing particle
and spectral resolutions into the dynamics of the 1D electrostatic
PIC model and by contrasting the spectral fidelity of the PIC model
with respect to the underlying physical system (or the gridless model).
The particle resolution can be either adopted for individual particle
or for the collective motion. The use of the particle and spectral
resolutions are not the only options for this purpose, but are suitable
ones as the particle dynamics consist of pair-wise interactions and
can also be viewed as collective wave-particle interactions. 

At the individual particle level, the charge deposition and field
interpolation operations of the PIC models exhibit systematic spectral
errors relative to the physical system (or gridless model) due to
the existence of the spatial alias modes from the use of the discrete
grid in conjunction with Lagrangian particles in continuous space.
In principle, these errors can be calculated for arbitrary particle
shape, and surprisingly, they have relatively compact analytic forms
shown in Eq. (\ref{eq:spect-err-B-spline}) for the B-spline particle
shapes and in Eq. (\ref{eq:spect-err-Gaussian}) for the Gaussian
particle shape. These forms allow us to understand the effects in
the spectral domain from using shaped particles in the PIC models.
It is observed that these systematic spectral errors depend on both
$k\Delta x$ and the particle's normalized position in the cell. These
errors are directly related to the instability growth rate but they
improve slowly with the increase of the order (smoothness) of the
B-spline particle shape. Another benefit of using higher order particle
shape is the stronger damping to the short-wavelength modes in the
fundamental Brillouin zone which can enlarge the stability domain.
Such a damping effect also applies to the physical modes and may be
achieved using a spatial mode filter on the grid which has a lower
computation cost and does not depend on individual particle's position.

At the collective motion level, as it is shown in Eqs. (\ref{eq:density-collective-coupling-PIC})
and (\ref{eq:field-collective-coupling-PIC}), the charge deposition
and field interpolation introduce unphysical mode couplings which
eventually result in amplitude and phase errors in the collective
modes. This is also due to the aliasing of the spatial modes. Within
the framework of collective wave-particle interactions, Eqs. (\ref{eq:density-collective-coupling}),
(\ref{eq:field-collective-coupling}) and (\ref{eq:eq-motion}) are
the coupled equations describing the evolution of the collective modes
in the physical system, while Eqs. (\ref{eq:density-collective-coupling-PIC})
and (\ref{eq:field-collective-coupling-PIC}) and the discrete time
version of Eq. (\ref{eq:eq-motion}) are the counterparts for the
PIC model, whose convergence with respect to time step can be verified
to rule out the time discretization as a possible origin of the FGI.
In fact, early works on FGI already indicated that the numerical heating
rate is independent of the time step with the leap-frog advance, while
our result also shows that the growth rates of the unstable modes
are not affected by the time step, making it clear the cause of FGI
can only be the lack of spectral fidelity of Eqs. (\ref{eq:density-collective-coupling-PIC})
and (\ref{eq:field-collective-coupling-PIC}) as compared to Eqs.
(\ref{eq:density-collective-coupling}) and (\ref{eq:field-collective-coupling}).
In this regard, the difference between the PIC model and the physical
system (or gridless model) is technical, not fundamental --- with
a properly-chosen particle shape such that $s(k)=0$ for $|k|>k_{g}/2$,
we recover Eqs. (\ref{eq:density-collective-coupling}) and (\ref{eq:field-collective-coupling})
from Eqs. (\ref{eq:density-collective-coupling-PIC}) and (\ref{eq:field-collective-coupling-PIC})
for $|k|\le k_{g}/2$ , thus making the PIC models effectively a gridless
model. This opens the question of the optimal compact particle shapes
for the PIC model. Finally, as with many physical instabilities which
result from the constructive feedback of modes, it can be understood
that the systematic spectral errors play important roles in the development
of FGI. This finding is verified in the simulation comparison with
the gridless model and may help the design of a PIC model with better
stability.

\section{Acknowledgements}

We are grateful to many of our colleagues for their valuable comments
to this work. One of the authors (C. H.) would like to thank Prof.
D. F. Escande for his lectures in Los Alamos National Laboratory (LANL)
which motivated the analysis using the collective modes. This work
is supported by the U.S. Department of Energy through the LDRD program
at LANL.

\appendix

\section{Particle shape function and interpolation function\label{sub:A1}}

The function $S(\bm{r})$ is used to define particle shape in phase
space (for PIC, the shape in velocity dimension is assumed to be a
Delta function) and the function $W(\bm{r_{g}},\,\bm{r})$ is used
in the charge deposition and field interpolation. The PIC method can
be derived from the Vlasov equation using the distribution function
of shaped particles \cite{Lapenta2006}, which reveals that any symmetric
function, $S(\bm{r})=S(-\bm{r})$, or $\tilde{S}(\bm{k})=\tilde{S}(-\bm{k})$,
can be used as long as the averaged fields on the particle is defined
as in Eq. (6) of \cite{Lapenta2006}. Self-force consideration and
momentum conservation require that the same function be used for charge
deposition and field interpolation. The other requirement on the shape
function is $\int S(\bm{r})d\bm{r}=1$, or $\tilde{S}(\bm{k}=0)=1$,
while for a valid interpolation function in PIC, charge conservation
requires $\underset{\bm{r_{g}}}{\sum}W(\bm{r_{g}},\,\bm{r})=1$. From
the Poisson summation %
, this requires 
\[
\underset{\bm{r_{g}}}{\sum}W(\bm{r_{g}},\,\bm{r})=\underset{\mathbf{\mathbf{q}}}{\sum}\tilde{W}(\mathbf{q}\cdot\bm{k_{g}})e^{i(\mathbf{q}\cdot\bm{k_{g}})\cdot\mathbf{r}}=1,
\]
which is equivalent to $\tilde{W}(\bm{k}=0)=1$ and $\tilde{W}(\mathbf{q}\cdot\bm{k_{g}})=0$
for $n^{2}+m^{2}+l^{2}\neq0$. Therefore a particle shape function
is not necessarily an interpolation function, e.g., a linear particle
shape function with a width of half the grid size is not a valid interpolation
function. But valid interpolation functions are true particle shape
functions, e.g., the family of B-spline interpolation functions used
in PIC. Furthermore, the convolution of an arbitrary shape function
$S(\bm{r})$ with these B-spline interpolation functions results in
valid interpolation functions. In this paper, we will not distinguish
the particle shape function and the interpolation function unless
necessary and will use $S(\bm{r})$ for clarity.

\section{Aliasing and phase factor\label{sub:A2}}

From the Poisson summation formula $\stackrel[n=-\infty]{\infty}{\sum}e^{ik_{0}n\Delta x}=\stackrel[q=-\infty]{\infty}{\sum}\delta(q-k_{0}\Delta x/2\pi)=\frac{2\pi}{\Delta x}\stackrel[q=-\infty]{\infty}{\sum}\delta(k_{0}-q\frac{2\pi}{\Delta x})$,
we have, 

\begin{equation}
\underset{\bm{r_{f}}}{\sum}e^{i(\bm{k'}-\bm{k})\cdot\bm{r_{f}}}=8\pi^{3}\underset{\mathbf{q}}{\sum}\delta(\bm{k'}-\bm{k_{q}}).\label{eq:Poisson-summation}
\end{equation}
We have also used $\Delta x\Delta y\Delta z=1$. The infinite sum
in Eq. (\ref{eq:rho-fourier}) is related to the infinite sum on the
reference grid, Eq. (\ref{eq:Poisson-summation}), by a phase factor,

\begin{equation}
\underset{\bm{r_{\rho}}}{\sum}e^{i(\bm{k'}-\bm{k})\cdot\bm{r_{\rho}}}=e^{i(\bm{k'}-\bm{k})\cdot\Delta\bm{r_{\rho}}}\cdot\underset{\bm{r_{f}}}{\sum}e^{i(\bm{k'}-\bm{k})\cdot\bm{r_{f}}}=8\pi^{3}\underset{\mathbf{q}}{\sum}\psi(\mathbf{q}\cdot\bm{k_{g}},\,\Delta\bm{r_{\rho}})\delta(\bm{k'}-\bm{k_{q}})\label{eq:Poisson-summation-grid-1}
\end{equation}
where $\bm{r_{\rho}}=\bm{r_{f}}+\Delta\bm{r_{\rho}}$ and $\psi(\mathbf{q}\cdot\bm{k_{g}},\,\Delta\bm{r_{\rho}})=e^{i(\mathbf{q}\cdot\bm{k_{g}})\cdot\Delta\bm{r_{\rho}}}$
is the phase factor. For zero offset $\Delta\bm{r_{\rho}}=(0,\,0,\,0)$,
$\psi=1$; while for a half cell offset, $\psi$ is either $1$ or
$-1$, e.g. if $\Delta\bm{r_{\rho}}=(\Delta x/2,\,0,\,0)$, $\psi=(-1)^{n}$.

\section{Linear dispersion of electrostatic PIC models\label{sec:A3}}

The linear numerical dispersion of the momentum conserving and energy
conserving PIC models are given in \cite{Birdsall-Langdon} for finite
time step. Here we gives a similar derivation but without introducing
temporal aliasing, as all PIC quantities, including the distribution
function, are discrete in time. The perturbed discrete (in time) distribution
function is derived in a way similar to Eq. (30) of \cite{Meyers2015}
but with one full position update instead of two half updates,

\begin{equation}
\tilde{f_{1}}(\bm{k},\bm{v},\omega)=\dfrac{-i\Delta te^{-i(\omega-\bm{k}\cdot\bm{v})\Delta t/2}}{2}\text{csc}[(\omega-\bm{k}\cdot\bm{v})\Delta t/2]\tilde{\bm{F}}(\bm{k},\omega)\cdot\nabla_{\bm{v}}f_{0},\label{eq:discrete-leap-frog}
\end{equation}

\noindent for which a phase factor $e^{i\omega\Delta t/2}$ is dropped
because $f_{1}$ is now needed at the same time step as the electric
field. 

When the leap-frog update is viewed as a continuous time integration
of the Newton's laws but with a force sampled on the time step, temporal
aliasing is introduced. In such case the distribution function of
the eigen mode is defined on the continuous time, and the discrete
distribution function in PIC is its time sample. The linearization
is performed on the former and before the time sampling. Using $\underset{q}{\sum}\frac{1}{\omega-\bm{k}\cdot\bm{v}-2\pi q/\Delta t}=\frac{\Delta t}{2}\text{cot}[(\omega-\bm{k}\cdot\bm{v})\Delta t/2]$
to evaluate the infinite sum from aliasing \cite{Birdsall-Langdon},
one arrives at a similar expression to Eq. (\ref{eq:discrete-leap-frog})
for the perturbed discrete distribution function but with $e^{-i(\omega-\bm{k}\cdot\bm{v})\Delta t/2}\text{csc}[(\omega-\bm{k}\cdot\bm{v})\Delta t/2]=\text{cot}[(\omega-\bm{k}\cdot\bm{v})\Delta t/2]-i$
replaced by $\text{cot}[(\omega-\bm{k}\cdot\bm{v})\Delta t/2]$. This
difference does not manifest in the zeroth moment of $\tilde{f}_{1}$
but may appear in higher order moments. 

The normalized equations for the perturbed density $\tilde{\rho}_{1}(\bm{k},\omega)$,
the potential $\tilde{\phi}_{1}(\bm{k},\omega)$,the electric field
$\tilde{\bm{E}}_{1}(\bm{k},\omega)$ and the force $\tilde{\bm{F}}(\bm{k},\bm{v},\omega)$
are,

\begin{align}
\tilde{\rho}_{1}(\bm{k},\omega) & =\underset{\mathbf{q}}{\sum}\tilde{S}_{\rho}(\bm{k}_{q})\int\tilde{f_{1}}(\bm{k_{q}},\bm{v},\omega)d\bm{v},\label{eq:deposition-for-dispersion}\\
\tilde{\phi}_{1}(\bm{k},\omega) & =\tilde{\rho}_{1}(\bm{k},\omega)/[\bm{k}]{}^{2},\label{eq:poisson-for-dispersion}\\
\tilde{\bm{E}}_{1}(\bm{k},\omega) & =-i\bm{\kappa}(\bm{k})\tilde{\phi}_{1}(\bm{k},\omega),\label{eq:gradient-for-dispersion}\\
\tilde{\bm{F}}(\bm{k},\omega) & =\tilde{S}_{E}(\bm{k})\tilde{\bm{E}}_{1}(\bm{k},\omega),
\end{align}

\noindent where $[\bm{k}]{}^{2}$ is the Poisson operator and $\bm{\kappa}(\bm{k})$
is the gradient operator. All grid quantities are transformed according
to the transform defined in Eq. (\ref{eq:transform}). Eq. (\ref{eq:deposition-for-dispersion})
also implies that $\Delta\bm{r}_{\rho}=0$ is chosen in Eq. (\ref{eq:deposition-PIC-distribution}),
so $\psi(\mathbf{q}\cdot\bm{k_{g}},\,\Delta\bm{r_{\rho}})=1$.

Substituting Eq. (\ref{eq:discrete-leap-frog}) into the above equations,
the dispersion equation is,

\begin{equation}
1+\dfrac{1}{[\bm{k}]{}^{2}}\underset{\mathbf{q}}{\sum}\tilde{S}_{\rho}(\bm{k_{q}})\tilde{S}_{E}(\bm{k_{q}})\int\dfrac{\Delta te^{-i(\omega-\bm{k_{q}}\cdot\bm{v})\Delta t/2}}{2}\text{csc}[(\omega-\bm{k_{q}}\cdot\bm{v})\Delta t/2][\kappa]\cdot\nabla_{\bm{v}}f_{0}d\bm{v}=0.
\end{equation}

For a cold uniform beam with velocity $\bm{v}_{0}$, it reduces to,

\begin{equation}
1-\left(\dfrac{\Delta t}{2}\right)^{2}\dfrac{1}{[\bm{k}]{}^{2}}\underset{\mathbf{q}}{\sum}\dfrac{\tilde{S}_{\rho}(\bm{k_{q}})\tilde{S}_{E}(\bm{k_{q}})\bm{\kappa}(\bm{k_{q}})\cdot\bm{k_{q}}}{\text{sin}^{2}[(\omega-\bm{k_{q}}\cdot\bm{v}_{0})\Delta t/2]}=0.\label{eq:cold-dispersion-finite-dt}
\end{equation}

Eq. (\ref{eq:cold-dispersion-finite-dt}) has many resonances that
are folded into the fundamental Brillouin zone $-\pi<\omega\Delta t/2\le\pi$.
For two interacting resonances, the dispersion equation can be simplified
as $1-a[\text{sin}^{-2}(y)+b\text{sin}^{-2}(y+c)]=0$, $|b|<1$. Here
we use the 1D case and include the fundamental mode and the $q$th
($|q|>0$) alias mode as an example, $y=(\omega-k\cdot v_{0})\Delta t/2$,
$a=\Delta t^{2}\tilde{s}_{\rho}(k)\tilde{s}_{E}(k)\kappa(k)k/4[k]^{2}\ge0$,
$b=\tilde{s}_{\rho}(k_{q})\tilde{s}_{E}(k_{q})\kappa(k_{q})k_{q}/\tilde{s}_{\rho}(k)\tilde{s}_{E}(k)\kappa(k)k$,
and $c=(qk_{g}v_{0}\Delta t/2)\text{ mod }2\pi$, $-\pi<c\le\pi$.
When $\Delta t$ is sufficiently small, this equation can be approximated
by a quartic equation in $y$, $-ac^{2}+2acy+\left(-a-ab+c^{2}\right)y^{2}-2cy^{3}+y^{4}=0$,
for which a pair of complex conjugate roots (the other two roots are
always real) exist when the discriminant $D=-16a^{2}bc^{2}\left[(a(1+b)-c^{2})^{3}-27a^{2}bc^{2}\right]<0$.
When $D>0$, there are (1) four real roots, if $a>0$; or (2) two
pairs of complex conjugate roots, if $a<0$. Therefore, for the interaction
of any two nearby resonances, there are domains of stability/instability
defined by the parameters $a,b$ and $c$, depending on the signs
of $D$ and $a$. 

Two types of situation can be distinguished. When $b>0$, the simplified
dispersion equation resembles that of a physical two-stream instability.
Stability requires $(a(1+b)-c^{2})^{3}-27a^{2}bc^{2}\le0$ and $a>0$,
which gives a stable domain $0<a<a_{1}$ and unstable domains $a>a_{1}$
or $a<0$. Here $a_{1}$ is the only real root of $(a(1+b)-c^{2})^{3}-27a^{2}bc^{2}=0$.
For $|b|\ll1$, $a_{1}\approx\left(1-3b^{1/3}\right)c^{2}$. When
$b<0$, which corresponds to unusual two streams with densities of
opposite signs, the stable domain is $a>a_{1}$ and the unstable domain
is $a<a_{1}$.

The dispersion equation Eq. (\ref{eq:cold-dispersion-finite-dt})
applies to both the energy conserving and momentum conserving PIC
models. The major difference in Eq. (\ref{eq:cold-dispersion-finite-dt})
for these two PIC models is the operator $\kappa(k_{q})$. For the
energy conserving model, $\kappa(k_{q})=k_{q}$, so $b\sim k_{q}^{2}>0$.
Therefore, interaction of any pair of resonances is of the usual two-stream
type and there is a stable domain $0<a<a_{1}$. This can be written
explicitly, e.g., for the fundamental mode and the $q$th ($|q|>0$)
alias mode, $[(qk_{g}v_{0}\Delta t/2)\text{ mod }2\pi]^{2}>\frac{\Delta t^{2}}{4[k]^{2}}\tilde{s}_{\rho}(k)\tilde{s}_{E}(k)\kappa(k)k\left[1-3(\tilde{s}_{\rho}(k_{q})\tilde{s}_{E}(k_{q})\kappa(k_{q})k_{q}/\tilde{s}_{\rho}(k)\tilde{s}_{E}(k)k^{2})^{1/3}\right]^{-1}$.
For momentum conserving PIC model, $\kappa(k_{q})=k_{q}\text{sinc}(k_{q}\Delta x)$,
so $b\sim k_{q}^{2}\text{sinc}(k_{q}\Delta x)$ which switches sign
when $q$ switches sign. Therefore, interaction of any pair of resonances
can be of either the first or the second type.

\section*{References}

\bibliographystyle{elsarticle-num}
\bibliography{refs}

\end{document}